\documentstyle[epsfig,12pt]{article}
\begin{document}
\bigskip \begin{flushright}
WATPHYS-TH01/01
\end{flushright}

\onecolumn

\begin{titlepage}
\begin{center}
{\LARGE \bf Statistical Mechanics of Relativistic One-Dimensional Self-Gravitating Systems }
\\ \vspace{2cm}
R.B. Mann \footnotemark\footnotetext{email: 
mann@avatar.uwaterloo.ca} 
and P. Chak 
\\
\vspace{1cm}
Dept. of Physics,
University of Waterloo
Waterloo, ONT N2L 3G1, Canada\\
PACS numbers: 
5.20.-y, 04.40.-b 5.90.+m\\
\vspace{2cm}
\today\\
\end{center}

\begin{abstract}
We consider the statistical mechanics of a general relativistic one-dimensional 
self-gravitating system.   The system consists of $N$-particles coupled to
lineal gravity and can be considered as a model of $N$ relativistically interacting
sheets of uniform mass.  The partition function and one-particle distitrubion functions
are computed to leading order in $1/c$ where $c$ is the speed of light; as 
$c\to\infty$ results for the non-relativistic one-dimensional self-gravitating system are
recovered.  We find that relativistic effects generally cause 
both position and momentum distribution functions to become more sharply 
peaked, and that the temperature of a relativistic gas is  smaller than its
non-relativistic counterpart at the same fixed energy.  We consider the large-$N$ 
limit of our results and compare this to the non-relativistic case.

\end{abstract}
\end{titlepage}\onecolumn

\section{Introduction}

One-dimensional systems of $N$ particles mutually interacting through
gravitational forces have been of interest in astrophysics for more than
three decades. \ While used primarily as prototypes for the behaviour of
gravity in higher dimensions, one-dimensional self-gravitating systems
(OGS's) also approximate the behaviour of some physical systems in 3 spatial
dimensions. These include the dynamics of stars in a direction orthogonal to
the plane of a highly flattened galaxy and the collisions of flat parallel
domain walls moving in directions perpendicular to their surfaces.
Furthermore, very long-lived core-halo structures in the OGS phase space are
known to exist, reminiscent of structures observed in globular clusters, in
which a dense massive core in near equilibrium is surrounded by a halo of
stars with high kinetic energy that interact only weakly with the core \cite%
{yawn}.

The statistical properties of the OGS are particularly intriguing. \ Despite
extensive study, many unanswered questions remain. For example it is not
clear if the OGS can attain a true equilibrium state from arbitrary initial
conditions. \ Its ergodic and equipartition properties are still not well
understood. \ This is primarily because the particle interactions of the OGS
(as with any self-gravitating system) are attractive and cumulatively long
range, in strong contrast to typical thermodynamic systems for which such
interactions are repulsive and short range. For the OGS the macroscopic \
dynamics does not decouple from the microscopic dynamics, and the usual
thermodynamic analysis does not apply. \ 

However there are some established features of the OGS. Rybicki \cite%
{Rybicki} derived in closed form the single-particle distribution function
in both the canonical and microcanonical ensembles. \ \ In the large $N$
limit these distribution functions reduce to the isothermal solution of the
Vlasov equation.

All studies to date have neglected relativistic effects. This limitation is
understandable since no relativistic $N$ particle Hamiltonian was available
for analysis. \ However this situation changed recently when a prescription
for obtaining the Hamiltonian for a relativistic one-dimensional
self-gravitating system (ROGS) was given by Mann and Ohta \cite{OR}. \ This
Hamiltonian can be rigorously derived from a generally covariant system
coupling relativistic gravity in one spatial dimension (i.e. a 1+1
dimensional theory of gravity \cite{r3}) to $N$ point particles. In the
non-relativistic limit, the Hamiltonian reduces to that of the OGS. \
Although not available in closed form, the Hamiltonian can be obtained as a
series expansion in inverse powers of the speed of light $\ c$ to arbitary
order. \ 

We consider in this paper the one-particle distribution function for the
ROGS. \ Our work here is a natural extension of previous work on the $N$%
-body problem in relativistic gravity. In three spatial dimensions an exact
solution to this problem is known for pure Newtonian gravity (and a series
solutions has been constructed for arbitrary $N$). In the general theory of
relativity dissipation of energy in the form of gravitational radiation has
obstructed progress toward obtaining exact solutions to the $N$-body problem
even when $N=2$. \ However for the\ ROGS an exact solution to the $2$-body
problem was recently obtained \cite{2bd}, and generalizations including a
cosmological constant and/or charge subsequently followed \cite%
{2bdcossh,2bdcoslo,2bdchglo}. These solutions include both an explicit
expression for the proper separtion of the two bodies as a function of time
and an explicit expression for the Hamiltonian for the $2$-body ROGS as a
function of the proper separation and the centre-of-inertia momentum of the
bodies.

Encouraged by these results, we here begin a first attempt to understand the
basic features of the $N$-body ROGS. \ We shall recapitulate the canonical
formalism used in ref. \cite{OR} to derive the Hamiltonian for the $N$-body
ROGS. \ We then compute the partition function and canonical distribution
functions.\ Using an integral transform we then calculate the microcanonical
distribution functions. All results are in closed form to leading order in $%
1/c$. We consider the limit of large $N$ and compare the ROGS and the OGS.
We close with a few remarks. Lengthy intermediate calculations are confined
to appendices.

\section{Canonical $N$-particle Hamiltonian of the ROGS}

\bigskip The OGS Hamiltonian is 
\begin{eqnarray}
H &=&\sum_{a}\frac{p_{a}^{2}}{2m_{a}}+\pi G\sum_{a}\sum_{b}m_{a}m_{b}\mid
z_{a}-z_{b}\mid  \nonumber \\
&=&\sum_{a}\frac{p_{a}^{2}}{2m_{a}}+2\pi G\sum_{a>b}m_{a}m_{b}\mid
z_{a}-z_{b}\mid  \label{ogs}
\end{eqnarray}%
where the summation is over all $N$ particles, located at positions $z_{a}$
along the spatial axis. \ The potential term straightforwardly follows upon
solving the Newtonian equation 
\begin{equation}
\nabla ^{2}\varphi =4\pi G\rho  \label{Neq}
\end{equation}%
in one spatial dimension, where $\rho =\sum_{a}m_{a}\delta (x-z_{a})$ is the
mass density of the $N$ point particles. Our task in this section is to find
a prescription for obtaining a relativistic generalization of (\ref{ogs}).

The Hamiltonian for the ROGS that we use is that of a $\left( 1+1\right) $
dimensional theory (a lineal gravity theory) that models $\left( 3+1\right) $
dimensional general relativity in that it sets the Ricci scalar equal to the
trace of the stress-energy of prescribed matter fields and sources. Hence
matter governs the evolution of spacetime curvature which reciprocally
governs the evolution of matter \cite{r3}. We refer to this theory as $R=T$
theory. Apart from being able to model a number of textbook scenarios in
general relativity \cite{Sharon}, it has the attractive feature of having a
consistent Newtonian limit \cite{r3}. This limit, essential for our
purposes, is problematic in a generic $(1+1)$-dimensional theory of gravity
theory \cite{jchan}.

Since the Einstein action is a topological invariant in $(1+1)$ dimensions,
a scalar (dilaton) field must be included in the action \cite{BanksMann}. \
Its coupling to the curvature is chosen so that only the trace of the stress
energy of matter ($N$ point particles here) is set equal to the Ricci
scalar. This action will form the basis for the ROGS we consider. Upon
canonical reduction of the action \cite{OR}, the ROGS Hamiltonian is given
in terms of a spatial integral of the second derivative of the dilaton
field, which is a function of the coordinates and momenta of the particles
and is determined from the constraint equations.

\bigskip

The action integral for the gravitational field coupled to $N$ point
particles is 
\begin{eqnarray}
I &=&\int d^{2}x\left[ \frac{1}{2\kappa }\sqrt{-g}g^{\mu \nu }\left\{ \Psi
R_{\mu \nu }+\frac{1}{2}\nabla _{\mu }\Psi \nabla _{\nu }\Psi \right\}
\right.  \nonumber \\
&&\makebox[2em]{}\left. +\sum_{a}\int d\tau _{a}\left\{ -m_{a}\left( -g_{\mu
\nu }(x)\frac{dz_{a}^{\mu }}{d\tau _{a}}\frac{dz_{a}^{\nu }}{d\tau _{a}}%
\right) ^{1/2}\right\} \delta ^{2}(x-z_{a}(\tau _{a}))\right] \;,
\label{act1}
\end{eqnarray}
where $\Psi $ is the dilaton field, $g_{\mu \nu }$ and $g$ are the metric
and its determinant, $R$ is the Ricci scalar, and $\tau _{a}$ is the proper
time of $a$-th particle whose mass is $m_{a}$, with $\kappa =8\pi G/c^{4}$.
We use $\nabla _{\mu }$ to denote the covariant derivative associated with $%
g_{\mu \nu }$.

\bigskip The field equations derived from the action (\ref{act1}) are 
\begin{eqnarray}
&&R-g^{\mu \nu }\nabla _{\mu }\nabla _{\nu }\Psi =0\;,  \label{eq-R} \\
&&\frac{1}{2}\nabla _{\mu }\Psi \nabla _{\nu }\Psi -\frac{1}{4}g_{\mu \nu
}\nabla ^{\lambda }\Psi \nabla _{\lambda }\Psi +g_{\mu \nu }\nabla ^{\lambda
}\nabla _{\lambda }\Psi -\nabla _{\mu }\nabla _{\nu }\Psi =\kappa T_{\mu \nu
}+\frac{1}{2}g_{\mu \nu }\Lambda \;,  \label{eq-Psi} \\
&&\frac{d}{d\tau _{a}}\left\{ g_{\mu \nu }(z_{a})\frac{dz_{a}^{\nu }}{d\tau
_{a}}\right\} -\frac{1}{2}g_{\nu \lambda ,\mu }(z_{a})\frac{dz_{a}^{\nu }}{%
d\tau _{a}}\frac{dz_{a}^{\lambda }}{d\tau _{a}}=0\;,  \label{eq-z}
\end{eqnarray}
where 
\begin{equation}
T_{\mu \nu }=\sum_{a}m_{a}\int d\tau _{a}\frac{1}{\sqrt{-g}}g_{\mu \sigma
}g_{\nu \rho }\frac{dz_{a}^{\sigma }}{d\tau _{a}}\frac{dz_{a}^{\rho }}{d\tau
_{a}}\delta ^{2}(x-z_{a}(\tau _{a}))\;,
\end{equation}
is the stress-energy due to the point masses. Eq.(\ref{eq-Psi}) guarantees
the conservation of $T_{\mu \nu }$. Inserting the trace of Eq.(\ref{eq-Psi})
into Eq.(\ref{eq-R}) yields 
\begin{equation}
R-\Lambda =\kappa T_{\;\;\mu }^{\mu }\;.  \label{RT}
\end{equation}
Eqs. (\ref{eq-Psi}), (\ref{eq-z}) and (\ref{RT}) form a closed sytem of
equations for gravity and matter.

In order to obtain the Hamiltonian in canonical form, we first decompose the
scalar curvature in terms of the extrinsic curvature $K$ via 
\begin{equation}
\sqrt{-g}R=-2\partial _{0}(\sqrt{\gamma }K)+2\partial _{1}[(N_{1}K-\partial
_{1}N_{0})/\sqrt{\gamma }]\;,  \label{extK}
\end{equation}
where the metric is 
\begin{equation}
ds^{2}=-N_{0}^{2}dt^{2}+\gamma \left( dx+\frac{N_{1}}{\gamma }dt\right)
^{2}\;,  \label{lineel}
\end{equation}
with $K=(2N_{0}\gamma )^{-1}(2\partial _{1}N_{1}-\gamma ^{-1}N_{1}\partial
_{1}\gamma -\partial _{0}\gamma )$, so that $\gamma
=g_{11},N_{0}=(-g^{00})^{-1/2}$ and $N_{1}=g_{10}$. Rewriting the action (%
\ref{act1}) in first-order form yields 
\begin{equation}
I=\int d^{2}x\left\{ \sum_{a}p_{a}\dot{z}_{a}\delta (x-z_{a}(t))+\pi \dot{%
\gamma}+\Pi \dot{\Psi}+N_{0}R^{0}+N_{1}R^{1}\right\}  \label{act2}
\end{equation}
where $\pi $ and $\Pi $ are the respective conjugate momenta to $\gamma $
and $\Psi $. Here 
\begin{eqnarray}
R^{0} &=&-\kappa \sqrt{\gamma }\gamma \pi ^{2}+2\kappa \sqrt{\gamma }\pi \Pi
+\frac{1}{4\kappa \sqrt{\gamma }}(\Psi ^{\prime })^{2}-\frac{1}{\kappa }%
\left( \frac{\Psi ^{\prime }}{\sqrt{\gamma }}\right) ^{\prime }  \nonumber \\
&&-\sum_{a}\sqrt{\frac{p_{a}^{2}}{\gamma }+m_{a}^{2}}\;\delta (x-z_{a}(t))\;,
\nonumber  \label{R0} \\
R^{1} &=&\frac{\gamma ^{\prime }}{\gamma }\pi -\frac{1}{\gamma }\Pi \Psi
^{\prime }+2\pi ^{\prime }+\sum_{a}\frac{p_{a}}{\gamma }\delta
(x-z_{a}(t))\;,  \label{R1}
\end{eqnarray}
with the symbols $(\;\dot{}\;)$ and $(\;^{\prime }\;)$ denoting $\partial
_{0}$ and $\partial _{1}$, respectively.

Variation of the action (\ref{act2}) yields the set of equations 
\begin{eqnarray}
\dot{\pi} &+&N_{0}\left\{ \frac{3\kappa }{2}\sqrt{\gamma }\pi ^{2}-\frac{%
\kappa }{\sqrt{\gamma }}\pi \Pi +\frac{1}{8\kappa \sqrt{\gamma }\gamma }%
(\Psi ^{\prime })^{2}-\sum_{a}\frac{p_{a}^{2}}{2\gamma ^{2}\sqrt{\frac{%
p_{a}^{2}}{\gamma }+m_{a}^{2}}}\;\delta (x-z_{a}(t))\right.  \nonumber
\label{e-pi} \\
&+&N_{1}\left\{ -\frac{1}{\gamma ^{2}}\Pi \Psi ^{\prime }+\frac{\pi ^{\prime
}}{\gamma }+\sum_{a}\frac{p_{a}}{\gamma ^{2}}\;\delta (x-z_{a}(t))\right\}
+N_{0}^{\prime }\frac{1}{2\kappa \sqrt{\gamma }\gamma }\Psi ^{\prime
}+N_{1}^{\prime }\frac{\pi }{\gamma }=0\;,
\end{eqnarray}%
\begin{eqnarray}
&&\dot{\gamma}-N_{0}(2\kappa \sqrt{\gamma }\gamma \pi -2\kappa \sqrt{\gamma }%
\Pi )+N_{1}\frac{\gamma ^{\prime }}{\gamma }-2N_{1}^{\prime }=0\;,
\label{e-gamma} \\
&&R^{0}=0\;,  \label{e-R0} \\
&&R^{1}=0\;,  \label{e-R1} \\
&&\dot{\Pi}+\partial _{1}(-\frac{1}{\gamma }N_{1}\Pi +\frac{1}{2\kappa \sqrt{%
\gamma }}N_{0}\Psi ^{\prime }+\frac{1}{\kappa \sqrt{\gamma }}N_{0}^{\prime
})=0\;,  \label{e-Pi} \\
&&\dot{\Psi}+N_{0}(2\kappa \sqrt{\gamma }\pi )-N_{1}(\frac{1}{\gamma }\Psi
^{\prime })=0\;,  \label{e-Psi} \\
&&\dot{p}_{a}+\frac{\partial N_{0}}{\partial z_{a}}\sqrt{\frac{p_{a}^{2}}{%
\gamma }+m_{a}^{2}}-\frac{N_{0}}{2\sqrt{\frac{p_{a}^{2}}{\gamma }+m_{a}^{2}}}%
\frac{p_{a}^{2}}{\gamma ^{2}}\frac{\partial \gamma }{\partial z_{a}}-\frac{%
\partial N_{1}}{\partial z_{a}}\frac{p_{a}}{\gamma }  \nonumber \\
&&\makebox[2em]{}+N_{1}\frac{p_{a}}{\gamma ^{2}}\frac{\partial \gamma }{%
\partial z_{a}}=0\;,  \label{e-p} \\
&&\dot{z_{a}}-N_{0}\frac{\frac{p_{a}}{\gamma }}{\sqrt{\frac{p_{a}^{2}}{%
\gamma }+m_{a}^{2}}}+\frac{N_{1}}{\gamma }=0\;.  \label{e-z}
\end{eqnarray}%
All metric components ($N_{0}$, $N_{1}$, $\gamma $) in equations (\ref{e-p})
and (\ref{e-z}) are evaluated at the point $x=z_{a}$, where 
\[
\frac{\partial f}{\partial z_{a}}\equiv \left. \frac{\partial f(x)}{\partial
x}\right| _{x=z_{a}}\;. 
\]%
The quantities $N_{0}$ and $N_{1}$ are Lagrange multipliers which yield the
constraint equations (\ref{e-R0}) and (\ref{e-R1}). The above set of
equations can be proved to be equivalent to the set of equations (\ref{eq-R}%
), (\ref{eq-Psi}) and (\ref{eq-z}) \cite{OR}. \ 

\bigskip

An examination of the generator of space and time transformations \cite%
{OR,2bd} indicates that we find that we can consistently choose the
coordinate conditions 
\begin{equation}
\gamma =1\makebox[2em]{}\mbox{and}\makebox[2em]{}\Pi =0\;  \label{cc}
\end{equation}%
upon which the action (\ref{act2}) reduces to 
\begin{equation}
I=\int d^{2}x\left\{ \sum_{a}p_{a}\dot{z}_{a}\delta (x-z_{a})-{\cal H}%
\right\} \;,
\end{equation}%
after elimination of the constraints, where 
\begin{equation}
H=\int dx{\cal H=-}\frac{1}{\kappa }\int {\cal \triangle }\Psi {\cal \;}
\label{ham1}
\end{equation}%
is the Hamiltonian for the ROGS.

The field $\Psi $ is no longer arbitrary, but is instead a function of $%
z_{a} $ and $p_{a}$ that is determined by solving the constraints which are
now 
\begin{equation}
\triangle \Psi -\frac{1}{4}(\Psi ^{\prime })^{2}+\kappa ^{2}\pi ^{2}+\kappa
\sum_{a=1}^{N}\sqrt{p_{a}^{2}+m_{a}^{2}}\delta (x-z_{a})=0\;,  \label{Psi}
\end{equation}
\begin{equation}
2\pi ^{\prime }+\sum_{a=1}^{N}p_{a}\delta (x-z_{a})=0\;.  \label{pi}
\end{equation}
once the coordinate conditions (\ref{cc}) are imposed. \ Equation (\ref{Psi}%
) is an energy-balance equation which states that the energy of the
particles plus the (negative) gravitational energy must vanish. \ Equation (%
\ref{pi}) states that the total momentum of the gravitational field and the
particles must vanish. \ The consistency of this canonical reduction was
proved in ref. \cite{OR} by showing that the canonical equations of motion
derived from the reduced Hamiltonian (\ref{ham1}) are identical with the
equations (\ref{e-p}) and (\ref{e-z}) .

Hence the Hamiltonian (\ref{ham1}) is a function only of the coordinates and
momenta of the $N$ particles in the system. \ We turn next to its evaluation.

\bigskip

\section{Computation of the ROGS Hamiltonian}

\bigskip

Although the constraint equations are straightforward to solve in the
regions between the particles, the matching conditions of these solutions at
the juncture of the particles are quite non-trivial. For the $2$-body ROGS
their enforcement yields an equation which determines the Hamiltonian in
terms of the remaining degrees of freedom of the system. While this
procedure holds in principle for the $N$-body ROGS, we have not found a
tractable means of obtaining an analogous determining equation for the
Hamiltonian.

However it is possible to straightforwardly and rigourously construct
approximation schemes for computing the ROGS Hamiltonian for $N$ particles.
For example, the post-linear approximation is an expansion of the
Hamiltonian in powers of the gravitational coupling $\kappa $, obtained by
writing 
\begin{eqnarray}
\Psi &=&\kappa \Psi ^{(1)}+\kappa ^{2}\Psi ^{(2)}+\cdot \cdot \cdot
\label{psiexp} \\
\chi &=&\chi ^{(0)}+\kappa \chi ^{(1)}+\cdot \cdot \cdot \;\;.
\label{chiexp}
\end{eqnarray}%
where$\chi $ is defined by $\chi ^{\prime }\equiv \pi $. Insertion of these
expansions into eqs. (\ref{Psi},\ref{pi}) yields 
\begin{eqnarray}
H^{(2)} &=&\sum_{a}\sqrt{p_{a}^{2}+m_{a}^{2}}+\frac{\kappa }{8}%
\sum_{a}\sum_{b}\left( \sqrt{p_{a}^{2}+m_{a}^{2}}\sqrt{p_{b}^{2}+m_{b}^{2}}%
-p_{a}p_{b}\right) \left| r_{ab}\right|  \nonumber \\
&&+\frac{\epsilon \kappa }{8}\sum_{a}\sum_{b}\left( \sqrt{p_{a}^{2}+m_{a}^{2}%
}\;p_{b}-p_{a}\sqrt{p_{b}^{2}+m_{b}^{2}}\right) r_{ab}  \nonumber \\
&&+\frac{1}{4}\left( \frac{\kappa }{4}\right) ^{2}\left\{ \sum_{a}\sqrt{%
p_{a}^{2}+m_{a}^{2}}\left[ \sum_{b}p_{b}\left| r_{ab}\right| +\epsilon
\;\sum_{b}\sqrt{p_{b}^{2}+m_{b}^{2}}r_{ab}\right] ^{2}\right.  \nonumber \\
&&\left. -\sum_{a}p_{a}\left[ \sum_{b}p_{b}\left| r_{ab}\right| +\epsilon
\;\sum_{b}\sqrt{p_{b}^{2}+m_{b}^{2}}r_{ab}\right] \left[ \sum_{c}\sqrt{%
p_{c}^{2}+m_{c}^{2}}\;\left| r_{ac}\right| +\epsilon \;\sum_{c}p_{c}r_{ac}%
\right] \right.  \nonumber \\
&&\left. +\sum_{a}\sum_{b}\left[ \sqrt{p_{a}^{2}+m_{a}^{2}}\sqrt{%
p_{b}^{2}+m_{b}^{2}}\;\left| r_{ab}\right| -\epsilon \;p_{a}\sqrt{%
p_{b}^{2}+m_{b}^{2}}r_{ab}\right] \right.  \nonumber \\
&&\makebox[8em]{}\left. \times \left[ \sum_{c}\sqrt{p_{c}^{2}+m_{c}^{2}}%
\;\left| r_{bc}\right| +\epsilon \;\sum_{c}p_{c}r_{bc}\right] \right. 
\nonumber \\
&&\left. -\sum_{a}\sum_{b}\left[ \sqrt{p_{a}^{2}+m_{a}^{2}}\;p_{b}\left|
r_{ab}\right| -\epsilon \;p_{a}p_{b}r_{ab}\right] \left[ \sum_{c}p_{c}\left|
r_{bc}\right| +\epsilon \;\sum_{c}\sqrt{p_{c}^{2}+m_{c}^{2}}r_{bc}\right]
\right\} \;\;  \nonumber \\
&&  \label{plinH}
\end{eqnarray}%
upon insertion into (\ref{ham1}), where $r_{ab}=z_{a}-z_{b}$ is the relative
separation between particles $a$ and $b$. It can be shown that the solutions
to eqs. (\ref{Psi},\ref{pi}) must satisfy the boundary condition $\Psi
^{2}-4\kappa ^{2}\chi ^{2}=0$ in the regions $\mid x\mid >>\mid z_{a}\mid $
in order for the Hamiltonian to be finite \cite{OR}.

The $\kappa $-expansion is appropriate for describing relativistic
fast-motion of the particles and can be carried out to any desired order.
However to compare the ROGS from $R=T$ theory with the OGS, we turn to the
post-Newtonian expansion, which is an expansion of the Hamiltonian in powers
of $c^{-1}$. Since both $p_{a}^{2}/m_{a}^{2}$ and $\sqrt{\kappa }$ are of
the order of $c^{-2}$ all terms up to the order of $c^{-4}$ are included in
the post-linear Hamiltonian (\ref{plinH}). \ The post-Newtonian Hamiltonian
to this order is therefore \cite{OR} 
\begin{eqnarray}
H &=&\sum_{a=1}^{N}m_{a}c^{2}+\sum_{a=1}^{N}\frac{p_{a}^{2}}{2m_{a}}+2\pi
G\sum_{a>b}^{N}m_{a}m_{b}\left| r_{ab}\right|  \nonumber \\
&&-\frac{1}{c^{2}}\sum_{a=1}^{N}\frac{p_{a}^{4}}{8m_{a}^{3}}+\frac{\pi G}{%
c^{2}}\sum_{a=1}^{N}\sum_{b=1}^{N}m_{a}\frac{p_{b}^{2}}{m_{b}}\left|
r_{ab}\right| -\frac{2\pi G}{c^{2}}\sum_{a>b}^{N}p_{a}p_{b}\left|
r_{ab}\right|  \nonumber \\
&&+\left( \frac{\pi G}{c}\right)
^{2}\sum_{a=1}^{N}\sum_{b=1}^{N}\sum_{c=1}^{N}m_{a}m_{b}m_{c}\left[ \left|
r_{ab}\right| \left| r_{ac}\right| -r_{ab}r_{ac}\right] +\cdots  \label{rogs}
\end{eqnarray}%
where the explicit powers of $c$ have been restored.

The first term in (\ref{rogs}) is the total rest energy of the particles,
and the second two terms are the OGS Hamiltonian (\ref{ogs}). \ The
remaining terms are all relativistic corrections to the OGS to order $c^{-2}$%
. \ The first of these corrections is a special-relativistic one, whereas
the remaining corrections are due to relativistic gravity in one spatial
dimension. \ Note that gravity not only modifies the potential to a
quadratic form, but also includes couplings between particle momenta and
their positions. These features -- modifications of the distance behaviour
of the potential and position-momentum couplings -- are fully analogous to
those in general relativity in three spatial dimensions.

We next find the equations of motion for the position of the $a$th particle.
This follows straightforwardly from Hamilton's principle. We have 
\begin{eqnarray}
\dot{z}_{a} &=&\frac{\partial H}{\partial p_{a}}  \nonumber \\
&=&\frac{p_{a}}{m_{a}}-\frac{1}{c^{2}}\left[ \frac{p_{a}^{3}}{2m_{a}^{3}}-%
\frac{\kappa }{4}\sum_{b=1}^{N}m_{b}\frac{p_{a}}{m_{a}}\left| r_{ab}\right| +%
\frac{\kappa }{4}\sum_{b=1}^{N}p_{b}\left| r_{ab}\right| \right]  \label{e2}
\end{eqnarray}
and 
\begin{eqnarray}
\dot{p}_{a} &=&-\frac{\partial H}{\partial z_{a}}  \nonumber \\
&=&-2\pi G\sum_{b=1}^{N}m_{a}m_{b}\varepsilon _{ab}-\frac{1}{c^{2}}\left[
\pi G\sum_{b=1}^{N}\left( m_{a}\frac{p_{b}^{2}}{m_{b}}+m_{b}\frac{p_{a}^{2}}{%
m_{a}}\right) \varepsilon _{ab}-2\pi G\sum_{b=1}^{N}p_{a}p_{b}\varepsilon
_{ab}\right.  \nonumber \\
&&\left. +2\left( \pi G\right)
^{2}\sum_{b=1}^{N}\sum_{c=1}^{N}m_{a}m_{b}m_{c}\left[ \varepsilon
_{ab}\left| r_{ac}\right| +\varepsilon _{ab}\left| r_{bc}\right| -r_{ab}%
\right] \right]  \label{e3}
\end{eqnarray}
where $\varepsilon _{ab}=\left\{ 
\begin{array}{cc}
1 & z_{a}>z_{b} \\ 
-1 & z_{a}<z_{b}%
\end{array}
\right. $ . \ We can solve (\ref{e2}) for $p_{a}$: 
\begin{equation}
p_{a}=m_{a}\dot{z}_{a}+\frac{1}{c^{2}}\left[ m_{a}\frac{\dot{z}_{a}^{3}}{2}%
+2\pi G\sum_{b=1}^{N}\left( \left( m_{b}m_{a}\dot{z}_{a}-m_{a}m_{b}\dot{z}%
_{b}\right) \left| r_{ab}\right| \right) \right] +\cdots  \label{e4}
\end{equation}
and then insert this into the equation (\ref{e3}) for $\dot{p}_{a}:$%
\begin{eqnarray}
m_{a}\ddot{z}_{a}+ &&\frac{1}{c^{2}}\left[ m_{a}\ddot{z}_{a}\frac{3\dot{z}%
_{a}^{2}}{2}-2\pi G\sum_{b=1}^{N}\left( \left( m_{b}m_{a}\ddot{z}%
_{a}-m_{a}m_{b}\ddot{z}_{b}\right) \left| r_{ab}\right| +\left( m_{b}m_{a}%
\dot{z}_{a}-m_{a}m_{b}\dot{z}_{b}\right) \varepsilon _{ab}\left( \dot{z}_{a}-%
\dot{z}_{b}\right) \right) \right]  \nonumber \\
&=&-2\pi G\sum_{b=1}^{N}m_{a}m_{b}\varepsilon _{ab}-\frac{1}{c^{2}}\left[
\pi G\sum_{b=1}^{N}\left( m_{a}m_{b}\left[ \dot{z}_{b}^{2}+\dot{z}_{a}^{2}%
\right] \right) \varepsilon _{ab}-2\pi G\sum_{b=1}^{N}m_{a}m_{b}\dot{z}_{a}%
\dot{z}_{b}\varepsilon _{ab}\right.  \nonumber \\
&&\left. +2\left( \pi G\right)
^{2}\sum_{b=1}^{N}\sum_{c=1}^{N}m_{a}m_{b}m_{c}\left[ \varepsilon
_{ab}\left| r_{ac}\right| +\varepsilon _{ab}\left| r_{bc}\right| -r_{ab}%
\right] \right]  \label{e5}
\end{eqnarray}
which simplifies to 
\begin{eqnarray}
m_{a}\ddot{z}_{a} &=&-2\pi G\sum_{b=1}^{N}m_{a}m_{b}\varepsilon _{ab}+\frac{%
\pi G}{c^{2}}\sum_{b=1}^{N}\left( m_{a}m_{b}\left\{ 3\dot{z}_{a}^{2}+\left[ 
\dot{z}_{b}-\dot{z}_{a}\right] ^{2}\right\} \varepsilon _{ab}\right) 
\nonumber \\
&&-\left( \frac{2\pi G}{c}\right)
^{2}\sum_{b=1}^{N}\sum_{c=1}^{N}m_{a}m_{b}m_{c}\left( \varepsilon
_{ac}-\varepsilon _{bc}\right) \left| r_{ab}\right|  \nonumber \\
&&-2\left( \frac{\pi G}{c}\right)
^{2}\sum_{b=1}^{N}\sum_{c=1}^{N}m_{a}m_{b}m_{c}\left[ \varepsilon
_{ab}\left| r_{ac}\right| +\varepsilon _{ab}\left| r_{bc}\right| -r_{ab}%
\right]  \label{e6}
\end{eqnarray}
upon an iterative substitution of $\ddot{z}_{a}$ in powers of $c^{-1}$.
These equations of motion reduce to those of the OGS in the limit $%
c\rightarrow \infty $ , and may be shown to be equivalent to the geodesic
equations to this order \cite{OR}.

\bigskip

We wish to investigate the instrinsic structure of the system described by
the Hamiltonian (\ref{rogs}) . \ However because of the translation
invariance of the system, two phase-space degrees of freedom are redundant,
and so must be factored out; otherwise certain average properties such as
density would be uniform throughout space.

Using equation(\ref{e3}), it is straightforward to show that 
\begin{equation}
\sum_{a=1}^{N}\dot{p}_{a}=0  \label{e7}
\end{equation}
and (using also (\ref{e2}) ) that the Hamiltonian is time-independent. \
This means that we can perform the phase-space integration subject to the
constraint 
\begin{equation}
\bar{p}=0\mbox{\ \ \ \ \ \ \ \ \ \ \ where \ \ \ \ \ \ }\overline{p}\equiv
\sum_{a=1}^{N}p_{a}  \label{e8}
\end{equation}
since we can choose a frame of reference in which the centre of inertia is
constant.

Removing the redundant position degree-of-freedom is somewhat more delicate.
Although the system is invariant under the translation $z_{a}\rightarrow
z_{a}+\hat{z}$, eq. (\ref{e4}) implies 
\begin{equation}
\overline{p}\equiv \frac{1}{N}\mbox{\ }\sum_{a=1}^{N}p_{a}=\frac{1}{N}\mbox{%
\ }\sum_{a=1}^{N}\left[ m_{a}\dot{z}_{a}\left( 1+\frac{1}{2}\frac{\dot{z}%
_{a}^{2}}{c^{2}}\right) \right]  \label{e9}
\end{equation}
which cannot be written as a total time derivative. \ Physically, the centre
of inertia is the relativistically well-defined concept, whereas the centre
of mass is not. \ However we can deal with this problem by inserting a
factor of unity in all phase-space averages in the form 
\[
\int_{-\infty }^{\infty }dL\,\delta \left( \overline{z}-L\right) 
\]
where $\overline{z}\equiv \frac{1}{M}\sum_{a=1}^{N}m_{a}z_{a}$ , with $M=$ $%
\sum_{a=1}^{N}m_{a}$.\ \ If the $L$-dependence of any integral trivially
factors out (or can be removed by a shift of variable in the integrand),
then we regard the remaining quantity as the physically relevant one to
describe the system.

For a canonical ensemble, all phase-space averages are carried out with a
weighting function $\exp \left[ -\beta H\right] $, where $k_{B}T=\beta ^{-1}$
is the temperature multiplied by Boltzmann's constant $k_{B}$. \ For the
microcanonical ensemble, an additional constraint of fixed total energy 
\[
H(z_{a},p_{a})=E 
\]
must be included, consistent with the time-independence of the Hamiltonian (%
\ref{rogs}) . \ Since the system is in momentum isolation, it \ is difficult
to see how it can be in energy contact with a heat bath, and so the physical
relevance of the canonical ensemble is somewhat unclear. \ However an
evaluation of quantities within the canonical ensemble is instructive in its
own right and is a necessary preliminary to computing quantities in the more
realistic microcanonical ensemble, and so we include it in the present
discussion.

Henceforth we set $m_{a}=m$, so that $M=Nm$.

\section{The Canonical Ensemble}

\bigskip

We consider in this section the relativistic corrections to the canonical
one-particle distribution function $f_{c}^{R}(p,z)$, which is defined to be
the phase-space average of the quantity 
\begin{equation}
\frac{1}{N}\sum_{a}\delta \left( z-z_{a}\right) \delta \left( p-p_{a}\right)
\label{can1}
\end{equation}%
weighted by $\exp \left( -\beta H\right) $ with the constraint $\hat{p}=0.$
Hence 
\begin{equation}
f_{c}^{R}(p,z)=\frac{1}{{\cal Z}N!}\int \int d{\bf p}d{\bf z}\delta \left( 
\overline{p}\right) \int_{-\infty }^{\infty }dL\,\delta \left( \overline{z}%
-L\right) \exp \left( -\beta H\right) N^{-1}\sum_{a}\delta \left(
z-z_{a}\right) \delta \left( p-p_{a}\right)  \label{can2}
\end{equation}%
where 
\begin{equation}
{\cal Z}=\frac{1}{N!}\int \int d{\bf p}d{\bf z}\delta \left( \overline{p}%
\right) \int_{-\infty }^{\infty }dL\,\delta \left( \overline{z}-L\right)
\exp \left( -\beta H\right)  \label{can4}
\end{equation}%
is the partition function and where the 2nd line in (\ref{can2}) follows
from the indistinguishability of the particles. \ Note that a shift of
integration variable 
\[
z_{a}^{\prime }=z_{a}+L 
\]%
renders the partition function in the form 
\begin{equation}
{\cal Z}=\frac{1}{N!}\int_{-\infty }^{\infty }dL\,\int \int d{\bf p}d{\bf z}%
^{\prime }\delta \left( \overline{p}\right) \delta \left( \overline{z}%
^{\prime }\right) \exp \left( -\beta H\right)  \label{can4a}
\end{equation}%
where the $L$-dependence is seen to trivially factor out. It can therefore
be dropped (along with the prime notation) from further consideration in the
evaluation of ${\cal Z}$. Similarly, the single-particle distribution
function becomes 
\begin{equation}
f_{c}^{R}(p,z)=\frac{1}{{\cal Z}N!}\int_{-\infty }^{\infty }dL\,\int \int d%
{\bf p}d{\bf z}\delta \left( \overline{p}\right) \delta \left( \overline{z}%
^{\prime }\right) \exp \left( -\beta H\right) N^{-1}\sum_{a}\delta \left(
z-L-z_{a}^{\prime }\right) \delta \left( p-p_{a}\right)  \label{can2a}
\end{equation}%
which is of the form $\int_{-\infty }^{\infty }dL\,f_{c}^{\prime R}(p,z-L)$.
\ We therefore regard $\,f_{c}^{\prime R}(p,z-L)$ as the physically relevant
quantity, where 
\begin{equation}
f_{c}^{\prime R}(p,z)=\frac{1}{{\cal Z}N!}\int \int d{\bf p}d{\bf z}\delta
\left( \overline{p}\right) \delta \left( \overline{z}^{\prime }\right) \exp
\left( -\beta H\right) N^{-1}\sum_{a}\delta \left( z-z_{a}^{\prime }\right)
\delta \left( p-p_{a}\right)  \label{can2b}
\end{equation}%
and where the primes will hencefore be dropped.

Unlike the non-relativistic case, neither the partition function nor $%
f_{c}^{R}(p,z)$ are separable. We proceed by first evaluating the partition
function. \ 

We first write the Hamiltonian (\ref{rogs}) in the following form

\begin{eqnarray}
H &=&Mc^{2}+H_{0}+\frac{1}{c^{2}}H_{R}  \label{can5} \\
H_{0} &=&\sum_{a=1}^{N}\frac{p_{a}^{2}}{2m}+2\pi Gm^{2}\sum_{a>b}^{N}\left|
r_{ab}\right|  \label{can6} \\
H_{R} &=&-\sum_{a=1}^{N}\frac{p_{a}^{4}}{8m^{3}}+\pi
G\sum_{a=1}^{N}\sum_{b=1}^{N}p_{b}^{2}\left| r_{ab}\right| -2\pi
G\sum_{a>b}^{N}p_{a}p_{b}\left| r_{ab}\right|  \nonumber \\
&&+\left( \pi G\right) ^{2}\sum_{a=1}^{N}\sum_{b=1}^{N}\sum_{c=1}^{N}m^{3} 
\left[ \left| r_{ab}\right| \left| r_{ac}\right| -r_{ab}r_{ac}\right]
\label{can7}
\end{eqnarray}%
so that 
\begin{equation}
\exp \left( -\beta H\right) =e^{-\beta Mc^{2}}e^{-\beta H_{0}}\left( 1-\frac{%
\beta }{c^{2}}H_{R}\right) +{\cal O}\left( \frac{\beta ^{2}}{c^{4}}\right)
\label{can8}
\end{equation}%
which is valid to the order in which we are working. \ Writing 
\begin{equation}
\delta \left( \overline{p}\right) =\frac{1}{2\pi }\int dk\exp \left[
ik\sum_{a=1}^{N}p_{a}\right]  \label{can9}
\end{equation}%
we have 
\begin{equation}
{\cal Z=}\frac{e^{-\beta Mc^{2}}}{N!}\,\int d{\bf z}\delta \left( \overline{z%
}\right) \exp \left( -2\pi G\beta m^{2}\sum_{a>b}^{N}\left| r_{ab}\right|
\right) \int \frac{dk}{2\pi }\int d{\bf p}\exp \left(
ik\sum_{a=1}^{N}p_{a}-\beta \sum_{a=1}^{N}\frac{p_{a}^{2}}{2m}\right) \left(
1-\frac{\beta }{c^{2}}H_{R}\right)  \label{can10}
\end{equation}

\subsection{\protect\bigskip The Partition Function}

Consider first the integral 
\begin{equation}
\int \frac{dk}{2\pi }\int d{\bf p}\exp \left( ik\sum_{a=1}^{N}p_{a}-\beta
\sum_{a=1}^{N}\frac{p_{a}^{2}}{2m}\right) \left( 1-\frac{\beta }{c^{2}}%
H_{R}\right)  \label{can11}
\end{equation}%
which has integrals that are at most quartic in the momenta. Straightforward
Gaussian integration yields 
\begin{equation}
\int \frac{dk}{2\pi }\int d{\bf p}\exp \left( ik\sum_{c=1}^{N}p_{c}-\beta
\sum_{c=1}^{N}\frac{p_{c}^{2}}{2m}\right) \left\{ 
\begin{array}{c}
1 \\ 
p_{a}p_{b} \\ 
p_{a}^{4}%
\end{array}%
\right\} =\frac{1}{\sqrt{N}}\left( \frac{2\pi m}{\beta }\right) ^{\left(
N-1\right) /2}\left\{ 
\begin{array}{c}
1 \\ 
\left( \frac{m}{\beta }\right) \times \left\{ 
\begin{array}{c}
\frac{N-1}{N}\mbox{ \ \ \ }a=b \\ 
-\frac{1}{N}\mbox{ \ \ \ }a\neq b%
\end{array}%
\right. \\ 
\left( \frac{m}{\beta }\right) ^{2}\frac{3\left( N-1\right) ^{2}}{N^{2}}%
\end{array}%
\right\}  \label{can11a}
\end{equation}%
Hence we obtain 
\begin{eqnarray}
&&\int \frac{dk}{2\pi }\int d{\bf p}\exp \left( ik\sum_{a=1}^{N}p_{a}-\beta
\sum_{a=1}^{N}\frac{p_{a}^{2}}{2m}\right) \left( 1-\frac{\beta }{c^{2}}%
H_{R}\right)  \nonumber \\
&=&\frac{1}{\sqrt{N}}\left( \frac{2\pi m}{\beta }\right) ^{\left( N-1\right)
/2}\left\{ \left( 1-\frac{\beta }{c^{2}}\left( \pi G\right)
^{2}\sum_{a=1}^{N}\sum_{b=1}^{N}\sum_{c=1}^{N}m^{3}\left[ \left|
r_{ab}\right| \left| r_{ac}\right| -r_{ab}r_{ac}\right] \right) \right. 
\nonumber \\
&&\left. -\frac{\pi G\beta }{c^{2}}\left( \frac{m}{\beta }\right) \frac{N-1}{%
N}\sum_{a=1}^{N}\sum_{b=1}^{N}\left| r_{ab}\right| +\frac{2\pi G\beta }{c^{2}%
}\left( -\frac{m}{N\beta }\right) \sum_{a>b}^{N}\left| r_{ab}\right| +\frac{%
\beta }{8m^{3}c^{2}}\frac{3\left( N-1\right) ^{2}}{N^{2}}\left( \frac{m}{%
\beta }\right) ^{2}\left( \sum_{a=1}^{N}1\right) \right\}  \nonumber \\
&=&\left( 1-\frac{\beta \left( \pi G\right) ^{2}}{c^{2}}\sum_{a=1}^{N}%
\sum_{b=1}^{N}\sum_{c=1}^{N}m^{3}\left[ \left| r_{ab}\right| \left|
r_{ac}\right| -r_{ab}r_{ac}\right] -\frac{2\pi Gm}{c^{2}}\sum_{a>b}^{N}%
\left| r_{ab}\right| +\frac{3\left( N-1\right) ^{2}}{8N\beta mc^{2}}\right) 
\frac{1}{\sqrt{N}}\left( \frac{2\pi m}{\beta }\right) ^{\left( N-1\right) /2}
\nonumber \\
&&  \label{can13}
\end{eqnarray}

\bigskip

\bigskip We next consider the integration over the spatial variables.
Introducing 
\begin{equation}
u_{l}=z_{l+1}-z_{l}\mbox{ \ }1\leq l\leq N\mbox{ \ \ \ \ \ \ \ \ \ }u_{N}=%
\frac{1}{N}\sum_{m=1}^{N}z_{m}  \label{can14}
\end{equation}
we have 
\begin{equation}
\sum_{a>b}^{N}\left| r_{ab}\right| =\frac{1}{2}\sum_{a=1}^{N}\sum_{b=1}^{N}%
\left| r_{ab}\right| =\sum_{l=1}^{N-1}l\left( N-l\right) \left(
z_{l+1}-z_{l}\right) =\sum_{l=1}^{N-1}l\left( N-l\right) u_{l}  \label{can15}
\end{equation}
where without loss of generality the particles are ordered in the sequence $%
z_{1}\leq z_{2}\leq \cdots \leq z_{N}$, and the overall result is then
multiplied by ${\bf N!}$ \ This gives 
\begin{eqnarray}
\int d{\bf z\delta \left( \overline{z}\right) \,{\sf F}(z)} &=&{\bf N!}%
\int_{-\infty }^{\infty }dz_{1}\int_{z_{1}}^{\infty
}dz_{2}\int_{z_{2}}^{\infty }dz_{3}\cdots \int_{z_{N-1}}^{\infty
}dz_{N}\,\delta \left( \overline{z}\right) {\bf {\sf F}(z)}  \nonumber \\
&=&{\bf N!}\int_{-\infty }^{\infty }du_{N}\int_{0}^{\infty
}du_{1}\int_{0}^{\infty }du_{2}\cdots \int_{0}^{\infty }du_{N-1}\delta
\left( u_{N}\right) \,{\bf {\sf F}(u)}  \nonumber \\
&=&{\bf N!}\int d{\bf u\,{\sf F}(u)}  \label{can16}
\end{eqnarray}
provided the function ${\bf {\sf F}(z)}$ is symmetric under interchange of
any pair of variables, which is the case here. \ The inverse transformation
reads 
\begin{equation}
z_{n}=u_{N}-\frac{1}{N}\sum_{l=n}^{N-1}D_{n,l}\mbox{ \ \ \ where }%
D_{n,l}=\left\{ 
\begin{array}{l}
-l\mbox{ \ \ \ \ \ \ \ \ }n>l \\ 
N-l\mbox{ \ \ \ }n\leq l%
\end{array}
\right. \mbox{ }  \label{can17}
\end{equation}
and so we have 
\begin{eqnarray}
\sum_{a=1}^{N}\sum_{b=1}^{N}\sum_{c=1}^{N}\left[ \left| r_{ab}\right| \left|
r_{ac}\right| -r_{ab}r_{ac}\right] &=&2\left(
\sum_{b>a>c}^{N}r_{ba}r_{ac}+\sum_{b>a>c}^{N}r_{ca}r_{ab}\right)  \nonumber
\\
&=&4\sum_{k=1}^{N-2}\sum_{l=k+1}^{N-1}\left( N-l\right) \left( l-k\right)
ku_{l}u_{k}  \label{can18}
\end{eqnarray}

\bigskip

\bigskip Consequently the partition function is 
\begin{eqnarray}
{\cal Z} &=&\frac{e^{-\beta Mc^{2}}}{\sqrt{N}}\left( \frac{2\pi m}{\beta }%
\right) ^{\left( N-1\right) /2}\int_{0}^{\infty }du_{1}\int_{0}^{\infty
}du_{2}\cdots \int_{0}^{\infty }du_{N-1}\exp \left( -2\pi G\beta
m^{2}\sum_{n=1}^{N-1}n\left( N-n\right) u_{n}\right)  \nonumber \\
&&\mbox{ }\times \left( 1-\frac{4\beta m^{3}}{c^{2}}\left( \pi G\right)
^{2}\sum_{k=1}^{N-2}\sum_{l=k+1}^{N-1}\left( N-l\right) \left( l-k\right)
ku_{l}u_{k}-\frac{2\pi Gm}{c^{2}}\sum_{l=1}^{N-1}l\left( N-l\right) u_{l}+%
\frac{3\left( N-1\right) ^{2}}{8N\beta mc^{2}}\right)  \nonumber \\
&&  \label{can19}
\end{eqnarray}%
The three basic integrals in (\ref{can19}) are 
\begin{eqnarray}
\int d{\bf u}\exp \left( -\lambda \sum_{n=1}^{N-1}n\left( N-n\right)
u_{n}\right) &=&\frac{1}{\lambda ^{N-1}\left[ \left( N-1\right) !\right] ^{2}%
}  \label{can20a} \\
\int d{\bf u}\sum_{l=1}^{N-1}k\left( N-k\right) u_{k}\exp \left( -\lambda
\sum_{n=1}^{N-1}n\left( N-n\right) u_{n}\right) &=&\frac{N-1}{\lambda ^{N}%
\left[ \left( N-1\right) !\right] ^{2}}  \label{can20} \\
\int d{\bf u}\sum_{k=1}^{N-2}\sum_{l=k+1}^{N-1}\left( N-l\right) \left(
l-k\right) ku_{l}u_{k}\exp \left( -\lambda \sum_{n=1}^{N-1}n\left(
N-n\right) u_{n}\right) &=&\frac{\sum_{k=1}^{N-2}\sum_{l=k+1}^{N-1}\frac{%
\left( l-k\right) }{l\left( N-k\right) }}{\lambda ^{N+1}\left[ \left(
N-1\right) !\right] ^{2}}\mbox{ \ \ \ \ \ \ \ }  \label{can20c}
\end{eqnarray}%
yielding 
\begin{eqnarray}
{\cal Z} &=&e^{-\beta Mc^{2}}\frac{1}{\sqrt{N}}\left( \frac{2\pi m}{\beta }%
\right) ^{\left( N-1\right) /2}\left( \frac{1+\frac{3\left( N-1\right) ^{2}}{%
8N\beta mc^{2}}}{\left( 2\pi G\beta m^{2}\right) ^{N-1}\left[ \left(
N-1\right) !\right] ^{2}}\right.  \nonumber \\
&&\mbox{ \ \ \ }\left. -\frac{4\beta }{c^{2}}\frac{\left( \pi G\right)
^{2}m^{3}}{\left( 2\pi G\beta m^{2}\right) ^{N+1}\left[ \left( N-1\right) !%
\right] ^{2}}\sum_{k=1}^{N-1}\sum_{l=k+1}^{N-1}\frac{\left( l-k\right) }{%
l\left( N-k\right) }-\frac{2\pi Gm}{c^{2}}\frac{N-1}{\left( 2\pi G\beta
m^{2}\right) ^{N}\left[ \left( N-1\right) !\right] ^{2}}\right)  \nonumber \\
&=&\frac{e^{-\beta Mc^{2}}\left( \frac{2\pi m}{\beta }\right) ^{\left(
N-1\right) /2}}{\sqrt{N}\left( 2\pi G\beta m^{2}\right) ^{N-1}\left[ \left(
N-1\right) !\right] ^{2}}\left[ 1-\frac{1}{\beta mc^{2}}\left\{ \frac{\left(
5N+3\right) \left( N-1\right) +8N\sum_{k=1}^{N-1}\sum_{l=k+1}^{N-1}\frac{%
\left( l-k\right) }{l\left( N-k\right) }}{8N}\right\} \right]  \nonumber \\
&=&\frac{\exp \left[ -\beta Mc^{2}-\frac{3\left( N-1\right) }{2}\ln \left(
\beta mc^{2}\right) -\left\{ \frac{\left( 5N+3\right) \left( N-1\right)
+8N\sum_{k=1}^{N-1}\sum_{l=k+1}^{N-1}\frac{\left( l-k\right) }{l\left(
N-k\right) }}{8N\beta mc^{2}}\right\} \right] }{\sqrt{N}\left( \sqrt{2\pi }%
G/c^{3}\right) ^{\left( N-1\right) }\left[ \left( N-1\right) !\right] ^{2}}%
\mbox{ }  \label{can21}
\end{eqnarray}%
which is the partition function to lowest relativistic order. \ The average
energy is 
\begin{eqnarray}
\left\langle E\right\rangle &=&-\frac{\partial }{\partial \beta }\ln {\cal Z}
\nonumber \\
&=&Mc^{2}+\frac{3}{2}\left( N-1\right) \beta ^{-1}-\left\{ \frac{\left(
5N+3\right) \left( N-1\right) +8N\sum_{k=1}^{N-1}\sum_{l=k+1}^{N-1}\frac{%
\left( l-k\right) }{l\left( N-k\right) }}{8Mc^{2}}\right\} \beta ^{-2}\mbox{
\ \ \ \ \ \ \ \ \ \ \ \ \ \ \ }  \label{can22}
\end{eqnarray}%
to the relevant order in $c^{-2}$. The relativistic correction grows
quadratically with $N$ (for fixed $M=Nm$) and is negative. \ Hence the
average energy of \ a relativistic gravitating system is lower than its
non-relativistic counterpart at the same temperature. \ \ 

\subsection{\protect\bigskip The Single-particle Distribution Function}

Consider next the one-particle distribution function, which is 
\begin{equation}
f_{c}^{R}(p,z)=\frac{1}{N}\sum_{n=1}^{N}f_{cn}^{R}(p,z)  \label{can23}
\end{equation}%
where 
\begin{eqnarray}
f_{cn}^{R}(p,z) &=&\frac{e^{-\beta Mc^{2}}}{{\cal Z}N!}\int d{\bf z}\delta
\left( \overline{z}\right) \exp \left( -2\pi G\beta
m^{2}\sum_{a>b}^{N}\left| r_{ab}\right| \right) \delta \left( z-z_{n}\right)
\nonumber \\
&&\mbox{ \ \ \ \ \ }\times \int \frac{dk}{2\pi }\int d{\bf p}\exp \left(
ik\sum_{a=1}^{N}p_{a}-\beta \sum_{a=1}^{N}\frac{p_{a}^{2}}{2m}\right) \left(
1-\frac{\beta }{c^{2}}H_{R}\right) \delta \left( p-p_{n}\right)  \nonumber \\
&=&\frac{e^{-\beta Mc^{2}}}{{\cal Z}}\int_{-\infty }^{\infty
}du_{N}\int_{0}^{\infty }du_{1}\int_{0}^{\infty }du_{2}\cdots
\int_{0}^{\infty }du_{N-1}\delta \left( u_{N}\right) \exp \left( -2\pi
G\beta m^{2}\sum_{l=1}^{N-1}l\left( N-l\right) u_{l}\right)  \nonumber \\
&&\mbox{ \ \ \ \ \ \ \ \ \ \ \ \ \ \ }\times \delta \left( z-u_{N}+\frac{1}{N%
}\sum_{l=1}^{N-1}D_{n,l}u_{l}\right) \theta _{cn}(p,{\bf z})  \label{can24}
\\
&=&\frac{e^{-\beta Mc^{2}}}{{\cal Z}}\int_{0}^{\infty }du_{1}\cdots
\int_{0}^{\infty }du_{N-1}\exp \left( -2\pi G\beta
m^{2}\sum_{l=1}^{N-1}C_{l}u_{l}\right) \delta \left( z+\frac{1}{N}%
\sum_{l=1}^{N-1}D_{n,l}u_{l}\right) \theta _{cn}(p,{\bf z})  \nonumber
\end{eqnarray}%
in which 
\begin{equation}
\theta _{cn}(p,{\bf z})=\int \frac{dk}{2\pi }\int d{\bf p}\exp \left(
ik\sum_{a=1}^{N}p_{a}-\beta \sum_{a=1}^{N}\frac{p_{a}^{2}}{2m}\right) \left(
1-\frac{\beta }{c^{2}}H_{R}\right) \delta \left( p-p_{n}\right)
\label{can25}
\end{equation}%
and 
\begin{equation}
C_{l}=l\left( N-l\right) \mbox{ \ \ \ \ \ \ \ \ \ \ \ \ \ \ \ }%
D_{n,l}=\left\{ 
\begin{array}{l}
\left( N-l\right) \mbox{ \ \ \ \ \ }n\leq l \\ 
-l\mbox{\ \ \ \ \ \ \ \ \ \ \ \ \ \ \ \ }n>l%
\end{array}%
\right.  \label{can25a}
\end{equation}%
where eq. (\ref{can14}) has been used to express 
\begin{equation}
z_{n}=u_{N}+\frac{1}{N}\sum_{l=1}^{N-1}lu_{l}-\sum_{l=n}^{N-1}u_{l}=u_{N}-%
\frac{1}{N}\sum_{l=1}^{N-1}D_{n,l}u_{l}  \label{can25b}
\end{equation}

\bigskip Evaluation of $\theta _{cn}(p,{\bf z})$ is somewhat lengthy, and so
we relegate its computation to the appendix. \ We obtain 
\begin{eqnarray}
\theta _{cn}(p,{\bf z}) &=&\frac{1}{\sqrt{N-1}}\exp \left( -\frac{N\beta
p^{2}}{2m\left( N-1\right) }\right) \left( \frac{2\pi m}{\beta }\right)
^{\left( N-2\right) /2}  \nonumber \\
&&\times \left\{ \left( 1-\frac{\beta m\left( \pi Gm\right) ^{2}}{c^{2}}%
\sum_{a=1}^{N}\sum_{b=1}^{N}\sum_{c=1}^{N}\left[ \left| r_{ab}\right| \left|
r_{ac}\right| -r_{ab}r_{ac}\right] \right) \right.  \nonumber \\
&&+\frac{1}{2\beta mc^{2}}\left[ \frac{\beta ^{2}p^{4}N\left(
N^{2}-3N+3\right) }{(2m)^{2}\left( N-1\right) ^{3}}+\frac{3\beta p^{2}\left(
N-2\right) }{2m\left( N-1\right) ^{5/2}}+\frac{3\left( N-2\right) ^{2}}{%
4\left( N-1\right) ^{3/2}}\right]  \label{can25bb} \\
&&+\left( \frac{\mbox{2}\pi mG}{c^{2}}\right) \left[ -\frac{1}{2}%
\sum_{a=1}^{N}\sum_{b=1}^{N}\left| r_{ab}\right| -\left( \frac{N^{2}\beta
p^{2}}{2m\left( N-1\right) ^{2}}-\frac{N}{2(N-1)}\right) \left(
\sum_{c=1}^{N}\left| r_{cn}\right| \right) \right]  \nonumber
\end{eqnarray}%
or alternatively, in terms of the ${\bf u}$ variables 
\begin{eqnarray}
\theta _{cn}(p,{\bf u}) &=&\frac{1}{\sqrt{N-1}}\exp \left( -\frac{N\beta
p^{2}}{2m\left( N-1\right) }\right) \left( \frac{2\pi m}{\beta }\right)
^{\left( N-2\right) /2}  \nonumber \\
&&\times \left\{ \left( 1-\frac{4\beta m\left( \pi Gm\right) ^{2}}{c^{2}}%
\sum_{k=1}^{N-2}\sum_{l=k+1}^{N-1}\left( N-l\right) \left( l-k\right)
ku_{l}u_{k}\right) \right.  \nonumber \\
&&+\frac{1}{2\beta mc^{2}}\left[ \frac{\beta ^{2}p^{4}N\left(
N^{2}-3N+3\right) }{(2m)^{2}\left( N-1\right) ^{3}}+\frac{3\beta p^{2}\left(
N-2\right) }{2m\left( N-1\right) ^{5/2}}+\frac{3\left( N-2\right) ^{2}}{%
4\left( N-1\right) ^{3/2}}\right]  \label{can25bc} \\
&&-\left( \frac{2\pi mG}{c^{2}}\right) \left[ \sum_{l=1}^{N-1}l\left(
N-l\right) u_{l}+\left( \frac{N^{2}\beta p^{2}}{2m\left( N-1\right) ^{2}}-%
\frac{N}{2(N-1)}\right) \left(
\sum_{s=1}^{n-1}su_{s}+\sum_{s=n}^{N-1}(N-s)u_{s}\right) \right]  \nonumber
\end{eqnarray}

Now consider eq. (\ref{can24}), which can be rewritten as 
\begin{equation}
f_{cn}^{R}(p,z)=\frac{e^{-\beta Mc^{2}}}{{\cal Z}}\int \frac{dk}{2\pi }\int d%
{\bf u}\exp \left( -ikz-\lambda \beta \sum_{l=1}^{N-1}\left( C_{l}+i\alpha
D_{nl}\right) u_{l}\right) \theta _{cn}(p,{\bf u})  \label{can32}
\end{equation}%
where 
\begin{equation}
\lambda =2\pi Gm^{2}\mbox{ \ \ \ \ }\alpha =\frac{k}{N\beta \lambda }\mbox{ }
\label{can33}
\end{equation}%
The integration now involves straightforward integrations over the ${\bf u}$
variables, after which an evaluation of the $k$-integral using Jordan's
lemma must be performed. \ This involves some rather tedious manipulations
which we describe in appendix ??? \ The final result is 
\begin{eqnarray}
f_{cn}(p,z) &=&\frac{\left( 2\pi Gm^{2}\right) \left( N\beta \right) ^{3/2}}{%
\sqrt{2\pi m\left( N-1\right) }}\exp \left[ \frac{1}{\beta mc^{2}}\left\{ 
\frac{(5N+3)\left( N-1\right) }{8N}+\sum_{k=1}^{N-1}\sum_{l=k+1}^{N-1}\frac{%
\left( l-k\right) }{l\left( N-k\right) }\right\} \right]  \nonumber \\
&&\mbox{ \ }\times \sum_{l=1}^{N-1}\left[ \left\{ A_{l}^{N}\left( 1+\frac{1}{%
2\beta mc^{2}}\left( \frac{\beta ^{2}p^{4}\left( 1+\left( N-1\right)
^{3}\right) }{\left( 2m\right) ^{2}\left( N-1\right) ^{3}}+\frac{3\left(
N-2\right) \beta p^{2}}{2m\left( N-1\right) ^{2}}+\frac{3\left( N-2\right)
^{2}}{4\left( N-1\right) }\right) \right) \right. \right.  \nonumber \\
&&\mbox{\ }-\frac{1}{\beta mc^{2}}\left[ B_{l}^{N}-A_{l}^{N}\left(
1-2N\left( \beta \pi Gm^{2}\right) l\left| z\right| \right) \right] 
\nonumber \\
&&+\frac{2}{\beta mc^{2}}\left( \frac{N}{2\left( N-1\right) }-\frac{\beta
p^{2}}{2m}\left[ \frac{N^{2}}{\left( N-1\right) ^{2}}\right] \right) \mbox{\ 
}\left[ C_{l}^{N}-\frac{1}{l}A_{l}^{N}\left( 1-2N\left( \beta \pi
Gm^{2}\right) l\left| z\right| \right) \right]  \label{can34} \\
&&\mbox{ \ \ \ \ }\left. \left. -\frac{1}{\beta mc^{2}}\left[
D_{l}^{N}+K_{l}^{N}\left( 1-2N\left( \beta \pi Gm^{2}\right) l\left|
z\right| \right) \right] \right\} \exp \left( -\frac{N\beta p^{2}}{2m\left(
N-1\right) }-2\pi GN\beta m^{2}l\left| z\right| \right) \right] \mbox{ \ \ }
\nonumber
\end{eqnarray}%
Integration over $p$ yields the canonical density distribution function 
\begin{eqnarray}
\rho _{c}\left( z\right) &=&\int_{-\infty }^{\infty }dpf_{cn}(p,z)  \nonumber
\\
&=&\left( 2\pi Gm^{2}N\beta \right) \exp \left[ \frac{1}{\beta mc^{2}}%
\left\{ \frac{(5N+3)\left( N-1\right) }{8N}+\sum_{k=1}^{N-1}%
\sum_{l=k+1}^{N-1}\frac{\left( l-k\right) }{l\left( N-k\right) }\right\} %
\right]  \nonumber \\
&&\times \sum_{l=1}^{N-1}\left\{ A_{l}^{N}+\frac{1}{\beta mc^{2}}\left( 
\frac{3}{8}\frac{\left( N-1\right) ^{2}}{N}A_{l}^{N}-B_{l}^{N}-D_{l}^{N}%
\right) \right.  \nonumber \\
&&\mbox{ \ \ \ \ \ \ \ \ \ \ \ \ \ }\left. +\frac{1}{\beta mc^{2}}\left[
A_{l}^{N}-K_{l}^{N}\right] \left( 1-2\pi Gm^{2}N\beta l\left| z\right|
\right) \right\} \exp \left( -2\pi Gm^{2}N\beta l\left| z\right| \right) 
\mbox{ \ \ \ \ \ }  \label{can35}
\end{eqnarray}%
whereas integration over $z$ yields the canonical momentum distribution
function 
\begin{eqnarray}
\vartheta _{cn}(p) &=&\int_{-\infty }^{\infty }dz\,f_{cn}(p,z)  \nonumber \\
&=&\sqrt{\frac{\left( N\beta \right) }{2\pi m\left( N-1\right) }}\exp \left[
-\frac{N\beta p^{2}}{2m\left( N-1\right) }\right]  \nonumber \\
&&\times \left( 1+\frac{1}{\beta mc^{2}}\left( \frac{N\beta ^{2}p^{4}\left(
N^{2}-3N+3\right) }{8m^{2}\left( N-1\right) ^{3}}-\frac{\beta p^{2}\left(
4N^{2}-7N+6\right) }{4m\left( N-1\right) ^{2}}+\frac{5N(N-1)+3}{8N\left(
N-1\right) }\right) \right) \mbox{ \ \ \ \ }  \label{can36}
\end{eqnarray}

\bigskip From the results in the appendix, it can be shown that 
\begin{equation}
\sum_{l=1}^{N-1}\frac{1}{l}A_{l}^{N}=\frac{1}{2}\mbox{ \ \ \ \ \ }%
\sum_{l=1}^{N-1}\frac{1}{l}B_{l}^{N}=\frac{1}{2}\mbox{ }\left( N-1\right) 
\mbox{ \ \ \ \ \ }\sum_{l=1}^{N-1}\frac{1}{l}C_{l}^{N}=\frac{\left(
N-1\right) }{N}\mbox{\ \ \ \ \ \ }\sum_{l=1}^{N-1}\frac{1}{l}D_{l}^{N}=\frac{%
1}{2}\sum_{k=1}^{N-1}\sum_{l=k+1}^{N-1}\frac{\left( l-k\right) }{l\left(
N-k\right) }  \label{can37}
\end{equation}
which can be used to show that 
\begin{equation}
\int_{-\infty }^{\infty }dz\,\rho _{c}\left( z\right) =1\mbox{ \ \ \ and \ \
\ }\int_{-\infty }^{\infty }dp\vartheta _{cn}(p)=1  \label{can38}
\end{equation}

\bigskip We also have the relations 
\begin{equation}
\sum_{l=1}^{N-1}A_{l}^{N}=\frac{1}{2}\frac{N-1}{2N-3}\mbox{ \ \ \ \ \ \ \ \
\ \ \ }\sum_{l=1}^{N-1}B_{l}^{N}=\frac{1}{2}\frac{\left( N-1\right) ^{2}}{%
2N-3}  \label{ABsumrules}
\end{equation}
the first of which was demonstrated by Rybicki \cite{Rybicki}.

\bigskip

\section{\protect\bigskip The Microcanonical Ensemble}

The results for the canonical ensemble obtained in the previous section are
for a system of relativistic gravitating particles coupled to a heat bath
which keeps the system at a constant temperature $T=\beta ^{-1}$. \ \ In
such a situation the energy of the system ill-defined, and undergoes
fluctuations of the order $kT$. \ An isolated system, on the other hand,
would have its total energy conserved, and this is the more realistic
astrophysical case. \ This entails usage of the microcanonical ensemble, in
which phase space integrations are carried out by constraining the total
energy to be $E.$ The weighting function $e^{-\beta H}$ in the phase space
integral is therefore replaced with $\delta \left( E-H\right) $.

\bigskip

Fortunately it is straightforward to compute the relevant microcanonical
quantities from the canonical ones. \ Using the same reasoning that led to (%
\ref{can2a}), the microcanonical single-particle distribution function is 
\begin{equation}
f_{mc}^{\prime R}(p,z)=\frac{1}{\Omega N!}\int \int d{\bf p}d{\bf z}\delta
\left( \overline{p}\right) \delta \left( \overline{z}\right) \delta \left(
E-H\right) N^{-1}\sum_{a}\delta \left( z-z_{a}\right) \delta \left(
p-p_{a}\right)  \label{mi1}
\end{equation}
where 
\begin{equation}
\Omega =\frac{1}{N!}\,\int \int d{\bf p}d{\bf z}\delta \left( \overline{p}%
\right) \delta \left( \overline{z}\right) \delta \left( E-H\right)
\label{mi2}
\end{equation}

Note that $\Omega $ and ${\cal Z}$ are related by the Laplace transforms 
\begin{eqnarray}
{\cal Z} &=&\int_{0}^{\infty }dE\,e^{-\beta H}\Omega \left( E\right)
\label{mi3a} \\
\Omega &=&\frac{1}{2\pi i}\int_{{\cal C}}d\beta \,e^{\beta E}{\cal Z}\left(
\beta \right)  \label{mi3b}
\end{eqnarray}%
where in contour ${\cal C}$ in the latter integral extends from $-i\infty $
to $+i\infty $ to the right of all singularities. \ Using the general result
that 
\begin{equation}
\frac{\left( w\right) _{+}^{\varsigma -1}}{\Gamma \left( \varsigma \right) }=%
\frac{1}{2\pi i}\int_{{\cal C}}d\beta \,e^{\beta w}\beta ^{-\varsigma }\mbox{
\ \ \ where \ \ \ \ }\left( w\right) _{+}=\left\{ 
\begin{array}{l}
w\mbox{ \ \ \ \ }w\geq 0 \\ 
0\mbox{\ \ \ \ \ }w<0%
\end{array}%
\right.  \label{mi4}
\end{equation}%
it is straightforward to obtain 
\begin{eqnarray}
\Omega &=&\frac{\left( 2\pi m\right) ^{\left( N-1\right) /2}\left(
E-Mc^{2}\right) ^{\left( 3N-5\right) /2}}{\sqrt{N}\left( \pi Gm^{2}\right)
^{N-1}\left[ \left( N-1\right) !\right] ^{2}\Gamma \left( \frac{3}{2}\left(
N-1\right) \right) }  \nonumber \\
&&\times \exp \left[ -\frac{\left( E-Mc^{2}\right) }{Mc^{2}}\left\{ \frac{%
\left( 5N+3\right) \left( N-1\right) +8N\sum_{k=1}^{N-1}\sum_{l=k+1}^{N-1}%
\frac{\left( l-k\right) }{l\left( N-k\right) }}{12(N-1)}\right\} \right]
\label{mi5}
\end{eqnarray}%
to leading order in $1/c$.

\bigskip Similarly using (\ref{mi4}) in (\ref{mi1}), we have 
\begin{equation}
\Omega \,f_{mc}^{\prime R}(p,z)=\frac{1}{2\pi i}\int_{{\cal C}}d\beta
\,e^{\beta E}\left( {\cal Z}f_{c}^{\prime R}(p,z)\right)   \label{mi6}
\end{equation}%
Using (\ref{can21}) and (\ref{can34}) we have 
\begin{eqnarray}
f_{cn}(p,z) &=&\frac{\left( 2\pi Gm^{2}\right) \left( N\beta \right) ^{3/2}}{%
\sqrt{2\pi m\left( N-1\right) }}\exp \left[ \frac{1}{\beta mc^{2}}\left\{ 
\frac{(5N+3)\left( N-1\right) }{8N}+\sum_{k=1}^{N-1}\sum_{l=k+1}^{N-1}\frac{%
\left( l-k\right) }{l\left( N-k\right) }\right\} \right]   \nonumber \\
&&\mbox{ \ }\times \sum_{l=1}^{N-1}\left[ \left\{ A_{l}^{N}\left( 1+\frac{1}{%
2\beta mc^{2}}\left( \frac{\beta ^{2}p^{4}\left( 1+\left( N-1\right)
^{3}\right) }{\left( 2m\right) ^{2}\left( N-1\right) ^{3}}+\frac{3\left(
N-2\right) \beta p^{2}}{2m\left( N-1\right) ^{2}}+\frac{3\left( N-2\right)
^{2}}{4\left( N-1\right) }\right) \right) \right. \right.   \nonumber \\
&&-\frac{1}{\beta mc^{2}}\left[ B_{l}^{N}-A_{l}^{N}\left( 1-2N\left( \beta
\pi Gm^{2}\right) l\left| z\right| \right) \right]   \nonumber \\
&&+\frac{2}{\beta mc^{2}}\left( \frac{N}{2\left( N-1\right) }-\frac{\beta
p^{2}}{2m}\left[ \frac{N^{2}}{\left( N-1\right) ^{2}}\right] \right) \mbox{\ 
}\left[ C_{l}^{N}-\frac{1}{l}A_{l}^{N}\left( 1-2N\left( \beta \pi
Gm^{2}\right) l\left| z\right| \right) \right]   \label{mi6a} \\
&&\mbox{\ \ \ }\left. \left. -\frac{1}{\beta mc^{2}}\left[
D_{l}^{N}+K_{l}^{N}\left( 1-2N\left( \beta \pi Gm^{2}\right) l\left|
z\right| \right) \right] \right\} \exp \left( -\frac{N\beta p^{2}}{2m\left(
N-1\right) }-2\pi GN\beta m^{2}l\left| z\right| \right) \right]   \nonumber
\end{eqnarray}%
\begin{eqnarray}
f_{mc}^{\prime R}(p,z) &=&\frac{2\pi Gm^{2}}{\sqrt{2\pi m\left( N-1)\right) }%
}\left( \frac{N}{\left( E-Mc^{2}\right) }\right) ^{3/2}\frac{\Gamma \left( 
\frac{3}{2}\left( N-1\right) \right) }{\Gamma \left( \frac{3}{2}\left(
N-2\right) \right) }  \nonumber \\
&&\times \exp \left[ \frac{\left( E-Mc^{2}\right) }{Mc^{2}}\left\{ \frac{%
\left( 5N+3\right) \left( N-1\right) +8N\sum_{k=1}^{N-1}\sum_{j=k+1}^{N-1}%
\frac{\left( j-k\right) }{j\left( N-k\right) }}{12(N-1)}\right\} \right]  
\nonumber \\
&&\times \sum_{l=1}^{N-1}\left\{ \left[ A_{l}^{N}+\left( \frac{A_{l}^{N}}{%
mc^{2}}\left[ \frac{3p^{2}\left( N-2\right) }{4m\left( N-1\right) ^{2}}%
\right] -\frac{C_{l}^{N}-\frac{1}{l}A_{l}^{N}}{mc^{2}}\left[ \frac{%
2N^{2}p^{2}}{2m\left( N-1\right) ^{2}}\right] \right) \right] \Upsilon
_{+}^{3N/2-4}\right.   \nonumber \\
&&+\left[ \frac{2N\left( \pi Gm^{2}\right) \left| z\right| }{mc^{2}}\left(
A_{l}^{N}\left( \frac{N}{\left( N-1\right) }-l\right) +lK_{l}^{N}\right) %
\right] \Upsilon _{+}^{3N/2-4}  \nonumber \\
&&\mbox{\ }+\frac{\left( \frac{3}{2}N-4\right) }{\left( E-Mc^{2}\right) }%
\frac{A_{l}^{N}}{2mc^{2}}\left[ \frac{p^{4}N\left( N^{2}-3N+3\right) }{%
(2m)^{2}\left( N-1\right) ^{3}}-\frac{2p^{2}}{m}\left[ \frac{2N^{3}\pi
Gm^{2}\left| z\right| }{\left( N-1\right) ^{2}}\right] \right] \Upsilon
_{+}^{3N/2-5}  \label{mi7} \\
&&\mbox{\ }\left. +\frac{\left( E-Mc^{2}\right) }{\left( \frac{3}{2}%
N-3\right) mc^{2}}\left( A_{l}^{N}\left[ \frac{3\left( N-2\right) ^{2}}{%
8\left( N-1\right) }+1\right] -B_{l}^{N}\mbox{\ }+\frac{N\left( C_{l}^{N}-%
\frac{1}{l}A_{l}^{N}\right) }{\left( N-1\right) }-D_{l}^{N}-K_{l}^{N}\right)
\Upsilon _{+}^{3N/2-3}\right\}   \nonumber
\end{eqnarray}%
as the expression for the relativistic microcanonical partition function,
valid to ${\cal O}\left( 1/c^{2}\right) $, where 
\begin{equation}
\Upsilon \left( p,z\right) \equiv 1-\frac{Np^{2}}{2m\left( N-1\right) \left(
E-Mc^{2}\right) }-\frac{2N\pi Gm^{2}}{\left( E-Mc^{2}\right) }l\left|
z\right|   \label{mi7a}
\end{equation}

{\bf \bigskip }Employing the expressions 
\begin{eqnarray}
\int_{-1}^{1}dy\left( 1-y^{2}\right) ^{\gamma } &=&\sqrt{\pi }\frac{\Gamma
\left( \gamma +1\right) }{\Gamma \left( \gamma +\frac{3}{2}\right) }
\label{mi8a} \\
\int_{-1}^{1}dy\,y^{2}\left( 1-y^{2}\right) ^{\gamma } &=&\sqrt{\pi }\frac{%
\Gamma \left( \gamma +1\right) }{2\Gamma \left( \gamma +\frac{5}{2}\right) }
\label{mi8b} \\
\int_{-1}^{1}dy\,y^{4}\left( 1-y^{2}\right) ^{\gamma } &=&\sqrt{\pi }\frac{%
3\Gamma \left( \gamma +1\right) }{4\Gamma \left( \gamma +\frac{7}{2}\right) }
\label{mi8c}
\end{eqnarray}%
the density distribution is 
\begin{eqnarray}
\rho _{mc}(z) &=&\int_{-\infty }^{\infty }dp\,f_{mc}^{\prime R}(p,z) 
\nonumber \\
&=&\left( \frac{2N\pi Gm^{2}}{\left( E-Mc^{2}\right) }\right) \exp \left[ 
\frac{\left( E-Mc^{2}\right) }{Mc^{2}}\left\{ \frac{\left( 5N+3\right)
\left( N-1\right) +8N\sum_{k=1}^{N-1}\sum_{j=k+1}^{N-1}\frac{\left(
j-k\right) }{j\left( N-k\right) }}{12(N-1)}\right\} \right]   \nonumber \\
&&\times \sum_{l=1}^{N-1}\left\{ \left( \frac{3N-5}{2}\right) \left[
A_{l}^{N}+\frac{2\left( \pi GM^{2}\right) \left| z\right| }{Mc^{2}}\left(
-lA_{l}^{N}+lK_{l}^{N}\right) \right] \left( \Upsilon \left( 0,z\right)
\right) _{+}^{3N/2-7/2}\right.   \nonumber \\
&&\mbox{ \ }+\left( \frac{\left( E-Mc^{2}\right) }{Mc^{2}}\right) \left(
\Upsilon \left( 0,z\right) \right) _{+}^{3N/2-5/2}\left[ \left( -\left(
C_{l}^{N}-\frac{1}{l}A_{l}^{N}\right) \left[ \frac{N^{2}}{\left( N-1\right) }%
\right] \right) \right.   \nonumber \\
&&\mbox{ \ \ \ \ }+A_{l}^{N}\left[ \frac{3\left( N^{2}-3N+3\right) }{8\left(
N-1\right) }+\frac{3\left( N-2\right) }{4\left( N-1\right) }+\frac{3N\left(
N-2\right) ^{2}}{8\left( N-1\right) }+N\right]   \nonumber \\
&&\mbox{\ \ \ \ \ \ }\left. \mbox{\ }+\left. \left( -NB_{l}^{N}\mbox{\ }+%
\frac{N^{2}\left( C_{l}^{N}-\frac{1}{l}A_{l}^{N}\right) }{\left( N-1\right) }%
-N\left( D_{l}^{N}+K_{l}^{N}\right) \right) \right] \right\}   \nonumber \\
&=&\left( \frac{2N\pi Gm^{2}}{\left( E-Mc^{2}\right) }\right) \exp \left[ 
\frac{\left( E-Mc^{2}\right) }{Mc^{2}}\left\{ \frac{\left( 5N+3\right)
\left( N-1\right) +8N\sum_{k=1}^{N-1}\sum_{j=k+1}^{N-1}\frac{\left(
j-k\right) }{j\left( N-k\right) }}{12(N-1)}\right\} \right]   \nonumber \\
&&\times \sum_{l=1}^{N-1}\left\{ \left( \frac{3N-5}{2}\right) \left[
A_{l}^{N}-\frac{2\left( \pi GM^{2}\right) l\left| z\right| }{Mc^{2}}\left(
A_{l}^{N}-K_{l}^{N}\right) \right] \left( \Upsilon \left( 0,z\right) \right)
_{+}^{3N/2-7/2}\right.   \nonumber \\
&&\left. +\left( \frac{\left( E-Mc^{2}\right) }{Mc^{2}}\right) \left(
\Upsilon \left( 0,z\right) \right) _{+}^{3N/2-5/2}\left( \left( \frac{%
3(N-1)^{2}}{8}\right) A_{l}^{N}-N\left( B_{l}^{N}-A_{l}^{N}\right) -N\left(
D_{l}^{N}+K_{l}^{N}\right) \right) \right\}   \nonumber \\
&&  \label{mi9}
\end{eqnarray}%
The normalization of the density $\int_{-\infty }^{\infty }dz\,\rho
_{mc}(z)=1$\ implies 
\begin{eqnarray}
&&2\sum_{l=1}^{N-1}\left\{ \frac{A_{l}^{N}}{l}+\left( \frac{\left(
E-Mc^{2}\right) }{Mc^{2}}\right) \left( \frac{A_{l}^{N}}{l}\left( \frac{%
\left( N-1\right) }{4}\right) -\frac{\mbox{2}NB_{l}^{N}}{3(N-1)l}-\frac{2N}{%
3\left( N-1\right) }\frac{D_{l}^{N}}{l}\right) \right\}   \nonumber \\
&=&\exp \left[ -\frac{\left( E-Mc^{2}\right) }{Mc^{2}}\left\{ \frac{\left(
5N+3\right) \left( N-1\right) +8N\sum_{k=1}^{N-1}\sum_{j=k+1}^{N-1}\frac{%
\left( j-k\right) }{j\left( N-k\right) }}{12(N-1)}\right\} \right] 
\label{mi10}
\end{eqnarray}%
which, \ using (\ref{can37}), is easily shown to be satisfied to first order
in $\zeta \equiv \frac{\left( E-Mc^{2}\right) }{Mc^{2}}$, where the latter
quantity is the dimensionless fraction of excess energy above the total rest
mass.

\bigskip 

The momentum distribution is 
\begin{eqnarray}
\vartheta _{mc}(p) &=&\int_{-\infty }^{\infty }dz\,f_{mc}^{\prime R}(p,z) 
\nonumber \\
&=&\left( \frac{N}{2\pi m\left( N-1\right) \left( E-Mc^{2}\right) }\right)
^{1/2}\frac{\Gamma \left( \frac{3}{2}\left( N-1\right) \right) }{\Gamma
\left( \frac{3}{2}N-2\right) }  \nonumber \\
&&\times \exp \left[ \frac{\left( E-Mc^{2}\right) }{Mc^{2}}\left\{ \frac{%
\left( 5N+3\right) \left( N-1\right) +8N\sum_{k=1}^{N-1}\sum_{j=k+1}^{N-1}%
\frac{\left( j-k\right) }{j\left( N-k\right) }}{12(N-1)}\right\} \right]  
\nonumber \\
&&\times \sum_{l=1}^{N-1}\left\{ \left[ \frac{2A_{l}^{N}}{l}+\left( \frac{%
2A_{l}^{N}}{lmc^{2}}\left[ \frac{3p^{2}\left( N-2\right) }{4m\left(
N-1\right) ^{2}}\right] -\frac{2\left( C_{l}^{N}-\frac{1}{l}A_{l}^{N}\right) 
}{lmc^{2}}\left[ \frac{N^{2}p^{2}}{m\left( N-1\right) ^{2}}\right] \right) %
\right] \left( \Upsilon \left( p,0\right) \right) _{+}^{3N/2-3}\right.  
\nonumber \\
\mbox{ \ \ \ \ \ } &&+\left[ \frac{4\left( E-Mc^{2}\right) }{mc^{2}\left(
3N-4\right) }\left( \frac{A_{l}^{N}}{l^{2}}\left( \frac{N}{\left( N-1\right) 
}-l\right) +\frac{K_{l}^{N}}{l}\right) \right] \left( \Upsilon \left(
p,0\right) \right) _{+}^{3N/2-2}  \nonumber \\
&&+\frac{2A_{l}^{N}}{l^{2}mc^{2}}\left[ -\frac{p^{2}}{m}\left[ \frac{N^{2}}{%
\left( N-1\right) ^{2}}\right] \right] \left( \Upsilon \left( p,0\right)
\right) _{+}^{3N/2-3}  \nonumber \\
&&+\frac{\left( \frac{3}{2}N-3\right) }{\left( E-Mc^{2}\right) }\frac{%
A_{l}^{N}}{lmc^{2}}\left[ \frac{p^{4}N\left( N^{2}-3N+3\right) }{%
(2m)^{2}\left( N-1\right) ^{3}}\right] \left( \Upsilon \left( p,0\right)
\right) _{+}^{3N/2-4}  \nonumber \\
&&+\frac{4\left( E-Mc^{2}\right) }{\left( 3N-4\right) mc^{2}}\left( \frac{%
A_{l}^{N}}{l}\left[ \frac{3\left( N-2\right) ^{2}}{8\left( N-1\right) }+1%
\right] -\frac{B_{l}^{N}}{l}\mbox{\ }\right.   \nonumber \\
&&\mbox{ \ \ \ \ }\left. \left. +\frac{N\left( C_{l}^{N}-\frac{1}{l}%
A_{l}^{N}\right) }{l\left( N-1\right) }-\frac{D_{l}^{N}+K_{l}^{N}}{l}\right)
\left( \Upsilon \left( p,0\right) \right) _{+}^{3N/2-2}\right\}   \nonumber
\\
&=&\left( \frac{N}{2\pi m\left( N-1\right) \left( E-Mc^{2}\right) }\right)
^{1/2}\frac{\Gamma \left( \frac{3}{2}\left( N-1\right) \right) }{\Gamma
\left( \frac{3}{2}N-2\right) }  \nonumber \\
&&\times \exp \left[ \frac{\left( E-Mc^{2}\right) }{Mc^{2}}\left\{ \frac{%
\left( 5N+3\right) \left( N-1\right) +8N\sum_{k=1}^{N-1}\sum_{j=k+1}^{N-1}%
\frac{\left( j-k\right) }{j\left( N-k\right) }}{12(N-1)}\right\} \right]  
\nonumber \\
&&\times \left\{ \left[ 1-\frac{N}{Mc^{2}}\left( \frac{p^{2}}{4m}\left[ 
\frac{8N^{2}-11N+6}{\left( N-1\right) ^{2}}\right] \right) \right] \left(
\Upsilon \left( p,0\right) \right) _{+}^{3N/2-3}\right.   \nonumber \\
&&+\frac{N}{\left( E-Mc^{2}\right) }\frac{3\left( N-2\right) }{4Mc^{2}}\left[
\frac{p^{4}N\left( N^{2}-3N+3\right) }{(2m)^{2}\left( N-1\right) ^{3}}\right]
\left( \Upsilon \left( p,0\right) \right) _{+}^{3N/2-4}  \nonumber \\
&&\left. -\frac{2N\left( E-Mc^{2}\right) }{\left( 3N-4\right) Mc^{2}}\left( 
\frac{\left( 5N^{2}-20N+12\right) }{8\left( N-1\right) }\mbox{\ }%
+\sum_{s=1}^{N-1}\sum_{t=s+1}^{N-1}\frac{\left( t-s\right) }{t\left(
N-s\right) }\right) \left( \Upsilon \left( p,0\right) \right)
_{+}^{3N/2-2}\right\}   \label{mi11}
\end{eqnarray}%
It is straightforward to show that $\int_{-\infty }^{\infty }dp\,\vartheta
_{mc}(p)=1$\ from (\ref{mi8a} -- \ref{mi8c}) to first order in $\zeta \equiv 
\frac{\left( E-Mc^{2}\right) }{Mc^{2}}$.

\bigskip

\section{\protect\bigskip The Large $N$ Limit}

\bigskip

For statisical systems (such as those of interest in stellar dynamics), the
large $N$ limit is of considerable physical interest. This is the limit in
which the total energy $E$ and total mass $M=Nm$ are fixed. \ In the
non-relativistic case, the single-particle distribution function approaches
the isothermal solution of the Vlasov equations in the large $N$ limit \cite%
{Rybicki}. \ However the relativistic case is somewhat more subtle, since
the expressions we have obtained are valid only if the speed of light is
sufficiently large relative to other quantities of the same dimension, and
as $N$ becomes large we must ensure that this approximation remains valid.

\ 

In order to investigate the large $N$ limit it is necessary to rewrite all
quantities in terms of $E$, $M$, and $N$. As in ref. \cite{Rybicki}, we
adopt the dimensionless variables 
\begin{equation}
\eta \equiv \frac{p}{mV}\mbox{ \ \ \ \ \ \ \ }\xi \equiv \frac{z}{L}\mbox{ }
\label{ln1}
\end{equation}
where 
\begin{equation}
L\equiv \frac{2\left( E-Mc^{2}\right) }{3\pi GM^{2}}=\frac{2\zeta c^{2}}{%
3\pi GM}\mbox{ \ \ \ \ \ \ \ }V^{2}\equiv \frac{4\left( E-Mc^{2}\right) }{3M}%
=\frac{4\zeta c^{2}}{3}  \label{ln2}
\end{equation}
are the characterisitic length and velocity scales of the system. \ The
scaled distributions functions are correspondingly defined 
\begin{equation}
\begin{array}{c}
\rho ^{\ast }(\xi )\equiv L\rho (L\xi ) \\ 
\vartheta ^{\ast }(\eta )\equiv mV\vartheta (mV\eta ) \\ 
f^{\prime \ast R}(\eta ,\xi )\equiv mVLf^{\prime R}(mV\eta ,L\xi )%
\end{array}
\label{ln3}
\end{equation}
so that 
\begin{equation}
\int \int d\eta \,d\xi f^{\prime \ast R}(\eta ,\xi )=\int d\xi \rho ^{\ast
}(\xi )=\int d\eta \,\vartheta ^{\ast }(\eta )=1  \label{ln4}
\end{equation}

\bigskip

\bigskip Consider first the partition function (\ref{can21}), which can be
rewritten as 
\begin{equation}
{\cal Z}=\frac{\exp \left[ -\beta Mc^{2}-\frac{1}{\beta Mc^{2}}\left\{ \frac{%
\left( 5N+3\right) \left( N-1\right) +8N\sum_{k=1}^{N-1}\sum_{l=k+1}^{N-1}%
\frac{\left( l-k\right) }{l\left( N-k\right) }}{8}\right\} \right] }{\sqrt{N}%
\left( \sqrt{2\pi }G/c^{3}\right) ^{\left( N-1\right) }\left[ \left(
N-1\right) !\right] ^{2}}\left( \frac{N}{\beta Mc^{2}}\right) ^{3\left(
N-1\right) /2}  \label{lnZ}
\end{equation}

\bigskip

The approximation (\ref{can8}) is valid provided 
\begin{equation}
\beta > \frac{\left( 5N+3\right) \left( N-1\right)
+8N\sum_{k=1}^{N-1}\sum_{l=k+1}^{N-1}\frac{\left( l-k\right) }{l\left(
N-k\right) }}{8Mc^{2}}  \label{lnZa}
\end{equation}
which sets an upper bound on the thermal energy $kT=\beta ^{-1}$ of the
system for a given value of $N$. \ For fixed $Mc^{2}$ the relativistic
corrections are valid only as the temperature becomes vanishingly small in
the limit of large $N$. The exponential approximation is slightly better
than the polynomial one because of the positivity of the partition function.

\begin{figure}
\begin{center}
\epsfig{file=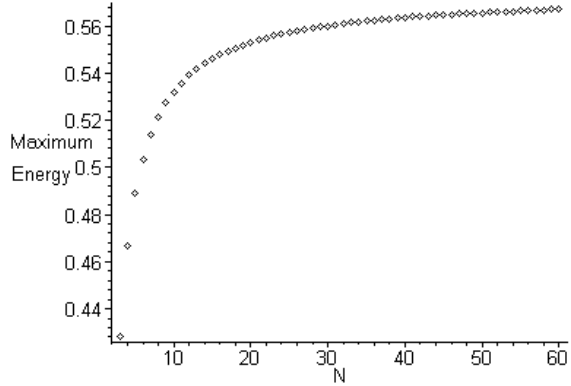,width=0.8\linewidth}
\end{center}
\caption{Maximum value of
the average relativistic energy as a function of $N$.}
\label{fig2}
\end{figure}

\bigskip Consider next the average energy in the canonical case as given by
eq. (\ref{can22}), which we rewrite as 
\begin{equation}
\left\langle \zeta \right\rangle =\frac{3}{2}\left( N-1\right) \frac{1}{%
\beta Mc^{2}}-\left\{ \frac{\left( 5N+3\right) \left( N-1\right)
+8N\sum_{k=1}^{N-1}\sum_{l=k+1}^{N-1}\frac{\left( l-k\right) }{l\left(
N-k\right) }}{8}\right\} \left( \frac{1}{\beta Mc^{2}}\right) ^{2}
\label{ln4a}
\end{equation}%
where $\zeta \equiv \frac{\left( E-Mc^{2}\right) }{Mc^{2}}$, as before. \
When the thermal energy $kT=\beta ^{-1}$ of the system is sufficiently small
relative to its rest energy $Mc^{2}$, the expression for the average energy $%
\left\langle \zeta \right\rangle $ does not differ much from the
non-relativistic value given by its first term. As the thermal energy grows
(i.e. as $\beta $ decreases) the value of $\left\langle \zeta \right\rangle $
increases more slowly than its non-relativistic counterpart, reaching a
maximum when 
\begin{equation}
\beta =\frac{\left( 5N+3\right) \left( N-1\right)
+8N\sum_{k=1}^{N-1}\sum_{l=k+1}^{N-1}\frac{\left( l-k\right) }{l\left(
N-k\right) }}{6\left( N-1\right) Mc^{2}}\equiv \beta _{\mbox{m}}
\label{ln4b}
\end{equation}%
after which the average energy decreases with decreasing $\beta $, becoming
negative when $\beta =\frac{1}{2}\beta _{\mbox{m}}$. \ As $N$ becomes large, 
$\beta _{\mbox{m}}\rightarrow \left( \frac{7}{2}-\frac{2\pi ^{2}}{9}\right) 
\frac{N}{Mc^{2}}$. \ At $\beta =\beta _{\mbox{m}}$ the average energy has
half the value of its non-relativistic counterpart. In figure\ref{fig2} we
plot the maximum value of\ $\ \left\langle \zeta \right\rangle $ as a
function of $N$.
The curve asymptotes to the
constant value of $\left\langle \zeta \right\rangle =.573940872$ as $%
N\rightarrow \infty $. 

\begin{figure}
\begin{center}
\epsfig{file=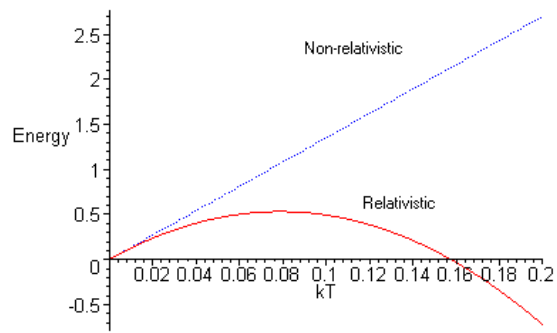,width=0.8\linewidth}
\end{center}
\caption{Average Energy $\left\langle \protect\zeta \right\rangle $as a function
of $kT$ for $N=10$ for the non-relativistic and relativistic cases. Axes are
in units of $Mc^{2}$.}
\label{fig-1}
\end{figure}

\begin{figure}
\begin{center}
\epsfig{file=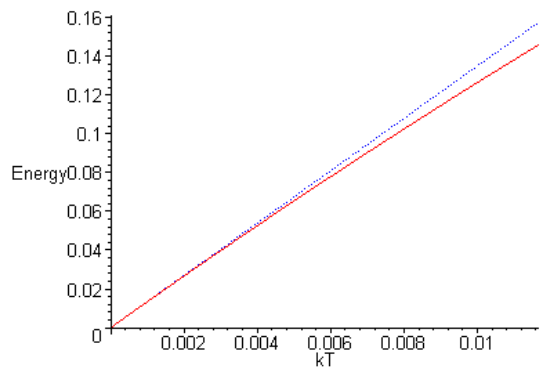,width=0.8\linewidth}
\end{center}
\caption{Average energy $\left\langle 
\protect\zeta \right\rangle $ for $N=10$ over the allowed range of $kT$.}
\label{fig1a}
\end{figure}

Of course the relativistic expansion (\ref{can5}) breaks down well before $%
\beta $ reaches this point. \ In figure \ref{fig-1} we plot the average
energy $\left\langle \zeta \right\rangle $ as a function of $kT$ for $N=10$.
\ The relativistic case is clearly distinguishable from its non-relativistic
counterpart once $kT/Mc^{2}>.02$. 
However the upper bound on the thermal energy is $kT/Mc^{2}<.0117$ for $N=10$. 
In \ref{fig1a}\ we plot the average energy over the allowed range of $kT$%
illustrating
that the distinction between the two cases is about $8\%$ at most. \ For $%
N=1000$ the maximum difference between the two cases is less than one part
in a thousand over the allowed range of $\ kT$. \bigskip 

\bigskip

In the canonical case we take the energy $E$ to be the fixed total average
energy as given by eq. (\ref{can22}). \ Solving this equation for the
inverse temperature $\beta $ yields 
\begin{equation}
\beta =\frac{18\left( N-1\right) -\left[ \left( 5N+3\right) +\frac{8N}{%
\left( N-1\right) }\sum_{k=1}^{N-1}\sum_{l=k+1}^{N-1}\frac{\left( l-k\right) 
}{l\left( N-l\right) }\right] \zeta }{12\left( E-Mc^{2}\right) }=\frac{a_{N}%
}{Mc^{2}\zeta }\left( 1-\frac{b_{N}}{a_{N}^{2}}\zeta \right)   \label{ln5}
\end{equation}%
where 
\begin{equation}
a_{N}=\frac{3}{2}\left( N-1\right) \mbox{ \ \ \ \ \ \ \ }b_{N}=\frac{1}{8}%
\left( \left( 5N+3\right) \left( N-1\right)
+8N\sum_{k=1}^{N-1}\sum_{j=k+1}^{N-1}\frac{\left( j-k\right) }{j\left(
N-k\right) }\right)   \label{ln6}
\end{equation}%
are defined for convenience. \ The limit (\ref{lnZa}) implies that 
\begin{equation}
1 > \left( \frac{b_{N}}{a_{N}^{2}}+\frac{b_{N}}{a_{N}}\right) \zeta
\equiv \frac{\zeta }{\zeta _{\max }}\longrightarrow \left( \frac{7}{4}-\frac{%
\pi ^{2}}{9}\right) N\zeta   \label{ln6a}
\end{equation}%
where the latter limit holds for large $N$. \ Since the exponential forms a
better approximation than the polynomial one, the value of $\zeta _{\max }$
is probably a bit larger than what is given in eq. (\ref{ln6a}), although it
is not clear how much. We obtain 
\begin{eqnarray}
f_{c}^{\ast }(\eta ,\xi ) &=&\frac{M}{N}VLf^{\prime R}(mV\eta ,L\xi )=\frac{2%
}{3N\pi G}\sqrt{\frac{4}{3}}\left( \zeta c^{2}\right) ^{3/2}f^{\prime
R}(mV\eta ,L\xi )  \nonumber \\
&=&\frac{2\left( \frac{2}{3}a_{N}\right) ^{3/2}}{N\sqrt{\pi \left(
N-1)\right) }}\left( 1-\frac{b_{N}}{a_{N}^{2}}\zeta \right) ^{3/2}\exp \left[
\frac{b_{N}}{a_{N}}\zeta \left( 1-\frac{b_{N}}{a_{N}^{2}}\zeta \right) ^{-1}%
\right]   \nonumber \\
&&\times \sum_{l=1}^{N-1}\left[ \left\{ A_{l}^{N}-\frac{N\zeta }{a_{N}}\left[
B_{l}^{N}-A_{l}^{N}\left( 1-\frac{4a_{N}}{3N}\left( 1-\frac{b_{N}}{a_{N}^{2}}%
\zeta \right) l\left| \xi \right| \right) \right] \left( 1-\frac{b_{N}}{%
a_{N}^{2}}\zeta \right) ^{-1}\right. \right.   \nonumber \\
&&+A_{l}^{N}\frac{N\zeta }{2a_{N}}\left[ a_{N}^{2}\frac{4\eta ^{4}\left(
N^{2}-3N+3\right) }{9N\left( N-1\right) ^{3}}\left( 1-\frac{b_{N}}{a_{N}^{2}}%
\zeta \right) \right]   \nonumber \\
&&+A_{l}^{N}\frac{N\zeta }{2a_{N}}\left[ a_{N}\frac{\eta ^{2}\left(
N-2\right) }{N\left( N-1\right) ^{2}}+\frac{3\left( N-2\right) ^{2}}{4\left(
N-1\right) }\left( 1-\frac{b_{N}}{a_{N}^{2}}\zeta \right) ^{-1}\right]  
\nonumber \\
&&+\frac{2N\zeta }{a_{N}}\left( \frac{N}{2\left( N-1\right) }\left( 1-\frac{%
b_{N}}{a_{N}^{2}}\zeta \right) -\frac{2a_{N}\eta ^{2}}{3N}\left[ \frac{N^{2}%
}{\left( N-1\right) ^{2}}\right] \right)   \nonumber \\
&&\mbox{ \ \ \ \ \ \ \ \ \ \ \ \ \ \ \ \ \ \ \ \ \ \ \ \ \ }\times \left[
C_{l}^{N}-\frac{1}{l}A_{l}^{N}\left( 1-\frac{4a_{N}}{3N}\left( 1-\frac{b_{N}%
}{a_{N}^{2}}\zeta \right) l\left| \xi \right| \right) \right]   \nonumber \\
&&\left. -\frac{N\zeta }{a_{N}}\left[ D_{l}^{N}+K_{l}^{N}\left( 1-\frac{%
4a_{N}}{3N}\left( 1-\frac{b_{N}}{a_{N}^{2}}\zeta \right) l\left| \xi \right|
\right) \right] \left( 1-\frac{b_{N}}{a_{N}^{2}}\zeta \right) ^{-1}\right\} 
\nonumber \\
&&\left. \mbox{ \ \ \ \ \ \ \ \ \ \ \ \ \ \ \ \ \ \ \ \ }\times \exp \left(
-\left( 1-\frac{b_{N}}{a_{N}^{2}}\zeta \right) \left( \eta ^{2}+\frac{4a_{N}%
}{3N}l\left| \xi \right| \right) \right) \right]   \label{ln7}
\end{eqnarray}%
which is valid only to first order in $\zeta $. \ 

\begin{figure}
\begin{center}
\epsfig{file=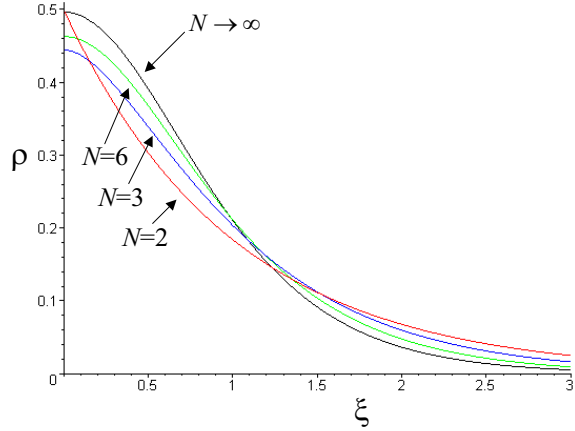,width=0.8\linewidth}
\end{center}
\caption{The non-relativistic canonical density function for various values of 
$N$.}
\label{fig3}
\end{figure}

The canonical density (\ref{can35}) becomes 
\begin{eqnarray}
\rho _{c}^{\ast }\left( \xi \right)  &=&L\rho _{c}(L\xi )=\frac{2\zeta c^{2}%
}{3\pi GM}\rho _{c}(L\xi )  \nonumber \\
&=&\frac{4a_{N}}{3N}\exp \left[ \frac{b_{N}}{a_{N}}\zeta \left( 1-\frac{b_{N}%
}{a_{N}^{2}}\zeta \right) ^{-1}\right]   \nonumber \\
&&\times \sum_{l=1}^{N-1}\left\{ A_{l}^{N}\left( 1-\frac{b_{N}}{a_{N}^{2}}%
\zeta \right) +\frac{N\zeta }{a_{N}}\left( \frac{3}{8}\frac{\left(
N-1\right) ^{2}}{N}A_{l}^{N}-B_{l}^{N}-D_{l}^{N}\right) \right\} 
\label{mirho} \\
&&\left. +\frac{N\zeta }{a_{N}}\left[ A_{l}^{N}-K_{l}^{N}\right] \left( 1-%
\frac{4a_{N}}{3N}\left( 1-\frac{b_{N}}{a_{N}^{2}}\zeta \right) l\left| \xi
\right| \right) \right\} \exp \left( -2\left( 1-1/N\right) \left( 1-\frac{%
b_{N}}{a_{N}^{2}}\zeta \right) l\left| \xi \right| \right)   \nonumber
\end{eqnarray}%
We plot in fig. \ref{fig3} the non-relativistic canonical single-particle
density function $\rho _{c}^{\ast }(\xi ,\zeta =0)$ \ for various values of $%
N$, recovering the results of Rybicki\cite{Rybicki}. With the exception of $%
N=2$, the central density grows with increasing $N$ and the distribution
becomes slightly more sharply peaked.
It can be shown that as $N\rightarrow \infty $ the canonical density 
$\rho _{c}^{\ast }(\xi ,\zeta=0)\rightarrow \frac{1}{2}\sec $h$^{2}\xi $, 
and the single particle
distribution function approaches the isothermal solution of the Vlasov
equation \cite{Rybicki}.

\bigskip

\begin{figure}
\begin{center}
\epsfig{file=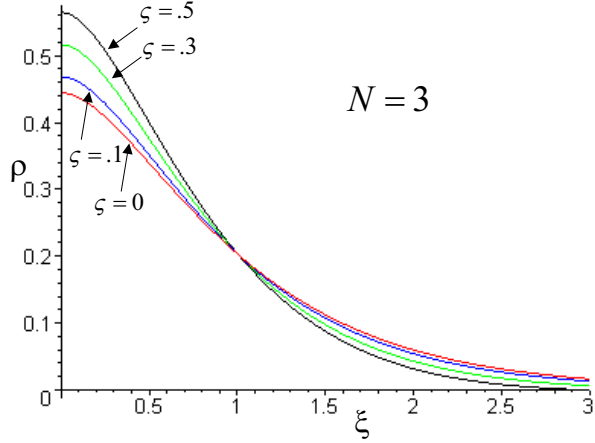,width=0.8\linewidth}
\end{center}
\caption{The canonical density
function for $N=3$ for various values of the relativistic paramter $\protect%
\zeta $. \ The non-relativistic red curve has $\protect\zeta =0$, and the
blue, green and black curves are respectively $\protect\zeta =0.1$, $\protect%
\zeta =0.3$, and $\protect\zeta =0.5$. \ Here eq. (\ref{ln6a}) $\ $\ yields $%
\protect\zeta _{\max }\approx 0.43$ as the limit for which the relativistic
expression is valid.}
\label{fig4}
\end{figure}

\begin{figure}
\begin{center}
\epsfig{file=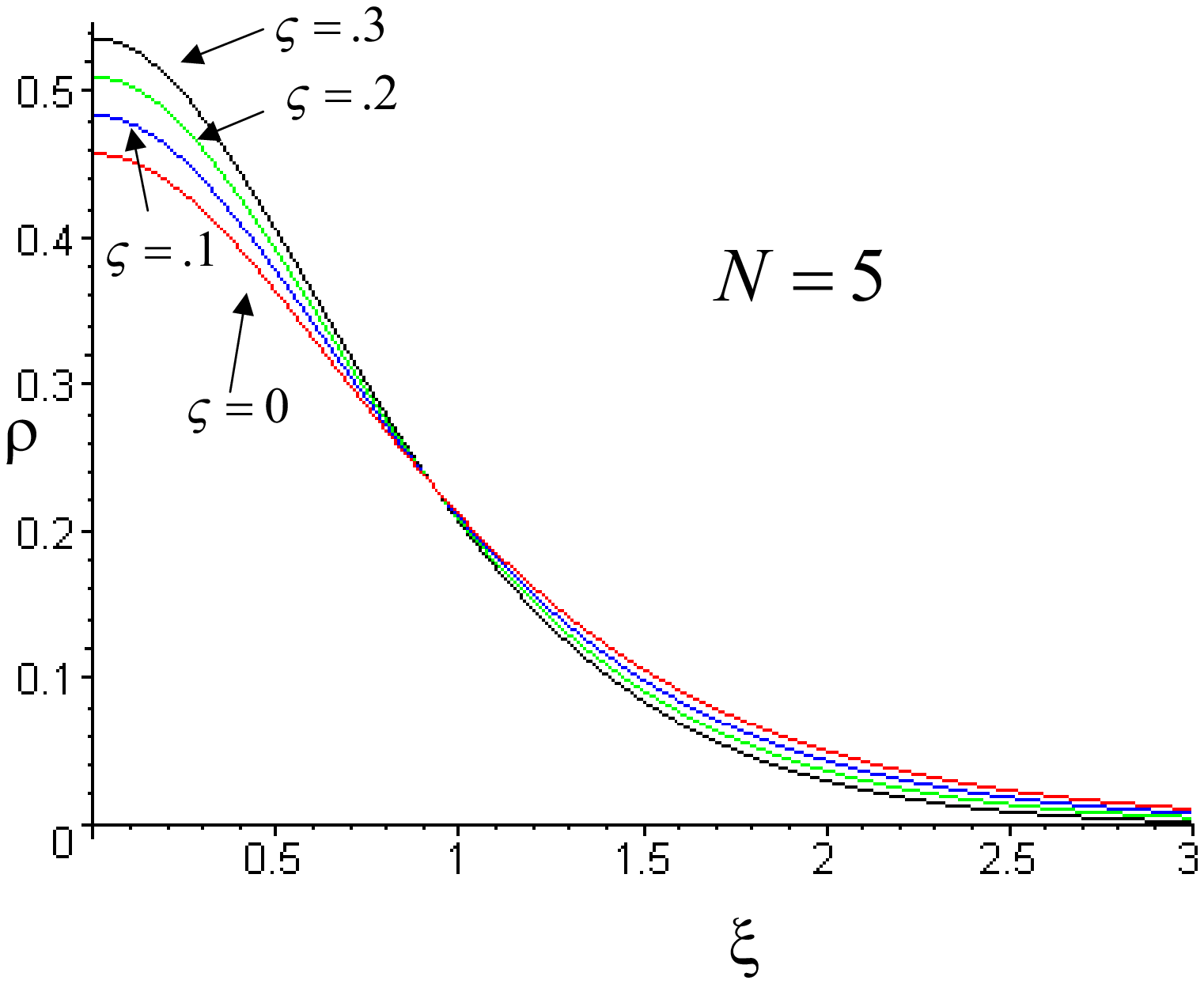,width=0.8\linewidth}
\end{center}
\caption{The canonical density function for $N=5$ where the red blue, green
and black curves respectively correspond to $\protect\zeta =0$, $0.1$, $0.2$%
, $0.3$, and \ $\protect\zeta _{\max }\approx 0.29$.}
\label{fig5}
\end{figure}

\begin{figure}
\begin{center}
\epsfig{file=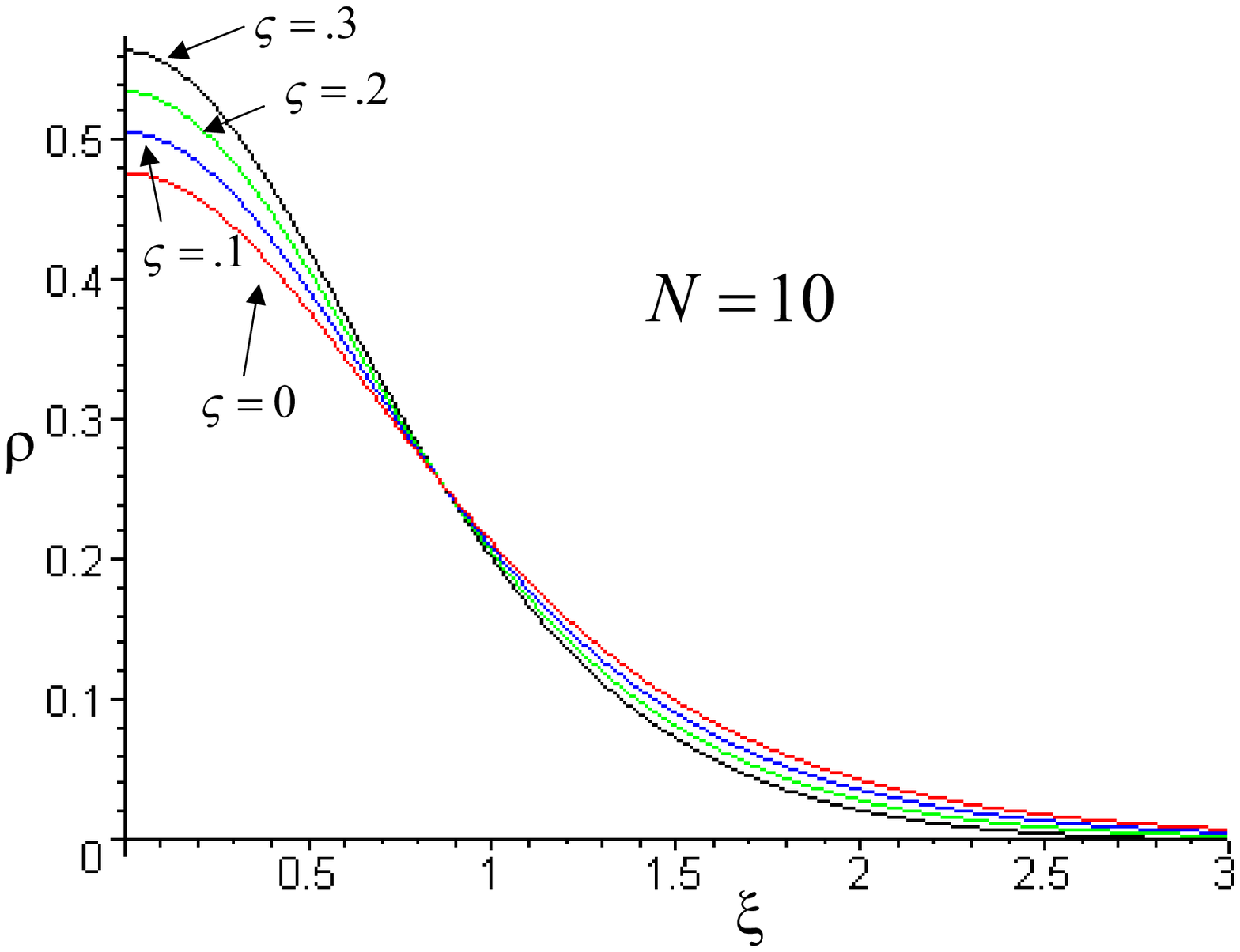,width=0.8\linewidth}
\end{center}
\caption{The canonical density function for $N=10$ where the red
blue, green and black curves respectively correspond to $\protect\zeta =0$, 
$0.1$, $0.2$, $0.3$, and \ $\protect\zeta_{\max }\approx 0.15$.}
\label{fig6}
\end{figure}

\begin{figure}
\begin{center}
\epsfig{ file=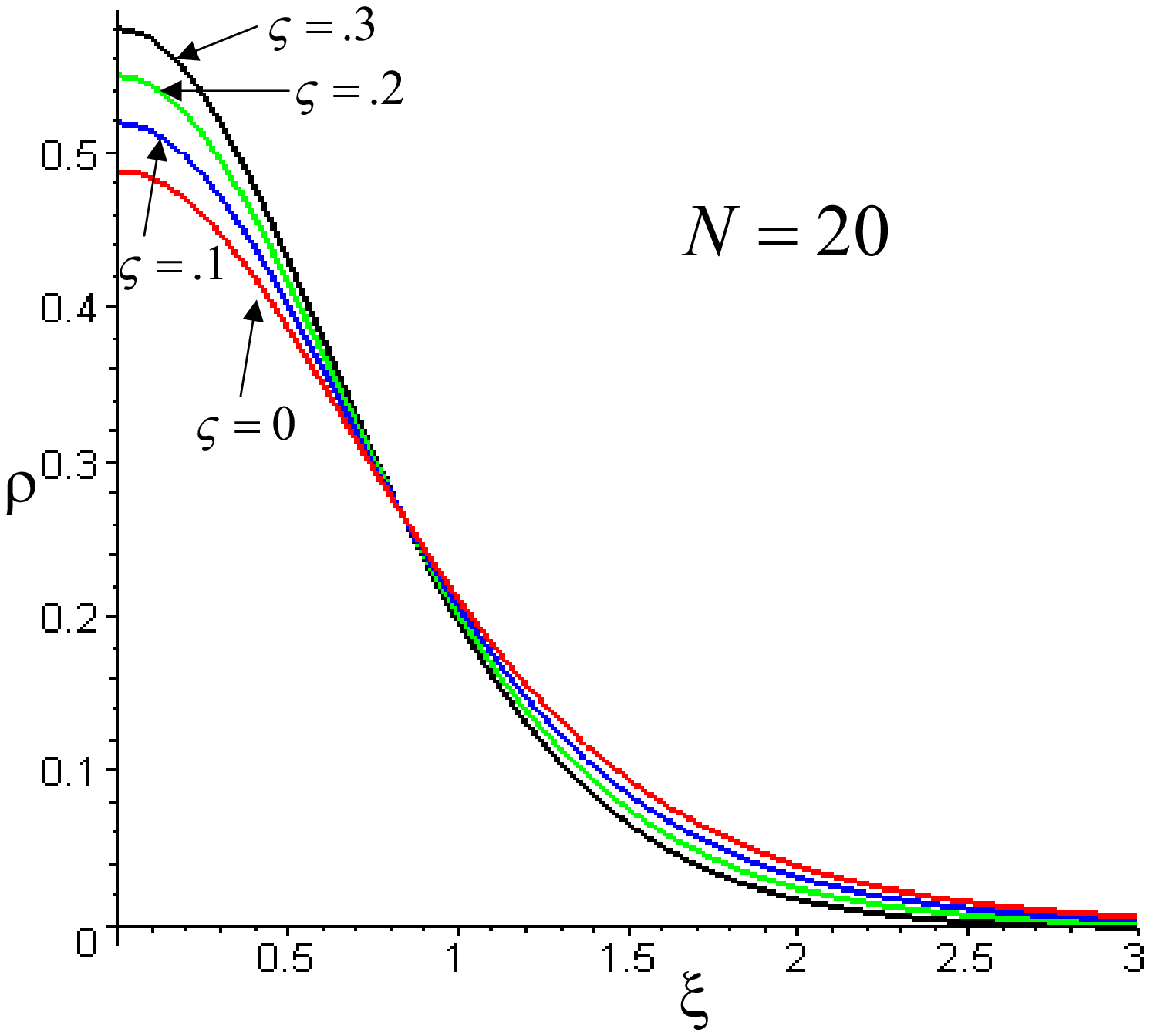,width=0.8\linewidth}
\end{center}
\caption{The canonical density
function for $N=20$ where the red blue, green and black curves respectively
correspond to $\protect\zeta =0$, $0.1$, $0.2$, $0.3$, and 
$\protect\zeta_{\max }\approx 0.15$.}
\label{fig7}
\end{figure}

In figs. \ \ref{fig4} -- \ref{fig7}\ we plot the relativistic canonical
function $\rho _{c}^{\ast }(\xi ;\zeta )$ for differing values of $N$. As is
clear from these figures, relativistic effects significantly enhance the
central density by as much as 30\% depending on the magnitude of $\zeta $.
Even for $\zeta =0.3$, the central density is larger than its value of $1/2$
in the non-relativistic large $N$ limit. The falloff of the relativistic
density functions is also more rapid than in the non-relativistic case. \ \ 

\begin{figure}
\begin{center}
\epsfig{file=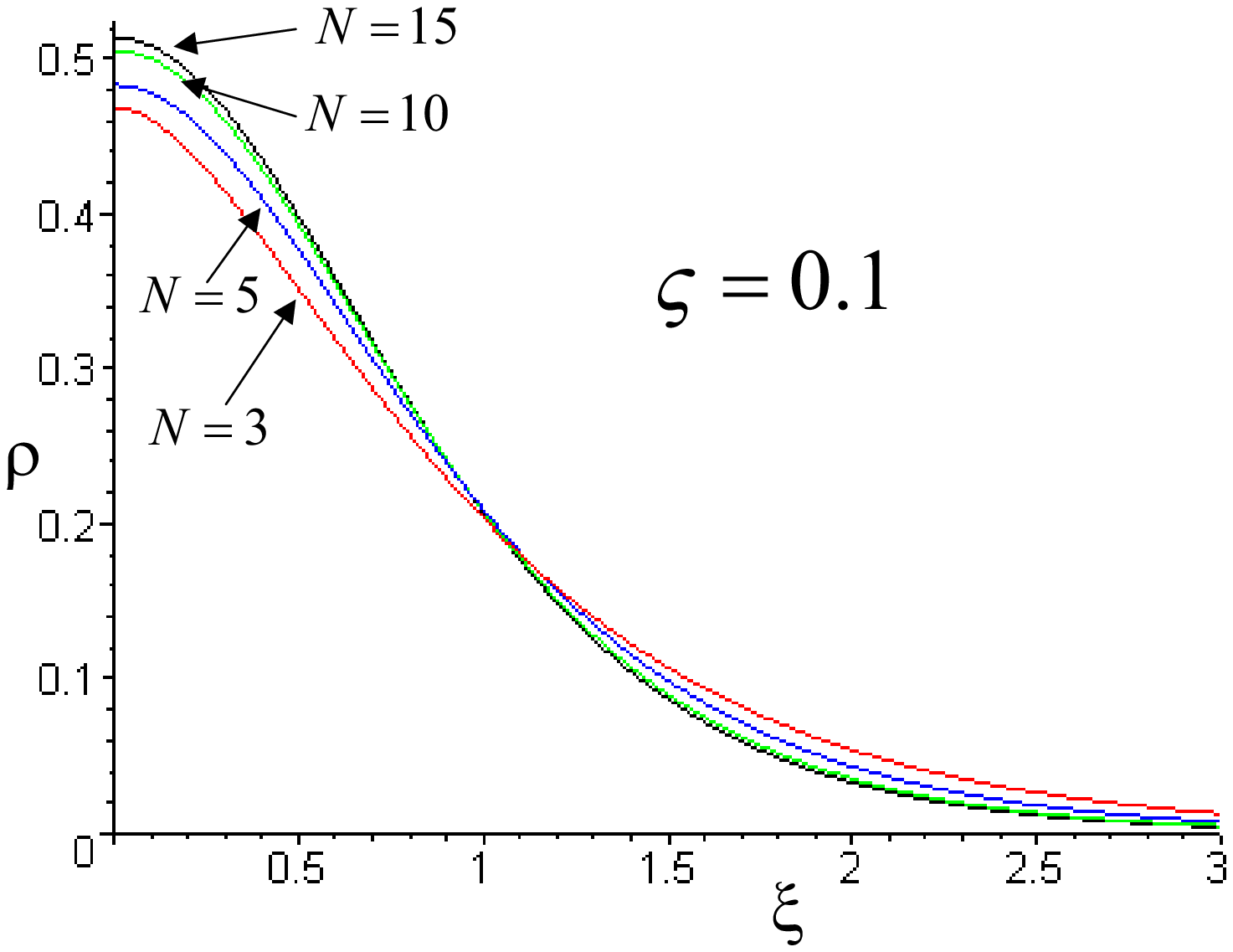,width=0.8\linewidth}
\end{center}
\caption{The
canonical density for different values of $N$ at $\protect\zeta =0.1$}
\label{fig8}
\end{figure}

\begin{figure}
\begin{center}
\epsfig{file=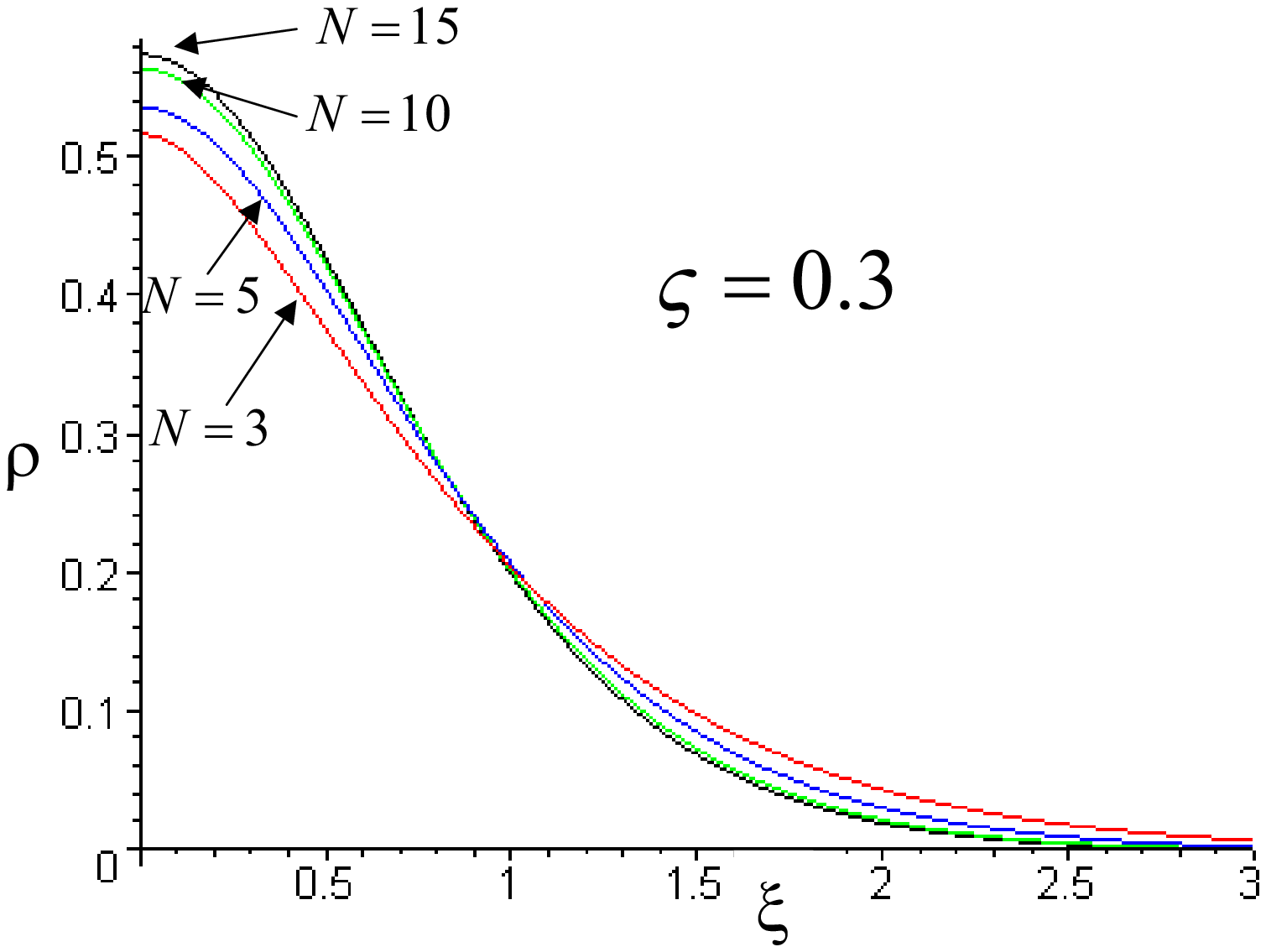,width=0.8\linewidth}
\end{center}
\caption{The
canonical density for different values of $N$ at $\protect\zeta =0.3$}
\label{fig9}
\end{figure}

Unfortunately there is no closed-form expression for the terms $K_{l}^{N}$
and $D_{l}^{N}$ \ and so it is not possible to evaluate an explicit
expression for either $\ \rho _{c}^{\ast }\left( \xi \right) $ or $\rho
_{mc}^{\ast }\left( \xi \right) $ in the large $N$ \ limit. \ Instead these
quantities must be computed using symbolic algebra; for $N>20$ this involve
the factorization of thousands of terms, and computer memory limitations
make this a prohibitive task. However the large $N$ behaviour should not be
too different from the $N=20$ case, at least for small values of $\zeta $. \
Figures \ref{fig8} and \ref{fig9} plot the density for different $N$ at two
different values of $\zeta $. 

The canonical momentum distribution function (\ref{can36}) is
straightforwardly evaluated to be 
\begin{eqnarray}
\vartheta _{cn}^{\ast }(\eta ) &=&mV\vartheta _{cn}(\sigma V\eta )=m\sqrt{%
\frac{4\zeta c^{2}}{3}}\vartheta _{cn}(\sigma V\eta )  \nonumber \\
&=&\frac{1}{\sqrt{\pi }}\left( 1-\frac{b_{N}}{a_{N}^{2}}\zeta \right)
^{1/2}\exp \left[ -\eta ^{2}\left( 1-\frac{b_{N}}{a_{N}^{2}}\zeta \right) %
\right] \left( 1+\frac{N\zeta }{a_{N}}\frac{\left( N^{2}-3N+3\right) }{%
2N\left( N-1\right) }\left( 1-\frac{b_{N}}{a_{N}^{2}}\zeta \right) \eta
^{4}\right.   \nonumber \\
&&\times \left. \frac{N\zeta }{a_{N}}\left( -\frac{\eta ^{2}\left(
4N^{2}-7N+6\right) }{2N\left( N-1\right) }+\frac{5N(N-1)+3}{8N\left(
N-1\right) }\left( 1-\frac{b_{N}}{a_{N}^{2}}\zeta \right) ^{-1}\right)
\right)   \label{mitheta}
\end{eqnarray}%
where each of (\ref{mirho},\ref{mitheta}) are also valid to first order in $%
\zeta $. Here we can easily take the large $N$ limit, which is 
\begin{equation}
\vartheta _{c}^{\ast }(\eta )\rightarrow \frac{1}{\sqrt{\pi }}(1+\frac{\zeta 
}{54}(18\eta ^{4}-9\eta ^{2}-9-2\pi ^{2}\left( 2\eta ^{2}-1\right) )e^{-\eta
^{2}}  \label{ln14}
\end{equation}%
to leading order in $\zeta $. \ 

\bigskip

\begin{figure}
\begin{center}
\epsfig{file=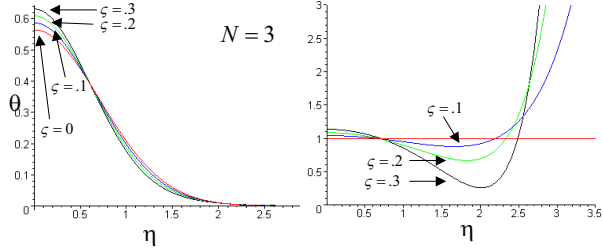,width=0.8\linewidth}
\end{center}
\caption{The canonical momentum
density as a function of $\protect\eta $ for $N=3.$ On the left hand side is
the behaviour of $\ \protect\theta _{c}^{\ast }\left( \protect\eta ;\protect%
\zeta \right) $\ and on the right hand side is its behaviour relative to the
non-relativistic density $\protect\theta _{c}^{\ast }\left( \protect\eta %
;0\right) $. The curves are respectively red, blue, green and black for $%
\protect\zeta =0,0.1,0.2,0.3$.}
\label{fig10}
\end{figure}

\begin{figure}
\begin{center}
\epsfig{file=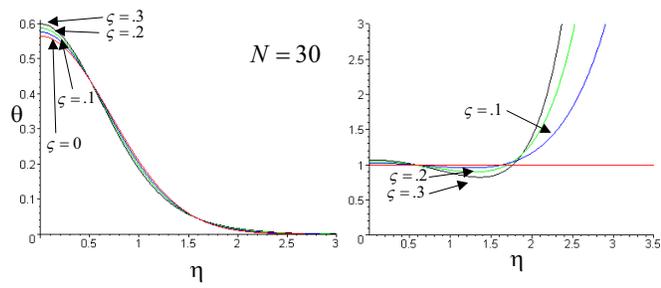,width=0.8\linewidth}
\end{center}
\caption{The canonical momentum
density as a function of $\protect\eta $ for $N=30.$ Notation is as in fig %
\ref{fig10}.}
\label{fig11}
\end{figure}

\begin{figure}
\begin{center}
\epsfig{file=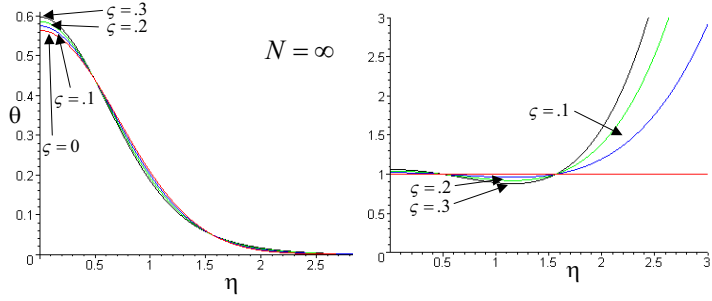,width=0.8\linewidth}
\end{center}
\caption{The canonical
momentum density as a function of $\protect\eta $ for $N=\infty .$ Notation
is as in fig \ref{fig10}.}
\label{fig12}
\end{figure}

We plot in figs. \ref{fig10} -- \ref{fig12} the behaviour of the canonical
momentum density as a function of the rescaled momentum $\eta $ for
differing values of $N.$ The central momentum density increases with
increasing $\zeta $, and falls off more rapidly than in the non-relativistic
case. \ 

However for $\eta >2$, the
momentum density grows relative to its non-relativistic counterpart,
overtaking this value for large enough $\eta $ . The relative growth is
exponential, although the overall momentum density is exponentially damped
for any $\zeta $.
As $N$ increases, the differences
between the non-relativistic and relativistic cases become less pronounced,
although the basic features remain the same even in the limit that $%
N\rightarrow \infty $.

The microcanonical results are
\begin{eqnarray}
&&f_{mc}^{\prime \ast R}(\eta ,\xi )  \nonumber \\
&=&\frac{2}{3N\pi G}\sqrt{\frac{4}{3}}\left( \zeta c^{2}\right)
^{3/2}f^{\prime R}(mV\eta ,L\xi )  \nonumber \\
&=&\frac{2}{3N}\sqrt{\frac{2}{3\pi \left( N-1\right) }}\frac{\Gamma \left( 
\frac{3}{2}\left( N-1\right) \right) }{\Gamma \left( \frac{3}{2}\left(
N-2\right) \right) }\exp \left[ \frac{b_{N}}{a_{N}}\zeta \right]
\sum_{l=1}^{N-1}\left\{ A_{l}^{N}\left( 1-\frac{2\eta ^{2}}{3\left(
N-1\right) }-\frac{4l\left| \xi \right| }{3N}\right) _{+}^{3N/2-4}\right.  
\nonumber \\
&&+\zeta \eta ^{2}\left( A_{l}^{N}\left[ \frac{\left( N-2\right) }{\left(
N-1\right) ^{2}}\right] -\left( C_{l}^{N}-\frac{1}{l}A_{l}^{N}\right) \left[ 
\frac{4N^{2}}{3\left( N-1\right) ^{2}}\right] \right) \left( 1-\frac{2\eta
^{2}}{3\left( N-1\right) }-\frac{4l\left| \xi \right| }{3N}\right)
_{+}^{3N/2-4}  \nonumber \\
&&+\left( \frac{3}{2}N-4\right) \zeta A_{l}^{N}\left[ \frac{2\eta ^{4}\left(
N^{2}-3N+3\right) }{9\left( N-1\right) ^{3}}-\frac{16N\left| \xi \right|
\eta ^{2}}{9\left( N-1\right) ^{2}}\right] \left( 1-\frac{2\eta ^{2}}{%
3\left( N-1\right) }-\frac{4l\left| \xi \right| }{3N}\right) _{+}^{3N/2-5} 
\nonumber \\
&&+\left[ \frac{4\zeta }{3}\left| \xi \right| \left( A_{l}^{N}\left( \frac{N%
}{\left( N-1\right) }-l\right) +lK_{l}^{N}\right) \right] \left( 1-\frac{%
2\eta ^{2}}{3\left( N-1\right) }-\frac{4l\left| \xi \right| }{3N}\right)
_{+}^{3N/2-4}  \nonumber \\
&&+\frac{N\zeta }{\left( \frac{3}{2}N-3\right) }\left( A_{l}^{N}\left[ \frac{%
3\left( N-2\right) ^{2}}{8\left( N-1\right) }+1\right] -B_{l}^{N}\mbox{\ }+%
\frac{N\left( C_{l}^{N}-\frac{1}{l}A_{l}^{N}\right) }{\left( N-1\right) }%
-D_{l}^{N}-K_{l}^{N}\right)   \nonumber \\
&&\mbox{ \ \ \ \ \ \ \ \ \ \ \ \ \ \ \ \ \ \ \ \ \ \ \ \ \ \ \ \ \ \ \ \ \ \
\ }\left. \times \left( 1-\frac{2\eta ^{2}}{3\left( N-1\right) }-\frac{%
4l\left| \xi \right| }{3N}\right) _{+}^{3N/2-3}\right\}   \label{ln10}
\end{eqnarray}

\begin{eqnarray}
\rho _{mc}^{\ast }(\xi ) &\equiv &L\rho _{mc}(L\xi )  \nonumber \\
&=&2\exp \left[ \frac{b_{N}}{a_{N}}\zeta \right] \sum_{l=1}^{N-1}\left\{
\left( \frac{3N-5}{3N}\right) \left[ A_{l}^{N}-\frac{4l\left| \xi \right| }{3%
}\zeta \left( A_{l}^{N}-K_{l}^{N}\right) \right] \left( 1-\frac{4l\left| \xi
\right| }{3N}\right) _{+}^{3N/2-7/2}\right.   \nonumber \\
&&\left. +\frac{2}{3}\zeta \left( 1-\frac{4l\left| \xi \right| }{3N}\right)
_{+}^{3N/2-5/2}\left( \left( \frac{3(N-1)^{2}}{8N}\right) A_{l}^{N}-\left(
B_{l}^{N}-A_{l}^{N}\right) -\left( D_{l}^{N}+K_{l}^{N}\right) \right)
\right\} \mbox{ \ \ \ \ \ \ \ \ \ \ \ \ }  \label{ln11}
\end{eqnarray}%
and \label{thetmic} 
\begin{eqnarray}
\vartheta _{mc}^{\ast }(\eta ) &\equiv &mV\vartheta _{mc}(mV\eta )  \nonumber
\\
&=&\sqrt{\frac{2}{3\pi \left( N-1\right) }}\frac{\Gamma \left( \frac{3}{2}%
\left( N-1\right) \right) }{\Gamma \left( \frac{3}{2}N-2\right) }\exp \left[ 
\frac{b_{N}}{a_{N}}\zeta \right]   \nonumber \\
&&\times \left\{ \left[ 1-\left( \frac{\eta ^{2}}{3}\left[ \frac{8N^{2}-11N+6%
}{\left( N-1\right) ^{2}}\right] \right) \zeta \right] \left( 1-\frac{2\eta
^{2}}{3\left( N-1\right) }\right) _{+}^{3N/2-3}\right.   \nonumber \\
&&\mbox{ \ \ \ \ \ \ \ \ \ \ \ }+\frac{\zeta \left( N-2\right) }{3}\left[ 
\frac{\eta ^{4}\left( N^{2}-3N+3\right) }{\left( N-1\right) ^{3}}\right]
\left( 1-\frac{2\eta ^{2}}{3\left( N-1\right) }\right) _{+}^{3N/2-4}
\label{ln12} \\
&&\mbox{\ }\left. -\frac{2N\zeta }{\left( 3N-4\right) }\left( \frac{\left(
5N^{2}-20N+12\right) }{8\left( N-1\right) }\mbox{\ }+\sum_{s=1}^{N-1}%
\sum_{t=s+1}^{N-1}\frac{\left( t-s\right) }{t\left( N-s\right) }\right)
\left( 1-\frac{2\eta ^{2}}{3\left( N-1\right) }\right) _{+}^{3N/2-2}\right\} 
\nonumber
\end{eqnarray}%
where each of (\ref{ln10},\ref{ln11},\ref{ln12}) are valid to first order in 
$\zeta $. \ As $N\rightarrow \infty $ we have 
\begin{equation}
\vartheta _{mc}^{\ast }(\eta )\rightarrow \frac{1}{\sqrt{\pi }}(1+\frac{%
\zeta }{54}(18\eta ^{4}-81\eta ^{2}+27-2\pi ^{2}\left( 2\eta ^{2}-1\right)
)e^{-\eta ^{2}}  \label{ln13}
\end{equation}%
Note that this differs from the canonical momentum density unless $\zeta =0.$

\begin{figure}
\begin{center}
\epsfig{file=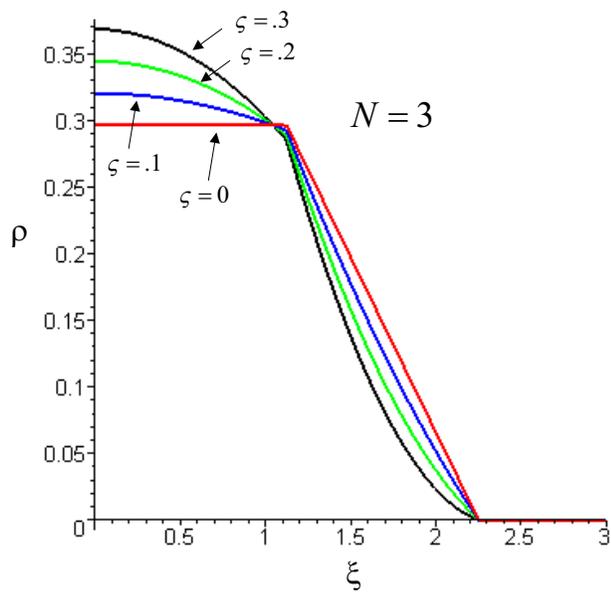,width=0.8\linewidth}
\end{center}
\caption{The microcanonical
density function for $N=3$ for various values of the relativistic paramter $%
\protect\zeta $. \ The non-relativistic red curve has $\protect\zeta =0$,
and the blue, green and black curves are respectively $\protect\zeta =0.1$, $%
\protect\zeta =0.2$, and $\protect\zeta =0.3$.}
\label{fig13}
\end{figure}

\begin{figure}
\begin{center}
\epsfig{file=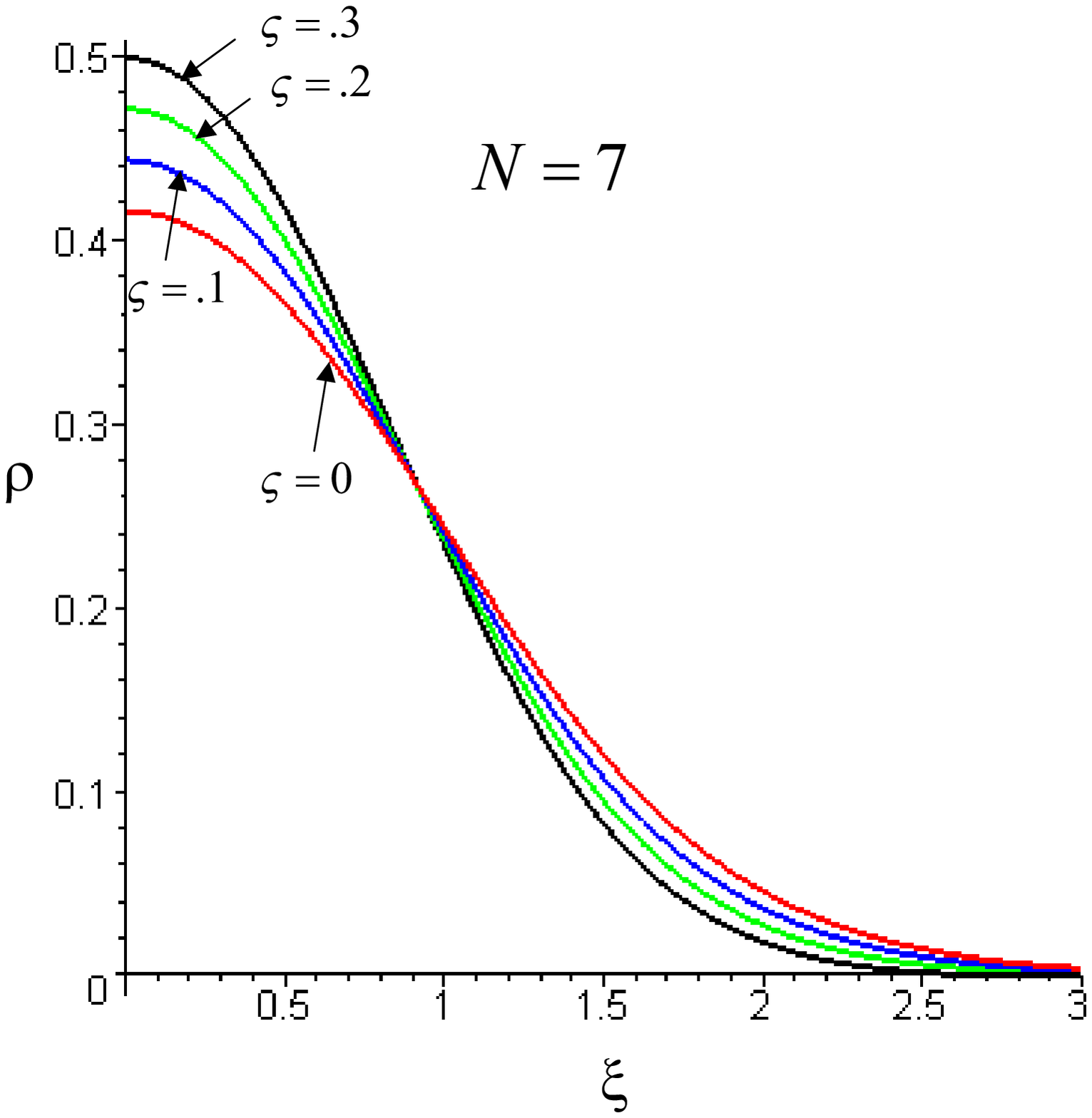,width=0.8\linewidth}
\end{center}
\caption{The microcanonical density function for $N=7$ for various 
values of the relativistic paramter $\protect\zeta $. 
Notation is as in fig. \ref{fig13}.}
\label{fig14}
\end{figure}

\begin{figure}
\begin{center}
\epsfig{file=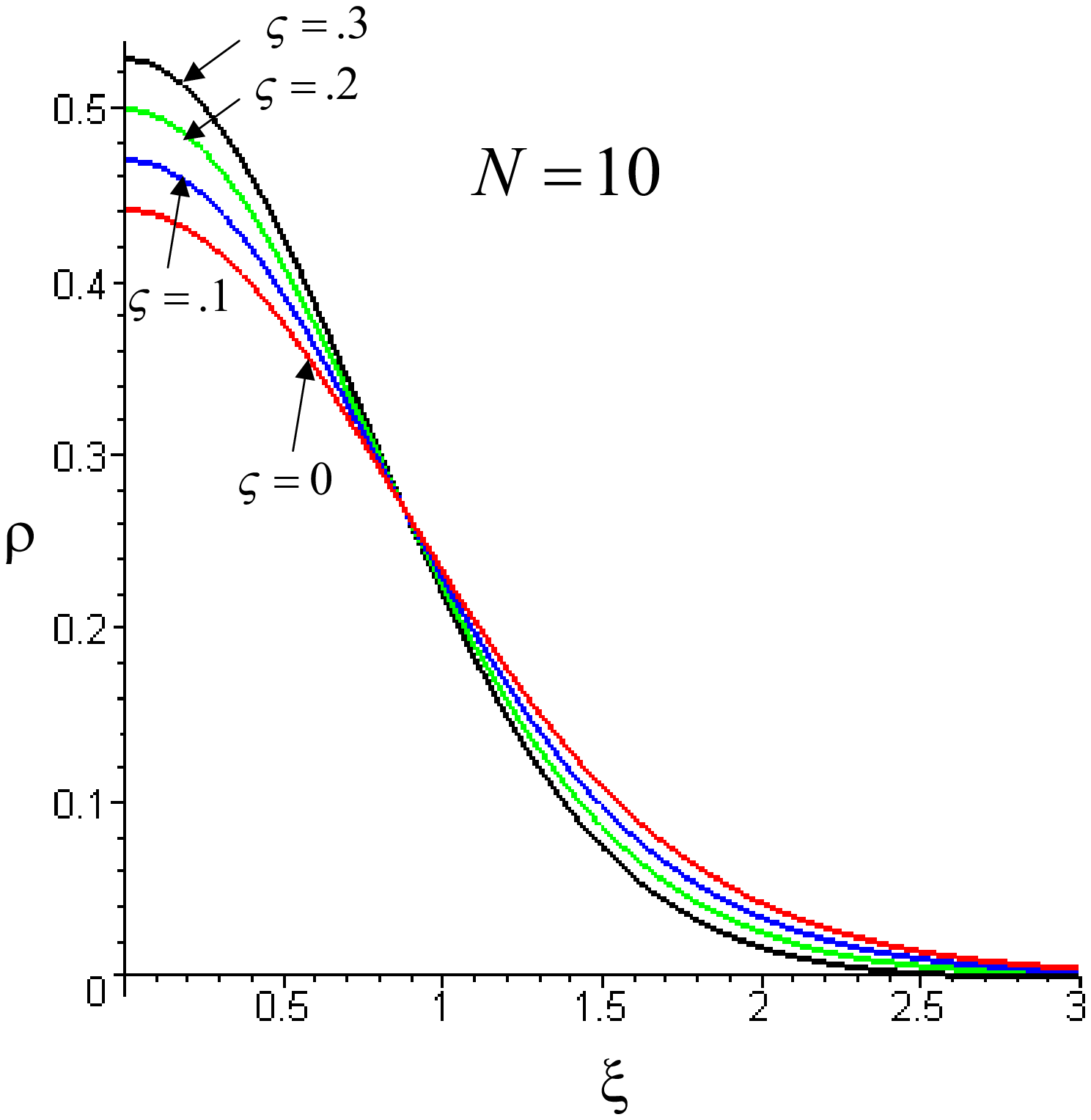,width=0.8\linewidth}
\end{center}
\caption{The microcanonical density function for $N=10$ for various 
values of the relativistic paramter $\protect\zeta $. 
Notation is as in fig. \ref{fig13}.}
\label{fig15}
\end{figure}

\begin{figure}
\begin{center}
\epsfig{file=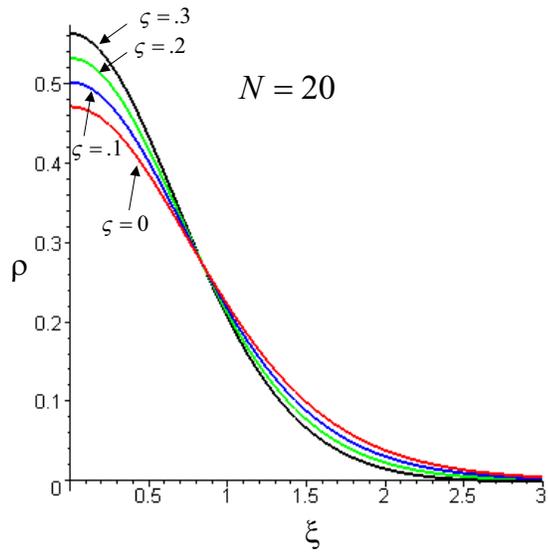,width=0.8\linewidth}
\end{center}
\caption{The microcanonical density function for $N=20$ for various 
values of the relativistic paramter $\protect\zeta $. 
Notation is as in fig. \ref{fig13}.}
\label{fig16}
\end{figure}

The microcanonical density function $\rho _{mc}^{\ast }(\xi ;\zeta )$ for
differing values of $N$ is plotted in figs \ref{fig13}--\ref{fig16}. For $N=3
$, the microcanonical density is uniform until $\xi =9/8$, after which it
falls linearly to zero. \ For $\xi <9/8$, at most two particles can
contribute to the density; in this region relativistic effects enhance their
contribution. \ However for $\xi >9/8$, at most one particle can contribute,
and relativistic effects suppress its contribution until $\xi =9/4$, after
which the density vanishes. 
For larger $N$, as in the
canonical case, relativistic effects significantly enhance the central
density by as much as 30\%, depending on the size of $\zeta $. \ Their
falloff is more rapid, and for sufficiently large $\xi $ the
non-relativistic density distribution is larger. \ As $N$ increases, the
density becomes more sharply peaked and the contrasts between the
non-relativistic and relativistic cases become less pronounced. 

\begin{figure}
\begin{center}
\epsfig{file=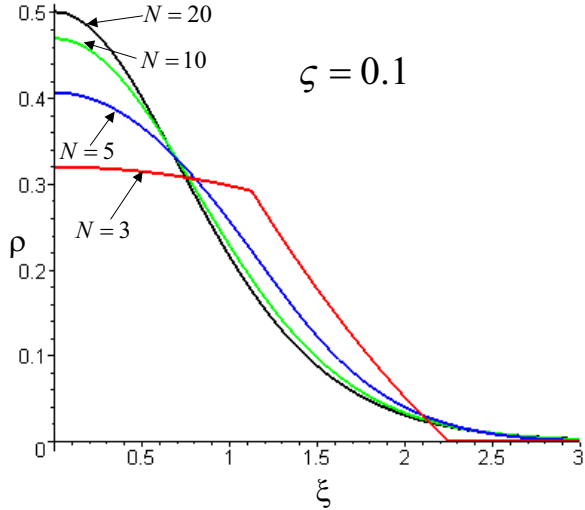,width=0.8\linewidth}
\end{center}
\caption{The
microcanonical density for different values of $N$ at $\protect\zeta =0.1$}
\label{fig17}
\end{figure}

\begin{figure}
\begin{center}
\epsfig{file=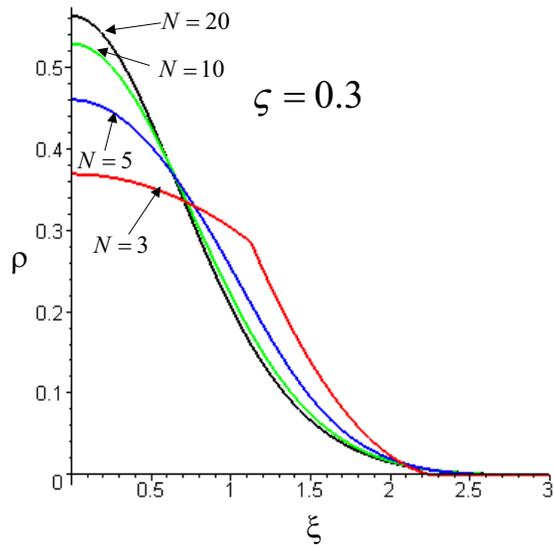,width=0.8\linewidth}
\end{center}
\caption{The
microcanonical density for different values of $N$ at $\protect\zeta =0.3$}
\label{fig18}
\end{figure}

In figs. \ref{fig17}, \ref{fig18} 
we plot the microcanonical density distribution for increasing values of $N
$ and fixed $\zeta $.

\begin{figure}
\begin{center}
\epsfig{file=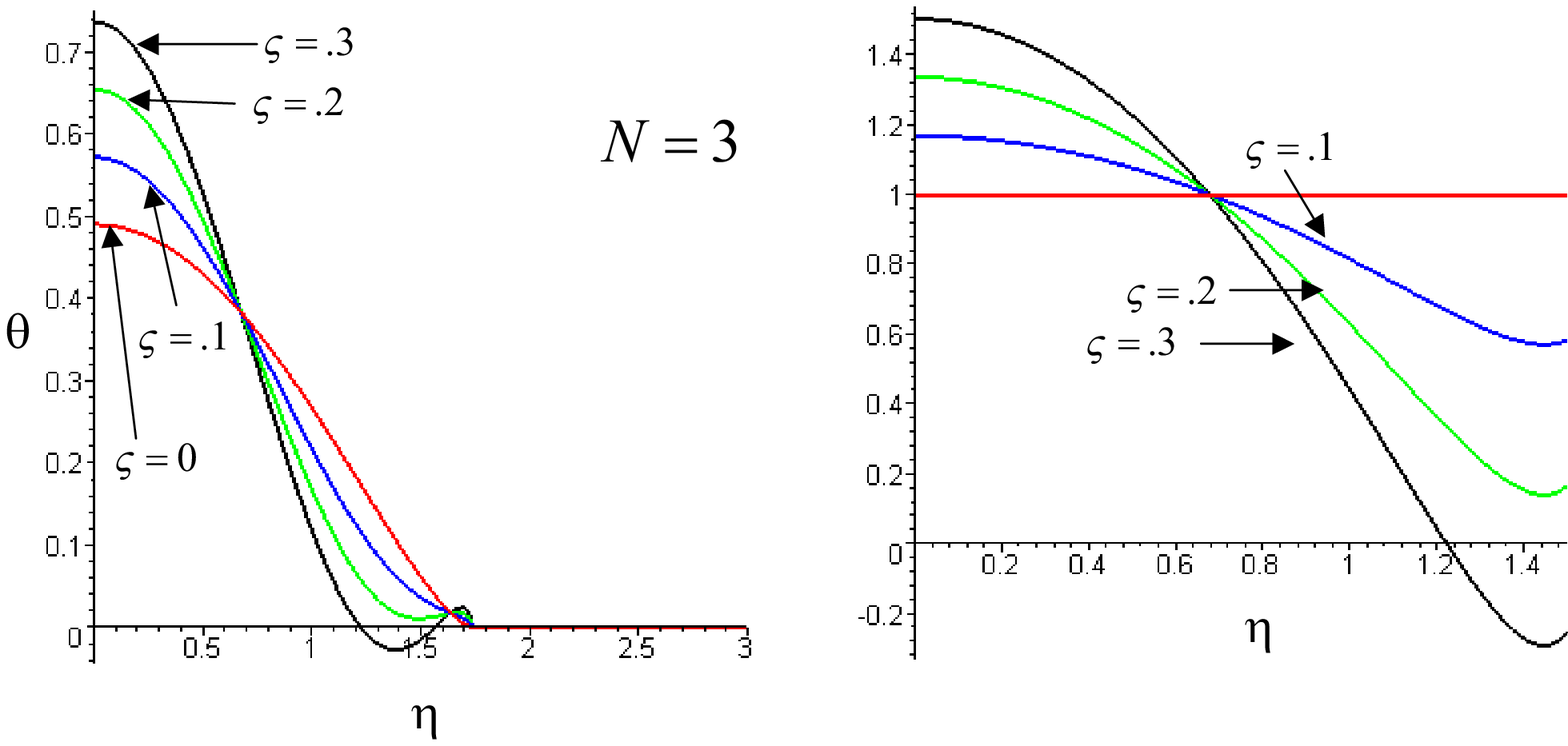,width=0.8\linewidth}
\end{center}
\caption{The microcanonical momentum density as a function of $%
\protect\eta $ for $N=3.$ On the left hand side is the behaviour of $\ 
\protect\theta _{mc}^{\ast }\left( \protect\eta ;\protect\zeta \right) $\
and on the right hand side is its behaviour relative to the non-relativistic
density $\protect\theta _{c}^{\ast }\left( \protect\eta ;0\right) $. The
curves are respectively red, blue, green and black for $\protect\zeta %
=0,0.1,0.2,0.3$.}
\label{fig19}
\end{figure}

We plot the microcanonical momentum distributions in figs. \ref{fig19}--\ref%
{fig21}. The results are qualitatively similar to the canonical case,
although the actual functional forms differ. 

For small $N$, the relativistic approximation breaks down even for $\zeta $
as small as $0.3$, and the momentum distribution function goes negative, as
shown in fig.\ref{fig19}. 
\begin{figure}
\begin{center}
\epsfig{file=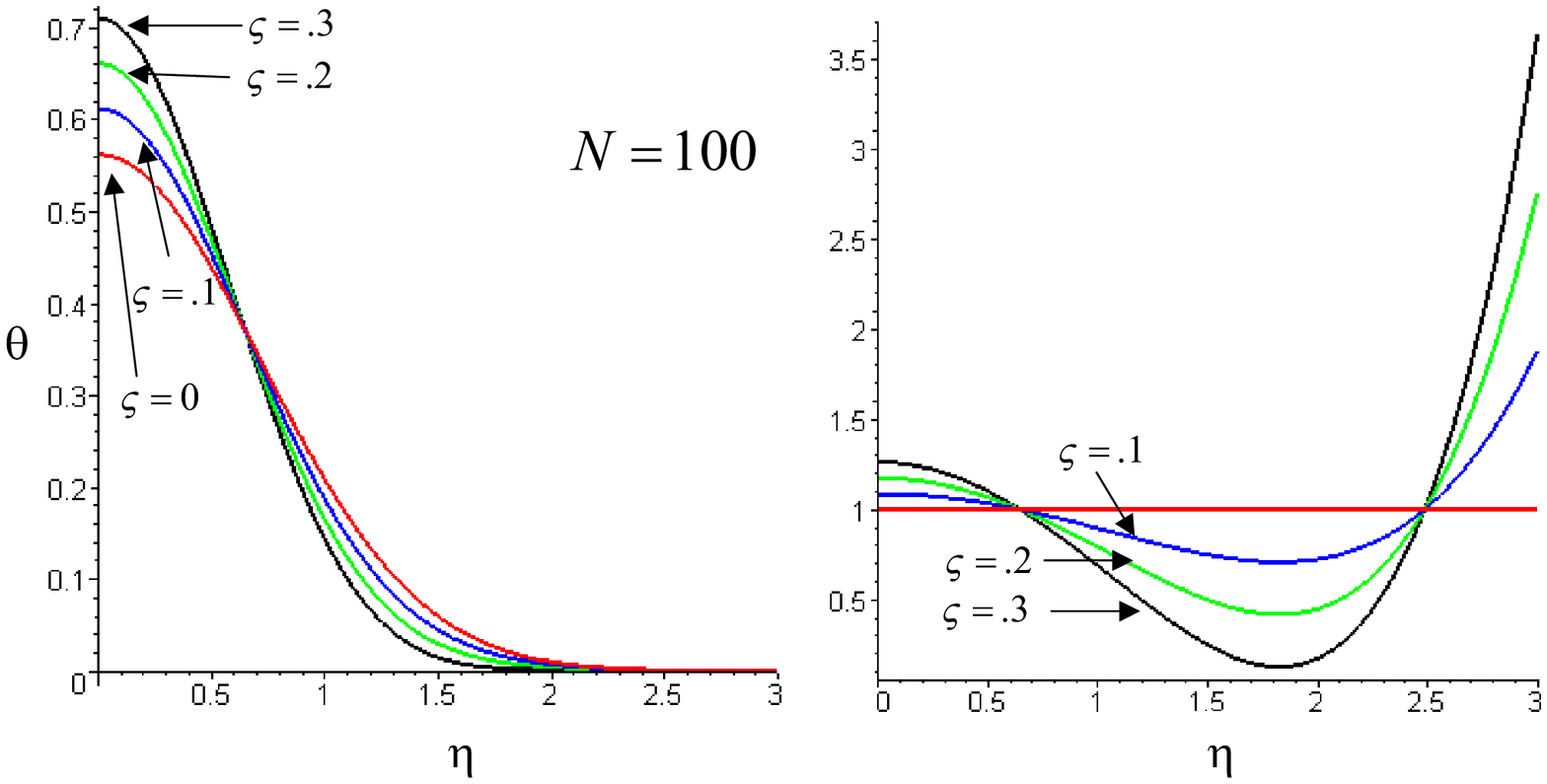,width=0.8\linewidth}
\end{center}
\caption{The
microcanonical momentum density as a function of $\protect\eta $ for $N=100$.
Notation is as in fig. \ref{fig19}}
\label{fig20}
\end{figure}
However for larger $N$ the
momentum distribution is positive for all $\zeta $ $\leq 0.3$, for example
as in fig. \ref{fig20}. \ The relativistic densities are more sharply peaked
and for sufficiently large momentum parameter $\eta $ are larger than their
non-relativistic counterparts.
\begin{figure}
\begin{center}
\epsfig{file=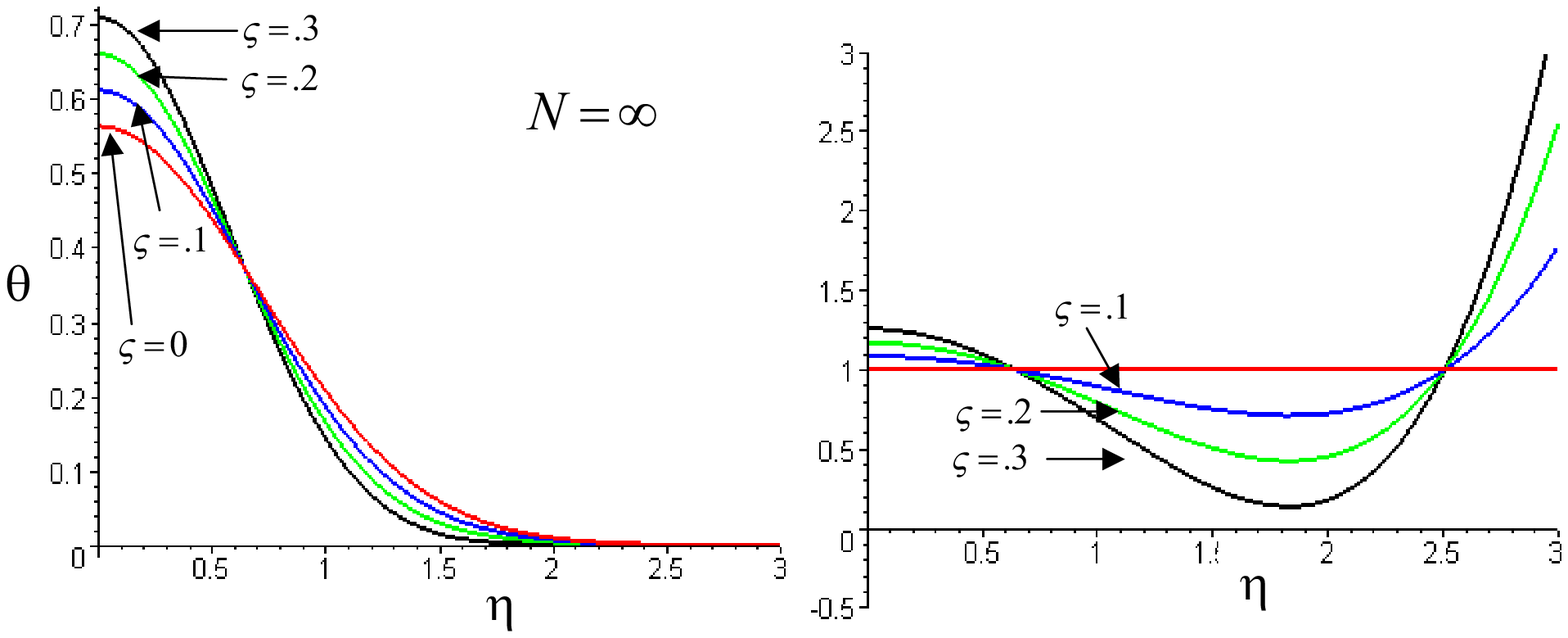,width=0.8\linewidth}
\end{center}
\caption{The microcanonical momentum density as a 
function of $\protect\eta$ for $N=\infty$. 
Notation is as in fig. \ref{fig19}}
\label{fig21}
\end{figure}

\section{Closing Remarks}

We have carried out the first analysis of the statistical behaviour of a
ROGS to leading order in $1/c$. \ The qualitative behaviour of the ROGS as
compared to its non-relativistic OGS counterpart \cite{Rybicki} is clear. \
At a given energy, the ROGS temperature is smaller than the OGS temperature;
relativistic effects cool the gas down. The one-particle distribution
functions become more sharply peaked in each case with increasing $N$. \ For
a given $N$, the ROGS density functions become more sharply peaked as the
relativistic parameter $\zeta $ increases. For both canonical and
microcanonical distribution functions, $\rho _{\mbox{OGS}}>\rho _{\mbox{ROGS}%
}$ for sufficiently large position parameter $\xi $. However for the
momentum densities, this is not true. Although $\vartheta _{\mbox{OGS}%
}>\vartheta _{\mbox{ROGS}}$ for intermediate values of the momentum
parameter $\eta $, once $\eta $ becomes large enough this inequality is
reversed. \ This behaviour is presumably due to quadratic character of the
ROGS potential relative to its linear OGS\ counterpart in the former case,
and from the $\ p^{4}$ corrections in the Hamiltonian (\ref{can7}) in the
latter situation. \ 

This work can be extended in several directions. \ It would be
straightforward to extend these results to the charged and cosmological
systems considered in refs. \cite{2bdcoslo,2bdchglo} to see what effects
these impose on the distribution functions. Extensions to unequal masses
would also be interesting, although considerably more difficult. It would be
very interesting to go beyond leading order in $1/c$ to investigate
non-perturbative effects of the\ ROGS.

Further understanding the ROGS will undoubtedly require numerical
experiments for various values of $N$. \ The equations of motion yield
quartic (as opposed to quadratic) time-dependence of the position variables
(to leading order in $1/c$), and so can be straightforwardly (although
somewhat tediously) integrated to investigate its equilibrium and
equipartion properties. Work on this is in progress.

\bigskip

{\huge Acknowledgements}

This work was supported by the Natural Sciences and Engineering Research
Council of Canada. We would like to thank T. Ohta for interesting discussion
and correspondence.

\section{Appendix}

\subsection{The Gaussian integrations}

The following Gaussian integrations were used to obtain (\ref{can11a}) 
\begin{eqnarray}
&&\int \frac{dk}{2\pi }\int d{\bf p}\exp \left( ik\sum_{a=1}^{N}p_{a}-\beta
\sum_{a=1}^{N}\frac{p_{a}^{2}}{2m}\right)   \nonumber \\
&=&\int \frac{dk}{2\pi }\prod_{a=1}^{N}\int dp_{a}e^{-\beta \frac{p_{a}^{2}}{%
2m}+ikp_{a}}=\left( \frac{2m}{\beta }\right) ^{N/2}\left( \frac{2m}{\beta }%
\right) ^{-1/2}\int \frac{d\tilde{k}}{2\pi }\prod_{a=1}^{N}\int d\tilde{p}%
_{a}e^{-\tilde{p}_{a}^{2}+i\tilde{k}\tilde{p}_{a}}  \nonumber \\
&=&\left( \frac{2m}{\beta }\right) ^{\left( N-1\right) /2}\pi ^{N/2}\int 
\frac{d\tilde{k}}{2\pi }\exp \left( -\frac{N\tilde{k}^{2}}{4}\right) =\frac{1%
}{\sqrt{N}}\left( \frac{2\pi m}{\beta }\right) ^{\left( N-1\right) /2}
\label{can12a}
\end{eqnarray}%
\begin{eqnarray}
&&\int \frac{dk}{2\pi }\int d{\bf p\,}p_{a}^{2}\exp \left(
ik\sum_{b=1}^{N}p_{b}-\beta \sum_{b=1}^{N}\frac{p_{b}^{2}}{2m}\right)  
\nonumber \\
&=&\left( \frac{2m}{\beta }\right) ^{(N+2)/2}\left( \frac{2m}{\beta }\right)
^{-1/2}\int \frac{d\tilde{k}}{2\pi }\int d\tilde{p}_{a{\bf \,}}\tilde{p}%
_{a}^{2}e^{-\tilde{p}_{a}^{2}+i\tilde{k}\tilde{p}_{a}}\prod_{b\neq
a}^{N}\int d\tilde{p}_{b}e^{-\tilde{p}_{b}^{2}+i\tilde{k}\tilde{p}_{b}} 
\nonumber \\
&=&\left( \frac{2m}{\beta }\right) ^{(N+1)/2}\pi ^{N/2}\int \frac{d\tilde{k}%
}{2\pi }\exp \left( -\frac{N\tilde{k}^{2}}{4}\right) \frac{1}{4}\left( 2-%
\tilde{k}^{2}\right)   \nonumber \\
&=&\pi ^{N/2}\left( \frac{2m}{\beta }\right) ^{\left( N+1\right) /2}\frac{N-1%
}{2\sqrt{\pi }N^{3/2}}=\frac{N-1}{2\pi N^{3/2}}\left( \frac{2\pi m}{\beta }%
\right) ^{\left( N+1\right) /2}  \label{can12appb}
\end{eqnarray}%
\begin{eqnarray}
&&\int \frac{dk}{2\pi }\int d{\bf p\,}p_{b}p_{c}\exp \left(
ik\sum_{a=1}^{N}p_{a}-\beta \sum_{a=1}^{N}\frac{p_{a}^{2}}{2m}\right)  
\nonumber \\
&=&\left( \frac{2m}{\beta }\right) ^{(N+2)/2}\left( \frac{2m}{\beta }\right)
^{-1/2}\int \frac{d\tilde{k}}{2\pi }\int d\tilde{p}_{b}\tilde{p}_{b}e^{-%
\tilde{p}_{b}^{2}+i\tilde{k}\tilde{p}_{b}}\int d\tilde{p}_{c}\tilde{p}%
_{c}e^{-\tilde{p}_{c}^{2}+i\tilde{k}\tilde{p}_{c}}\prod_{a\neq b,c}^{N}\int d%
\tilde{p}_{a{\bf \,}}\tilde{p}_{a}^{2}e^{-\tilde{p}_{a}^{2}+i\tilde{k}\tilde{%
p}_{a}}  \nonumber \\
&=&\left( \frac{2m}{\beta }\right) ^{(N+1)/2}\pi ^{N/2}\int \frac{d\tilde{k}%
}{2\pi }\exp \left( -\frac{N\tilde{k}^{2}}{4}\right) \left( \frac{-\tilde{k}%
^{2}}{4}\right) =-\frac{1}{2\pi N^{3/2}}\left( \frac{2\pi m}{\beta }\right)
^{\left( N+1\right) /2}  \label{can12appc}
\end{eqnarray}%
\begin{eqnarray}
&&\int \frac{dk}{2\pi }\int d{\bf p\,}p_{b}^{4}\exp \left(
ik\sum_{a=1}^{N}p_{a}-\beta \sum_{a=1}^{N}\frac{p_{a}^{2}}{2m}\right)  
\nonumber \\
&=&\left( \frac{2m}{\beta }\right) ^{(N+4)/2}\left( \frac{2m}{\beta }\right)
^{-1/2}\int \frac{d\tilde{k}}{2\pi }\int d\tilde{p}_{a{\bf \,}}\tilde{p}%
_{a}^{4}e^{-\tilde{p}_{a}^{2}+i\tilde{k}\tilde{p}_{a}}\prod_{b\neq
a}^{N}\int d\tilde{p}_{b}e^{-\tilde{p}_{b}^{2}+i\tilde{k}\tilde{p}_{b}} 
\nonumber \\
&=&\left( \frac{2m}{\beta }\right) ^{(N+3)/2}\pi ^{N/2}\int \frac{d\tilde{k}%
}{2\pi }\exp \left( -\frac{N\tilde{k}^{2}}{4}\right) \frac{1}{16}\left( 
\tilde{k}^{4}-12\tilde{k}^{2}+12\right)   \nonumber \\
&=&\left( \frac{2m}{\beta }\right) ^{(N+3)/2}\pi ^{N/2}\frac{3\left(
N-1\right) ^{2}}{4\sqrt{\pi }N^{5/2}}=\frac{3\left( N-1\right) ^{2}}{\left(
2\pi \right) ^{2}N^{5/2}}\left( \frac{2\pi m}{\beta }\right) ^{(N+3)/2}
\label{can12appd}
\end{eqnarray}

\bigskip

\subsection{The $\protect\theta (p,z)$\ term}

This term is 
\begin{equation}
\theta _{cn}(p,{\bf z})=\int \frac{dk}{2\pi }\int d{\bf p}\exp \left(
ik\sum_{a=1}^{N}p_{a}-\beta \sum_{a=1}^{N}\frac{p_{a}^{2}}{2m}\right) \left(
1-\frac{\beta }{c^{2}}H_{R}\right) \delta \left( p-p_{n}\right)
\label{appthetaA}
\end{equation}
where 
\begin{eqnarray}
H_{R} &=&-\lambda _{1}\sum_{a=1}^{N}\frac{p_{a}^{4}}{8m^{3}}+\pi G\lambda
_{2}\sum_{a=1}^{N}\sum_{b=1}^{N}p_{b}^{2}\left| r_{ab}\right| -2\pi G\lambda
_{3}\sum_{a>b}^{N}p_{a}p_{b}\left| r_{ab}\right|  \nonumber \\
&&+\lambda _{4}\left( \pi G\right)
^{2}\sum_{a=1}^{N}\sum_{b=1}^{N}\sum_{c=1}^{N}m^{3}\left[ \left|
r_{ab}\right| \left| r_{ac}\right| -r_{ab}r_{ac}\right]  \label{appthetaB}
\end{eqnarray}

\subsubsection{\protect\bigskip The classical and $\protect\lambda _{4}$ part%
}

Here we must compute 
\begin{equation}
\int \frac{dk}{2\pi }\int d{\bf p}\exp \left( ik\sum_{a=1}^{N}p_{a}-\beta
\sum_{a=1}^{N}\frac{p_{a}^{2}}{2m}\right) \delta \left( p-p_{n}\right)
\left( 1-\frac{\beta m^{3}}{c^{2}}\lambda _{4}\left( \pi G\right)
^{2}\sum_{a=1}^{N}\sum_{b=1}^{N}\sum_{c=1}^{N}\left[ \left| r_{ab}\right|
\left| r_{ac}\right| -r_{ab}r_{ac}\right] \right)  \label{appl4A}
\end{equation}
of which the relevant part is 
\begin{eqnarray}
&&\int \frac{dk}{2\pi }\int d{\bf p}\exp \left( ik\sum_{a=1}^{N}p_{a}-\beta
\sum_{a=1}^{N}\frac{p_{a}^{2}}{2m}\right) \delta \left( p-p_{n}\right) 
\nonumber \\
&=&\int \frac{dk}{2\pi }\exp \left( ikp-\frac{\beta p^{2}}{2m}\right) \int d%
{\bf p}\exp \left( ik\sum_{a\neq n}^{N}p_{a}-\beta \sum_{a\neq n}^{N}\frac{%
p_{a}^{2}}{2m}\right)  \nonumber \\
&=&\left( \frac{2m}{\beta }\right) ^{\left( N-2\right) /2}\pi ^{\left(
N-1\right) /2}\int \frac{d\tilde{k}}{2\pi }\exp \left( i\tilde{k}\tilde{p}-%
\tilde{p}^{2}\right) \exp \left( -\frac{\left( N-1\right) \tilde{k}^{2}}{4}%
\right)  \nonumber \\
&=&\left( \frac{2\pi m}{\beta }\right) ^{\left( N-2\right) /2}\frac{1}{\sqrt{%
N-1}}\exp \left( -\frac{\beta p^{2}N}{2m\left( N-1\right) }\right)
\label{appl4B}
\end{eqnarray}
and so 
\begin{eqnarray}
&&\int \frac{dk}{2\pi }\int d{\bf p}\exp \left( ik\sum_{a=1}^{N}p_{a}-\beta
\sum_{a=1}^{N}\frac{p_{a}^{2}}{2m}\right) \delta \left( p-p_{n}\right)
\left( 1-\frac{\beta }{c^{2}}\lambda _{4}\left( \pi G\right)
^{2}\sum_{a=1}^{N}\sum_{b=1}^{N}\sum_{c=1}^{N}m^{3}\left[ \left|
r_{ab}\right| \left| r_{ac}\right| -r_{ab}r_{ac}\right] \right)  \nonumber \\
&=&\left( \frac{2\pi m}{\beta }\right) ^{\left( N-2\right) /2}\frac{1}{\sqrt{%
N-1}}\exp \left( -\frac{\beta p^{2}N}{2m\left( N-1\right) }\right) \left( 1-%
\frac{4\beta \lambda _{4}m^{3}\left( \pi G\right) ^{2}}{c^{2}}%
\sum_{k=1}^{N-1}\sum_{l=k+1}^{N-1}\left( N-l\right) \left( l-k\right)
ku_{l}u_{k}\right)  \label{appl4C}
\end{eqnarray}
where (\ref{can18}) was used.

\subsection{The $\protect\lambda _{1}$\ part}

Now we have 
\begin{eqnarray}
&&\int \frac{dk}{2\pi }\int d{\bf p}\exp \left( ik\sum_{a=1}^{N}p_{a}-\beta
\sum_{a=1}^{N}\frac{p_{a}^{2}}{2m}\right) \delta \left( p-p_{n}\right)
\left( -\frac{\beta }{c^{2}}\right) \left( -\lambda _{1}\sum_{c=1}^{N}\frac{%
p_{c}^{4}}{8m^{3}}\right)   \nonumber \\
&=&\frac{\lambda _{1}\beta }{8m^{3}c^{2}}\sum_{c=1}^{N}\int \frac{dk}{2\pi }%
\int d{\bf p}p_{c}^{4}\exp \left( ik\sum_{a=1}^{N}p_{a}-\beta \sum_{a=1}^{N}%
\frac{p_{a}^{2}}{2m}\right) \delta \left( p-p_{n}\right)   \label{appl1A}
\end{eqnarray}%
and 
\begin{eqnarray}
&&\int \frac{dk}{2\pi }\int d{\bf p}p_{c}^{4}\exp \left(
ik\sum_{a=1}^{N}p_{a}-\beta \sum_{a=1}^{N}\frac{p_{a}^{2}}{2m}\right) \delta
\left( p-p_{n}\right)   \nonumber \\
&=&\int \frac{dk}{2\pi }\exp \left( ikp-\frac{\beta p^{2}}{2m}\right) \int d%
{\bf p}\exp \left( ik\sum_{a\neq n}^{N}p_{a}-\beta \sum_{a\neq n}^{N}\frac{%
p_{a}^{2}}{2m}\right) \left\{ 
\begin{array}{c}
p_{c}^{4}\mbox{\ \ \ \ \ \ \ \ \ \ \ }\left( c\neq n\right)  \\ 
p^{4}\mbox{\ \ \ \ \ \ \ \ \ \ \ }\left( c=n\right) 
\end{array}%
\right.   \nonumber \\
&=&\left( \frac{2\pi m}{\beta }\right) ^{\left( N+2\right) /2}\frac{1}{\pi
^{2}}\exp \left( -\frac{\beta p^{2}N}{2m\left( N-1\right) }\right) \left\{ 
\begin{array}{l}
\left( \frac{\beta ^{2}p^{4}}{\left( 2m\right) ^{2}\left( N-1\right) ^{9/2}}+%
\frac{3\left( N-2\right) \beta p^{2}}{2m\left( N-1\right) ^{7/2}}+\frac{%
3\left( N-2\right) ^{2}}{4\left( N-1\right) ^{5/2}}\right) \mbox{ \ \ \ \ \
\ }\left( c\neq n\right)  \\ 
\left( \frac{\beta ^{2}p^{4}}{\left( 2m\right) ^{2}\left( N-1\right) ^{1/2}}%
\right) \mbox{ \ \ \ \ \ \ \ \ \ \ \ \ \ \ \ \ }\left( c=n\right) \mbox{\ \
\ \ \ \ \ }%
\end{array}%
\right.   \label{appl1B}
\end{eqnarray}%
which gives 
\begin{eqnarray}
&&\int \frac{dk}{2\pi }\int d{\bf p}\exp \left( ik\sum_{a=1}^{N}p_{a}-\beta
\sum_{a=1}^{N}\frac{p_{a}^{2}}{2m}\right) \delta \left( p-p_{n}\right)
\left( -\frac{\beta }{c^{2}}\right) \left( -\lambda _{1}\sum_{c=1}^{N}\frac{%
p_{c}^{4}}{8m^{3}}\right)   \nonumber \\
&=&\left( \frac{2\pi m}{\beta }\right) ^{\left( N+2\right) /2}\frac{\lambda
_{1}\beta }{8m^{3}\pi ^{2}c^{2}}\exp \left( -\frac{\beta p^{2}N}{2m\left(
N-1\right) }\right)   \nonumber \\
&&\mbox{ \ \ \ \ }\times \left\{ \left( N-1\right) \left( \frac{\beta
^{2}p^{4}}{\left( 2m\right) ^{2}\left( N-1\right) ^{9/2}}+\frac{3\left(
N-2\right) \beta p^{2}}{2m\left( N-1\right) ^{7/2}}+\frac{3\left( N-2\right)
^{2}}{4\left( N-1\right) ^{5/2}}\right) +\frac{\beta ^{2}p^{4}}{\left(
2m\right) ^{2}\left( N-1\right) ^{1/2}}\right\}   \nonumber \\
&=&\left( \frac{2\pi m}{\beta }\right) ^{\left( N-2\right) /2}\frac{\exp
\left( -\frac{\beta p^{2}N}{2m\left( N-1\right) }\right) }{\sqrt{N-1}}\frac{%
\lambda _{1}}{2\beta mc^{2}}\left( \frac{\beta ^{2}p^{4}\left( 1+\left(
N-1\right) ^{3}\right) }{\left( 2m\right) ^{2}\left( N-1\right) ^{3}}+\frac{%
3\left( N-2\right) \beta p^{2}}{2m\left( N-1\right) ^{2}}+\frac{3\left(
N-2\right) ^{2}}{4\left( N-1\right) }\right)   \nonumber \\
&&  \label{appl1C}
\end{eqnarray}

\subsubsection{The $\protect\lambda _{2}$\ part}

Now we must do 
\begin{eqnarray}
&&\int \frac{dk}{2\pi }\int d{\bf p}\exp \left( ik\sum_{a=1}^{N}p_{a}-\beta
\sum_{a=1}^{N}\frac{p_{a}^{2}}{2m}\right) \delta \left( p-p_{n}\right)
\left( -\frac{\beta }{c^{2}}\right) \left( +\pi G\lambda
_{2}\sum_{a=1}^{N}\sum_{b=1}^{N}p_{b}^{2}\left| r_{ab}\right| \right) 
\nonumber \\
&=&-\frac{\lambda _{2}\pi G\beta }{c^{2}}\sum_{c=1}^{N}\sum_{b=1}^{N}\left|
r_{bc}\right| \int \frac{dk}{2\pi }\int d{\bf p\,}p_{b}^{2}\exp \left(
ik\sum_{a=1}^{N}p_{a}-\beta \sum_{a=1}^{N}\frac{p_{a}^{2}}{2m}\right) \delta
\left( p-p_{n}\right)  \label{appl2A}
\end{eqnarray}
and the integrals are 
\begin{eqnarray}
&&\int \frac{dk}{2\pi }\int d{\bf p}p_{c}^{2}\exp \left(
ik\sum_{a=1}^{N}p_{a}-\beta \sum_{a=1}^{N}\frac{p_{a}^{2}}{2m}\right) \delta
\left( p-p_{n}\right)  \nonumber \\
&=&\int \frac{dk}{2\pi }\exp \left( ikp-\frac{\beta p^{2}}{2m}\right) \int d%
{\bf p}\exp \left( ik\sum_{a\neq n}^{N}p_{a}-\beta \sum_{a\neq n}^{N}\frac{%
p_{a}^{2}}{2m}\right) \left\{ 
\begin{array}{c}
p_{c}^{2}\mbox{\ \ \ \ \ \ \ \ \ \ \ }\left( c\neq n\right) \\ 
p^{2}\mbox{\ \ \ \ \ \ \ \ \ \ \ }\left( c=n\right)%
\end{array}
\right.  \nonumber \\
&=&\left( \frac{2\pi m}{\beta }\right) ^{N/2}\frac{1}{\pi }\exp \left( -%
\frac{\beta p^{2}N}{2m\left( N-1\right) }\right) \left\{ 
\begin{array}{l}
\left( \frac{\beta p^{2}}{2m\left( N-1\right) ^{5/2}}+\frac{\left(
N-2\right) }{2\left( N-1\right) ^{3/2}}\right) \mbox{ \ \ \ \ \ \ }\left(
c\neq n\right) \\ 
\left( \frac{\beta p^{2}}{2m\left( N-1\right) ^{1/2}}\right) \mbox{ \ \ \ \
\ \ \ \ \ \ \ \ \ \ \ \ \ \ \ \ \ }\left( c=n\right) \mbox{\ \ \ \ \ }%
\end{array}
\right.  \label{appl2B}
\end{eqnarray}
giving 
\begin{eqnarray}
&&\int \frac{dk}{2\pi }\int d{\bf p}\exp \left( ik\sum_{a=1}^{N}p_{a}-\beta
\sum_{a=1}^{N}\frac{p_{a}^{2}}{2m}\right) \delta \left( p-p_{n}\right)
\left( -\frac{\beta }{c^{2}}\right) \left( +\pi G\lambda
_{2}\sum_{a=1}^{N}\sum_{b=1}^{N}p_{b}^{2}\left| r_{ab}\right| \right) 
\nonumber \\
&=&-\frac{\lambda _{2}\pi G\beta }{c^{2}}\left( \frac{2\pi m}{\beta }\right)
^{N/2}\frac{1}{\pi }\exp \left( -\frac{\beta p^{2}N}{2m\left( N-1\right) }%
\right)  \nonumber \\
&&\mbox{ \ }\times \left( \sum_{c=1}^{N}\sum_{b\neq n}^{N}\left|
r_{bc}\right| \left\{ \frac{\beta p^{2}}{2m\left( N-1\right) ^{5/2}}+\frac{%
\left( N-2\right) }{2\left( N-1\right) ^{3/2}}\right\} +\sum_{c=1}^{N}\left|
r_{cn}\right| \left( \frac{\beta p^{2}}{2m\left( N-1\right) ^{1/2}}\right)
\right)  \nonumber \\
&=&-\frac{\mbox{2}\lambda _{2}\pi Gm}{c^{2}}\frac{1}{\sqrt{N-1}}\left( \frac{%
2\pi m}{\beta }\right) ^{\left( N-2\right) /2}\exp \left( -\frac{\beta p^{2}N%
}{2m\left( N-1\right) }\right)  \nonumber \\
&&\mbox{ \ \ \ \ \ \ \ \ \ \ }\times \left( \sum_{c=1}^{N}\sum_{b\neq
n}^{N}\left| r_{bc}\right| \left\{ \frac{\beta p^{2}}{2m\left( N-1\right)
^{2}}+\frac{\left( N-2\right) }{2\left( N-1\right) }\right\}
+\sum_{c=1}^{N}\left| r_{cn}\right| \left( +\frac{\beta p^{2}}{2m}\right)
\right)  \label{appl2C}
\end{eqnarray}

Since $\sum_{b\neq n}^{N}\left| r_{cb}\right| =\sum_{b=1}^{N}\left|
r_{cb}\right| -\left| r_{cn}\right| $, this becomes 
\begin{eqnarray}
&&\int \frac{dk}{2\pi }\int d{\bf p}\exp \left( ik\sum_{a=1}^{N}p_{a}-\beta
\sum_{a=1}^{N}\frac{p_{a}^{2}}{2m}\right) \delta \left( p-p_{n}\right)
\left( -\frac{\beta }{c^{2}}\right) \left( +\pi G\lambda
_{2}\sum_{a=1}^{N}\sum_{b=1}^{N}p_{b}^{2}\left| r_{ab}\right| \right) 
\nonumber \\
&=&-\frac{\mbox{2}\lambda _{2}\pi Gm}{\sqrt{N-1}c^{2}}\left( \frac{2\pi m}{%
\beta }\right) ^{\left( N-2\right) /2}\exp \left( -\frac{\beta p^{2}N}{%
2m\left( N-1\right) }\right) \left\{ \sum_{c=1}^{N}\sum_{b=1}^{N}\left|
r_{bc}\right| \left\{ \frac{\beta p^{2}}{2m\left( N-1\right) ^{2}}+\frac{%
\left( N-2\right) }{2\left( N-1\right) }\right\} \right.  \nonumber \\
&&\mbox{ \ \ \ \ \ \ \ }\left. +\sum_{c=1}^{N}\left| r_{cn}\right| \left( +%
\frac{\beta p^{2}}{2m}\left[ \frac{N\left( N-2\right) }{\left( N-1\right)
^{2}}\right] -\frac{\left( N-2\right) }{2\left( N-1\right) }\right) \right\}
\nonumber \\
&=&-\frac{\mbox{2}\lambda _{2}\pi Gm}{\sqrt{N-1}c^{2}}\left( \frac{2\pi m}{%
\beta }\right) ^{\left( N-2\right) /2}\exp \left( -\frac{\beta p^{2}N}{%
2m\left( N-1\right) }\right) \left( 2\sum_{l=1}^{N-1}l\left( N-l\right)
u_{l}\left\{ \frac{\beta p^{2}}{2m\left( N-1\right) ^{2}}+\frac{\left(
N-2\right) }{2\left( N-1\right) }\right\} \right.  \nonumber \\
&&\mbox{ \ \ \ \ \ \ \ \ \ \ \ \ \ \ \ \ \ \ \ \ \ \ \ \ }+\left. \left(
\sum_{s=1}^{n-1}su_{s}+\sum_{s=n}^{N-1}(N-s)u_{s}\right) \left( +\frac{\beta
p^{2}}{2m}\left[ \frac{N\left( N-2\right) }{\left( N-1\right) ^{2}}\right] -%
\frac{\left( N-2\right) }{2\left( N-1\right) }\right) \right)  \label{appl2D}
\end{eqnarray}

\subsubsection{The $\protect\lambda _{3}$\ part}

Now we must do 
\begin{eqnarray}
&&\int \frac{dk}{2\pi }\int d{\bf p}\exp \left( ik\sum_{a=1}^{N}p_{a}-\beta
\sum_{a=1}^{N}\frac{p_{a}^{2}}{2m}\right) \delta \left( p-p_{n}\right)
\left( -\frac{\beta }{c^{2}}\right) \left( -2\pi G\lambda
_{3}\sum_{a>b}^{N}p_{a}p_{b}\left| r_{ab}\right| \right)   \label{appl3A} \\
&=&\frac{\lambda _{3}\pi G\beta }{c^{2}}\int \frac{dk}{2\pi }\int d{\bf p}%
\exp \left( ik\sum_{a=1}^{N}p_{a}-\beta \sum_{a=1}^{N}\frac{p_{a}^{2}}{2m}%
\right) \delta \left( p-p_{n}\right) \left( \sum_{c\neq n}^{N}\sum_{b\neq
n}^{N}p_{c}p_{b}\left| r_{cb}\right| +2p\sum_{c=1}^{N}\left| r_{cn}\right|
p_{c}\right)   \nonumber
\end{eqnarray}%
where the integrations are 
\begin{eqnarray}
&&\int \frac{dk}{2\pi }\int d{\bf p}p_{b}p_{c}\exp \left(
ik\sum_{a=1}^{N}p_{a}-\beta \sum_{a=1}^{N}\frac{p_{a}^{2}}{2m}\right) \delta
\left( p-p_{n}\right)   \nonumber \\
&=&\mbox{ }\left( \frac{2\pi m}{\beta }\right) ^{N/2}\frac{1}{\pi }\exp
\left( -\frac{\beta p^{2}N}{2m\left( N-1\right) }\right) \left\{ 
\begin{array}{l}
\left( \frac{\beta p^{2}}{2m\left( N-1\right) ^{5/2}}-\frac{1}{2\left(
N-1\right) ^{3/2}}\right) \mbox{ \ \ \ \ \ \ }\left( b\neq c\neq n\right) 
\\ 
\left( -\frac{\beta }{2m}\frac{p^{2}}{\left( N-1\right) ^{3/2}}\right) \mbox{
\ \ \ \ \ \ \ \ \ \ \ \ \ \ \ \ \ \ }\left( c\neq n,b=n\right) 
\end{array}%
\right.   \label{appl3B}
\end{eqnarray}%
giving 
\begin{eqnarray}
&&\int \frac{dk}{2\pi }\int d{\bf p}\exp \left( ik\sum_{a=1}^{N}p_{a}-\beta
\sum_{a=1}^{N}\frac{p_{a}^{2}}{2m}\right) \delta \left( p-p_{n}\right)
\left( -\frac{\beta }{c^{2}}\right) \left( -2\pi G\lambda
_{3}\sum_{a>b}^{N}p_{a}p_{b}\left| r_{ab}\right| \right)   \nonumber \\
&=&\frac{\lambda _{3}\pi G\beta }{c^{2}}\left( \frac{2\pi m}{\beta }\right)
^{N/2}\frac{1}{\pi }\exp \left( -\frac{\beta p^{2}N}{2m\left( N-1\right) }%
\right) \left\{ 2\sum_{c=1}^{N}\left| r_{cn}\right| \left( -\frac{\beta }{2m}%
\frac{p^{2}}{\left( N-1\right) ^{3/2}}\right) \right.   \nonumber \\
&&\mbox{ \ \ \ \ \ \ \ \ \ \ }\left. +\sum_{c\neq n}^{N}\sum_{b\neq
n}^{N}\left| r_{bc}\right| \left( \frac{\beta p^{2}}{2m\left( N-1\right)
^{5/2}}-\frac{1}{2\left( N-1\right) ^{3/2}}\right) \right\}   \label{appl3C}
\end{eqnarray}%
Using again $\sum_{b\neq n}^{N}\left| r_{cb}\right| =\sum_{b=1}^{N}\left|
r_{cb}\right| -\left| r_{cn}\right| $, this becomes 
\begin{eqnarray}
&&\int \frac{dk}{2\pi }\int d{\bf p}\exp \left( ik\sum_{a=1}^{N}p_{a}-\beta
\sum_{a=1}^{N}\frac{p_{a}^{2}}{2m}\right) \delta \left( p-p_{n}\right)
\left( -\frac{\beta }{c^{2}}\right) \left( -2\pi G\lambda
_{3}\sum_{a>b}^{N}p_{a}p_{b}\left| r_{ab}\right| \right)   \nonumber \\
&=&\frac{\lambda _{3}\pi G\beta }{c^{2}}\left( \frac{2\pi m}{\beta }\right)
^{N/2}\frac{1}{\pi }\exp \left( -\frac{\beta p^{2}N}{2m\left( N-1\right) }%
\right) \left\{ 2\sum_{c=1}^{N}\left| r_{cn}\right| \left( -\frac{\beta }{2m}%
\frac{p^{2}}{\left( N-1\right) ^{3/2}}\right) \right.   \nonumber \\
&&\mbox{ \ \ \ \ \ \ \ \ \ \ \ \ \ \ \ \ \ \ \ \ \ \ \ \ \ \ }\left. +\left(
\sum_{c=1}^{N}\sum_{b=1}^{N}\left| r_{bc}\right| -2\sum_{c=1}^{N}\left|
r_{cn}\right| \right) \left( \frac{\beta p^{2}}{2m\left( N-1\right) ^{5/2}}-%
\frac{1}{2\left( N-1\right) ^{3/2}}\right) \right\}   \nonumber \\
&=&\frac{\mbox{2}\lambda _{3}\pi Gm}{\sqrt{N-1}c^{2}}\left( \frac{2\pi m}{%
\beta }\right) ^{\left( N-2\right) /2}\exp \left( -\frac{\beta p^{2}N}{%
2m\left( N-1\right) }\right) \left\{ \sum_{c=1}^{N}\sum_{b=1}^{N}\left|
r_{bc}\right| \left( \frac{\beta p^{2}}{2m\left( N-1\right) ^{2}}-\frac{1}{%
2\left( N-1\right) }\right) \right.   \nonumber \\
&&\mbox{ \ \ \ \ \ \ \ \ \ \ \ \ \ \ \ \ \ \ \ \ }\left.
+\sum_{c=1}^{N}\left| r_{cn}\right| \left( \frac{1}{\left( N-1\right) }-%
\frac{\beta p^{2}}{2m}\left[ \frac{2N}{\left( N-1\right) ^{2}}\right]
\right) \right\}   \nonumber \\
&=&\frac{\mbox{2}\lambda _{3}\pi Gm}{\sqrt{N-1}c^{2}}\left( \frac{2\pi m}{%
\beta }\right) ^{\left( N-2\right) /2}\exp \left( -\frac{\beta p^{2}N}{%
2m\left( N-1\right) }\right) \left( 2\sum_{l=1}^{N-1}l\left( N-l\right)
u_{l}\left( \frac{\beta p^{2}}{2m\left( N-1\right) ^{2}}-\frac{1}{2\left(
N-1\right) }\right) \right.   \nonumber \\
&&\mbox{ \ \ \ \ \ \ \ \ \ \ \ \ \ \ \ \ \ \ \ \ }\left. +\left(
\sum_{s=1}^{n-1}su_{s}+\sum_{s=n}^{N-1}(N-s)u_{s}\right) \left( \frac{1}{%
\left( N-1\right) }-\frac{\beta p^{2}}{2m}\left[ \frac{2N}{\left( N-1\right)
^{2}}\right] \right) \right)   \nonumber \\
&&  \label{appl3D}
\end{eqnarray}

\subsubsection{\protect\bigskip The full expression for $\protect\theta %
_{cn}(p,{\bf z})$}

From eqs. (\ref{appl4C},\ref{appl1C},\ref{appl2D},\ref{appl3D}) we have 
\begin{eqnarray}
\theta _{cn}(p,{\bf z}) &=&\int \frac{dk}{2\pi }\int d{\bf p}\exp \left(
ik\sum_{a=1}^{N}p_{a}-\beta \sum_{a=1}^{N}\frac{p_{a}^{2}}{2m}\right) \left(
1-\frac{\beta }{c^{2}}H_{R}\right) \delta \left( p-p_{n}\right)  \nonumber \\
&=&\left( \frac{2\pi m}{\beta }\right) ^{\left( N-2\right) /2}\frac{1}{\sqrt{%
N-1}}\exp \left( -\frac{\beta p^{2}N}{2m\left( N-1\right) }\right)  \nonumber
\\
&&\times \left( 1+\frac{\lambda _{1}}{2\beta mc^{2}}\left( \frac{\beta
^{2}p^{4}\left( 1+\left( N-1\right) ^{3}\right) }{\left( 2m\right)
^{2}\left( N-1\right) ^{3}}+\frac{3\left( N-2\right) \beta p^{2}}{2m\left(
N-1\right) ^{2}}+\frac{3\left( N-2\right) ^{2}}{4\left( N-1\right) }\right)
\right.  \nonumber \\
&&\mbox{ \ \ }+\frac{\mbox{2}\lambda _{2}\pi Gm}{c^{2}}\left[
-2\sum_{l=1}^{N-1}l\left( N-l\right) u_{l}\left\{ \frac{\beta p^{2}}{%
2m\left( N-1\right) ^{2}}+\frac{\left( N-2\right) }{2\left( N-1\right) }%
\right\} \right.  \nonumber \\
&&\mbox{ \ \ \ \ \ \ \ \ \ \ \ \ \ \ \ \ \ }\left. +\left(
\sum_{s=1}^{n-1}su_{s}+\sum_{s=n}^{N-1}(N-s)u_{s}\right) \left( \frac{\left(
N-2\right) }{2\left( N-1\right) }-\frac{\beta p^{2}}{2m}\left[ \frac{N\left(
N-2\right) }{\left( N-1\right) ^{2}}\right] \right) \right]  \nonumber \\
&&\mbox{ \ \ }+\frac{\mbox{2}\lambda _{3}\pi Gm}{c^{2}}\left[
2\sum_{l=1}^{N-1}l\left( N-l\right) u_{l}\left( \frac{\beta p^{2}}{2m\left(
N-1\right) ^{2}}-\frac{1}{2\left( N-1\right) }\right) \right.  \nonumber \\
&&\mbox{ \ \ \ \ \ \ \ \ \ \ \ \ \ \ \ }\left. +\left(
\sum_{s=1}^{n-1}su_{s}+\sum_{s=n}^{N-1}(N-s)u_{s}\right) \left( \frac{1}{%
\left( N-1\right) }-\frac{\beta p^{2}}{2m}\left[ \frac{2N}{\left( N-1\right)
^{2}}\right] \right) \right]  \nonumber \\
&&\mbox{ \ \ \ }\left. -\frac{4\beta m^{3}}{c^{2}}\lambda _{4}\left( \pi
G\right) ^{2}\sum_{k=1}^{N-1}\sum_{l=k+1}^{N-1}\left( N-l\right) \left(
l-k\right) ku_{l}u_{k}\right)  \label{appfulA}
\end{eqnarray}
\bigskip or alternatively 
\begin{eqnarray}
\theta _{cn}(p,{\bf z}) &=&\left( \frac{2\pi m}{\beta }\right) ^{\left(
N-2\right) /2}\frac{1}{\sqrt{N-1}}\exp \left( -\frac{\beta p^{2}N}{2m\left(
N-1\right) }\right)  \nonumber \\
&&\mbox{ \ \ }\times \left[ 1+\frac{\lambda _{1}}{2\beta mc^{2}}\left( \frac{%
\beta ^{2}p^{4}\left( 1+\left( N-1\right) ^{3}\right) }{\left( 2m\right)
^{2}\left( N-1\right) ^{3}}+\frac{3\left( N-2\right) \beta p^{2}}{2m\left(
N-1\right) ^{2}}+\frac{3\left( N-2\right) ^{2}}{4\left( N-1\right) }\right)
\right.  \nonumber \\
&&\mbox{ \ \ \ \ }-\frac{4\beta m}{c^{2}}\lambda _{4}\left( \pi Gm\right)
^{2}\sum_{k=1}^{N-1}\sum_{l=k+1}^{N-1}\left( N-l\right) \left( l-k\right)
ku_{l}u_{k}  \nonumber \\
&&\mbox{ \ \ \ \ \ \ \ \ \ }+\frac{\mbox{2}\pi Gm}{c^{2}}\sum_{l=1}^{N-1}l%
\left( N-l\right) u_{l}\left( \frac{\left( 2\lambda _{3}-2\lambda
_{2}\right) \beta p^{2}}{2m\left( N-1\right) ^{2}}-\frac{\left( 2\lambda
_{3}+2(N-2)\lambda _{2}\right) }{2\left( N-1\right) }\right)  \nonumber \\
&&\mbox{ \ \ \ \ \ \ \ \ \ \ }\left. +\frac{\mbox{2}\left( 2\lambda
_{3}+(N-2)\lambda _{2}\right) \pi Gm}{c^{2}}\left(
\sum_{s=1}^{n-1}su_{s}+\sum_{s=n}^{N-1}(N-s)u_{s}\right) \left( \frac{N-1-%
\frac{N\beta p^{2}}{m}}{2\left( N-1\right) ^{2}}\right) \right]
\label{appfulB}
\end{eqnarray}
\bigskip

\bigskip When the $\lambda $'s are all equal to unity, this is 
\begin{eqnarray}
\theta _{cn}(p,{\bf u}) &=&\frac{1}{\sqrt{N-1}}\exp \left( -\frac{N\beta
p^{2}}{2m\left( N-1\right) }\right) \left( \frac{2\pi m}{\beta }\right)
^{\left( N-2\right) /2}  \label{appfulC} \\
&&\times \left\{ \left( 1-\frac{4\beta m\left( \pi Gm\right) ^{2}}{c^{2}}%
\sum_{k=1}^{N-1}\sum_{l=k+1}^{N-1}\left( N-l\right) \left( l-k\right)
ku_{l}u_{k}\right) \right.  \nonumber \\
&&+\frac{1}{2\beta mc^{2}}\left[ \frac{\beta ^{2}p^{4}N\left(
N^{2}-3N+3\right) }{(2m)^{2}\left( N-1\right) ^{3}}+\frac{3\beta p^{2}\left(
N-2\right) }{2m\left( N-1\right) ^{5/2}}+\frac{3\left( N-2\right) ^{2}}{%
4\left( N-1\right) ^{3/2}}\right]  \nonumber \\
&&+\left( \frac{\mbox{2}\pi mG}{c^{2}}\right) \left[ -\sum_{l=1}^{N-1}l%
\left( N-l\right) u_{l}-\left( \frac{N^{2}\beta p^{2}}{2m\left( N-1\right)
^{2}}-\frac{N}{2(N-1)}\right) \left(
\sum_{s=1}^{n-1}su_{s}+\sum_{s=n}^{N-1}(N-s)u_{s}\right) \right]  \nonumber
\end{eqnarray}

\subsection{An evaluation of $f_{cn}(p,z)$\ }

Now consider eq. (\ref{can24b}), which can be rewritten as 
\begin{equation}
f_{cn}^{R}(p,z)=\frac{e^{-\beta Mc^{2}}}{{\cal Z}}\int \frac{dk}{2\pi }\int d%
{\bf u}\exp \left( -ikz-\lambda \beta \sum_{l=1}^{N-1}\left( C_{l}+i\alpha
D_{nl}\right) u_{l}\right) \theta _{cn}(p,{\bf u})  \label{appfz1}
\end{equation}%
where 
\begin{equation}
\lambda =2\pi Gm^{2}\mbox{ \ \ \ \ }\alpha =\frac{k}{N\beta \lambda }\mbox{ }
\label{appfz2}
\end{equation}%
The integrals are now 
\begin{eqnarray}
I_{n}^{0} &=&\int d{\bf u}\exp \left( -\lambda \beta \sum_{l=1}^{N-1}\left(
C_{l}+i\alpha D_{nl}\right) u_{l}\right)   \nonumber \\
&=&\frac{1}{\left( 2\pi G\beta m^{2}\mbox{ }\right) ^{N-1}}\prod_{l=1}^{N-1}%
\frac{1}{C_{l}+i\alpha D_{nl}}  \label{appfz3} \\
I_{n,s}^{1} &=&\int d{\bf u}\exp \left( -\lambda \beta
\sum_{l=1}^{N-1}\left( C_{l}+i\alpha D_{nl}\right) u_{l}\right) u_{s} 
\nonumber \\
&=&\frac{1}{\left( 2\pi G\beta m^{2}\mbox{ }\right) ^{N}}\frac{1}{%
C_{s}+i\alpha D_{ns}}\prod_{l=1}^{N-1}\frac{1}{C_{l}+i\alpha D_{nl}}
\label{appfz4} \\
I_{n,s,t}^{2} &=&\int d{\bf u}\exp \left( -\lambda \beta
\sum_{l=1}^{N-1}\left( C_{l}+i\alpha D_{nl}\right) u_{l}\right) u_{s}u_{t} 
\nonumber \\
&=&\frac{1}{\left( 2\pi G\beta m^{2}\mbox{ }\right) ^{N+1}}\frac{1}{%
C_{s}+i\alpha D_{ns}}\frac{1}{C_{t}+i\alpha D_{nt}}\prod_{l=1}^{N-1}\frac{1}{%
C_{l}+i\alpha D_{nl}}  \label{appfz5}
\end{eqnarray}%
and we must divide by $N$ and sum over all values of $n$ to obtain the
correct result.

For example 
\begin{eqnarray}
\frac{1}{N}\sum_{n=1}^{N}I_{n}^{0} &=&\frac{1}{N\left( 2\pi G\beta m^{2}%
\mbox{ }\right) ^{N-1}}\sum_{n=1}^{N}\left[ \prod_{l=1}^{n-1}\frac{1}{%
l\left( N-l-i\alpha \right) }\prod_{l=n}^{N-1}\frac{1}{\left( N-l\right)
\left( l+i\alpha \right) }\right]   \nonumber \\
&=&\frac{1}{N\left( 2\pi G\beta m^{2}\mbox{ }\right) ^{N-1}}\sum_{n=1}^{N}%
\frac{1}{\left( n-1\right) !(N-n)!}\frac{\Gamma \left( N-n+1-i\alpha \right) 
}{\Gamma \left( N-i\alpha \right) }\frac{\Gamma \left( n+i\alpha \right) }{%
\Gamma \left( N+i\alpha \right) }  \nonumber \\
&=&\frac{1}{\Gamma \left( N-i\alpha \right) \Gamma \left( N+i\alpha \right)
\left( 2\pi G\beta m^{2}\mbox{ }\right) ^{N-1}}\sum_{n=1}^{N}\frac{\left(
N-1\right) !}{\left( n-1\right) !(N-n)!}\int_{0}^{1}dww^{N-n-i\alpha }\left(
1-w\right) ^{n+i\alpha -1}  \nonumber \\
&=&\frac{1}{\Gamma \left( N-i\alpha \right) \Gamma \left( N+i\alpha \right)
\left( 2\pi G\beta m^{2}\mbox{ }\right) ^{N-1}}\int_{0}^{1}dww^{-i\alpha
}\left( 1-w\right) ^{i\alpha }\left( w+1-w\right) ^{N-1}  \nonumber \\
&=&\frac{1}{\left( 2\pi G\beta m^{2}\mbox{ }\right) ^{N-1}}\frac{\Gamma
\left( 1-i\alpha \right) }{\Gamma \left( N-i\alpha \right) }\frac{\Gamma
\left( 1+i\alpha \right) }{\Gamma \left( N+i\alpha \right) }  \label{appfz6}
\end{eqnarray}%
\bigskip This function has single poles at $\alpha =in$, where $n$ is an
integer taking its values between $-N$ and $N$, except for $N=0$. Its
residues are 
\begin{equation}
\mbox{Residue}\left[ \frac{\Gamma \left( 1-i\alpha \right) }{\Gamma \left(
N-i\alpha \right) }\frac{\Gamma \left( 1+i\alpha \right) }{\Gamma \left(
N+i\alpha \right) }\right] _{a=in}=\frac{in\left( -1\right) ^{n}}{\Gamma
\left( N-n\right) \Gamma \left( N+n\right) }  \label{appfz7}
\end{equation}%
yielding 
\begin{equation}
\frac{1}{N}\sum_{n=1}^{N}\int \frac{dk}{2\pi }\int d{\bf u}\mbox{ }e^{\left(
-ikz-\lambda \beta \sum_{l=1}^{N-1}\left( C_{l}+i\alpha D_{nl}\right)
u_{l}\right) }=\frac{2N\beta \pi Gm^{2}}{\left( 2\pi G\beta m^{2}\mbox{ }%
\right) ^{N-1}\left[ \Gamma (N)\right] ^{2}}\sum_{l=1}^{N-1}A_{l}^{N}\exp
\left( -2N\left( \beta \pi Gm^{2}\right) l\left| z\right| \right) 
\label{appfz8}
\end{equation}%
for the leading non-relativistic term, where 
\begin{equation}
A_{n}^{N}=\frac{n\left( -1\right) ^{n+1}\left[ \Gamma (N)\right] ^{2}}{%
\Gamma \left( N-n\right) \Gamma \left( N+n\right) }  \label{appfzA}
\end{equation}

The remaining integrals are somewhat more difficult. We obtain 
\begin{eqnarray}
&&\frac{1}{N}\sum_{n=1}^{N}\int d{\bf u}e^{\left( -ikz-\lambda \beta
\sum_{l=1}^{N-1}\left( C_{l}+i\alpha D_{nl}\right) u_{l}\right)
}\sum_{s=1}^{N-1}s\left( N-s\right) u_{s}  \nonumber \\
&=&-\frac{1}{\lambda \beta N}\sum_{n=1}^{N}\frac{d}{d\sigma }\left[ \int d%
{\bf u}\exp \left( -\lambda \beta \sum_{l=1}^{N-1}\left( \sigma
C_{l}+i\alpha D_{nl}\right) u_{l}\right) \right] _{\sigma =1}  \nonumber \\
&=&-\frac{1}{\lambda \beta N}\frac{d}{d\sigma }\left[ \sum_{n=1}^{N}\frac{1}{%
\left( \lambda \beta \sigma \right) ^{N-1}}\prod_{l=1}^{N-1}\frac{1}{C_{l}+i%
\frac{\alpha }{\sigma }D_{nl}}\right] _{\sigma =1}  \nonumber \\
&=&-\frac{d}{d\sigma }\left[ \frac{\sigma ^{1-N}}{\left( \lambda \beta
\right) ^{N}}\frac{\Gamma \left( 1-i\alpha /\sigma \right) }{\Gamma \left(
N-i\alpha /\sigma \right) }\frac{\Gamma \left( 1+i\alpha /\sigma \right) }{%
\Gamma \left( N+i\alpha /\sigma \right) }\right] _{\sigma =1}  \nonumber \\
&=&\frac{1}{\left( 2\pi G\beta m^{2}\mbox{ }\right) ^{N}}\frac{\Gamma \left(
1-i\alpha \right) }{\Gamma \left( N-i\alpha \right) }\frac{\Gamma \left(
1+i\alpha \right) }{\Gamma \left( N+i\alpha \right) }  \nonumber \\
&&\times \left[ N-1+i\alpha \left( \Psi \left( N-i\alpha \right) -\Psi
\left( 1-i\alpha \right) -\Psi \left( N+i\alpha \right) +\Psi \left(
1+i\alpha \right) \right) \right]   \label{appfz9}
\end{eqnarray}%
where$\Psi \left( x\right) =\frac{d}{dx}\ln \Gamma (x)$ is the digamma
function. This function on the right-hand side of (\ref{appfz9}) has a
combination of single and double poles, each located at $\alpha =in$, where $%
n$ is an integer taking its values between $-N$ and $N$, except for $N=0$.
Writing 
\[
\Psi \left( N-i\alpha \right) -\Psi \left( 1-i\alpha \right)
=\sum_{s=1}^{N-1}\frac{1}{s-i\alpha }
\]%
we have the last line of (\ref{appfz9}) proportional to 
\begin{eqnarray*}
&&\frac{\Gamma \left( 1-i\alpha \right) }{\Gamma \left( N-i\alpha \right) }%
\frac{\Gamma \left( 1+i\alpha \right) }{\Gamma \left( N+i\alpha \right) }%
\left[ N-1+i\alpha \sum_{s=1}^{N-1}\left( \frac{1}{s-i\alpha }-\frac{1}{%
s+i\alpha }\right) \right]  \\
&=&\frac{\Gamma \left( 1-i\alpha \right) }{\Gamma \left( N-i\alpha \right) }%
\frac{\Gamma \left( 1+i\alpha \right) }{\Gamma \left( N+i\alpha \right) }%
\left[ N-3-2\alpha ^{2}\sum_{s\neq n}^{N-1}\left( \frac{1}{s^{2}+\alpha ^{2}}%
\right) +\left\{ \frac{2n^{2}}{\alpha ^{2}+n^{2}}\right\} \right] 
\end{eqnarray*}%
where the term in curly brackets contains the double-pole at $\alpha =in$.
Hence 
\begin{eqnarray}
&&\frac{1}{N}\sum_{n=1}^{N}\int \frac{dk}{2\pi }\int d{\bf u}\exp \left(
-ikz-\lambda \beta \sum_{l=1}^{N-1}\left( C_{l}+i\alpha D_{nl}\right)
u_{l}\right) \sum_{s=1}^{N-1}s\left( N-s\right) u_{s}  \label{appfz11} \\
&=&\frac{N\beta \lambda }{\left( \lambda \beta \right) ^{N}}\int \frac{%
d\alpha }{2\pi }e^{-i\alpha N\beta \lambda z}\frac{\Gamma \left( 1-i\alpha
\right) }{\Gamma \left( N-i\alpha \right) }\frac{\Gamma \left( 1+i\alpha
\right) }{\Gamma \left( N+i\alpha \right) }\left[ N-3-2\alpha
^{2}\sum_{s\neq n}^{N-1}\left( \frac{1}{s^{2}+\alpha ^{2}}\right) +\left\{ 
\frac{2n^{2}}{\alpha ^{2}+n^{2}}\right\} \right]   \nonumber
\end{eqnarray}%
The residue from the double pole at $\alpha =in$ is 
\begin{eqnarray}
&&-i\frac{d}{dx}\left[ \frac{2n^{2}x\left( -1\right) ^{n}\exp (N\beta
\lambda xz)}{\Gamma \left( N-x\right) \Gamma \left( N+x\right) \left(
x+n\right) }\right] _{x=n}  \nonumber \\
&=&\frac{-in\left( -1\right) ^{n}\exp (N\beta \lambda nz)}{2\Gamma \left(
N-n\right) \Gamma \left( N+n\right) }\left[ 1+2N\beta \lambda nz+2n\left(
\Psi \left( N-n\right) -\Psi \left( N+n\right) \right) \right] 
\label{appfz12}
\end{eqnarray}%
provided $z<0$, and so the total residue at $\alpha =in$ from (\ref{appfz11}%
) is 
\begin{eqnarray}
&&\frac{in\left( -1\right) ^{n}\exp (N\beta \lambda nz)}{\Gamma \left(
N-n\right) \Gamma \left( N+n\right) }\left[ N-\frac{5}{2}+2n^{2}\sum_{s\neq
n}^{N-1}\left( \frac{1}{s^{2}-n^{2}}\right) +n\left( \Psi \left( N+n\right)
-\Psi \left( N-n\right) \right) -\left( 1+N\beta \lambda nz\right) \right]  
\nonumber \\
&\equiv &\frac{-i\exp (N\beta \lambda nz)}{\left[ \Gamma (N)\right] ^{2}}%
\left( B_{n}^{N}-A_{n}^{N}\left( 1+N\beta \lambda nz\right) \right) 
\label{appfz13}
\end{eqnarray}%
giving finally 
\begin{eqnarray}
&&\frac{1}{N}\sum_{n=1}^{N}\int \frac{dk}{2\pi }\int d{\bf u}\exp \left(
-ikz-\lambda \beta \sum_{l=1}^{N-1}\left( C_{l}+i\alpha D_{nl}\right)
u_{l}\right) \sum_{s=1}^{N-1}s\left( N-s\right) u_{s}  \nonumber \\
&=&\frac{2N\beta \pi Gm^{2}}{\left( 2\pi G\beta m^{2}\mbox{ }\right) ^{N}%
\left[ \Gamma (N)\right] ^{2}}\sum_{l=1}^{N-1}\left[ B_{l}^{N}-A_{l}^{N}%
\left( 1-2N\left( \beta \pi Gm^{2}\right) l\left| z\right| \right) \right]
\exp \left( -2N\left( \beta \pi Gm^{2}\right) l\left| z\right| \right) 
\label{appfz14}
\end{eqnarray}%
where 
\begin{equation}
B_{n}^{N}=\frac{n\left( -1\right) ^{n+1}\left[ \Gamma (N)\right] ^{2}}{%
\Gamma \left( N-n\right) \Gamma \left( N+n\right) }\left[ N-\frac{5}{2}%
+2n^{2}\sum_{s\neq n}^{N-1}\left( \frac{1}{s^{2}-n^{2}}\right) +n\left( \Psi
\left( N+n\right) -\Psi \left( N-n\right) \right) \right]   \label{appfz15}
\end{equation}

\bigskip

\bigskip Using the relation 
\begin{equation}
\sum_{c=1}^{N}\left| r_{cn}\right|
=\sum_{l=1}^{n-1}lu_{l}+\sum_{l=n}^{N-1}(N-l)u_{l}  \label{appfz16}
\end{equation}%
we find 
\begin{eqnarray}
&&\frac{1}{N}\sum_{n=1}^{N}\int d{\bf u}\exp \left( -\lambda \beta
\sum_{l=1}^{N-1}\left( C_{l}+i\alpha D_{nl}\right) u_{l}\right)
\sum_{s=1}^{N}\left| r_{sn}\right|   \nonumber \\
&=&\frac{1}{N}\sum_{n=1}^{N}\left(
\sum_{s=1}^{n-1}sI_{n,s}^{1}+\sum_{s=n}^{N-1}(N-s)I_{n,s}^{1}\right)  
\nonumber \\
&=&\frac{1}{N\left( 2\pi G\beta m^{2}\mbox{ }\right) ^{N}}\sum_{n=1}^{N}%
\left[ \sum_{s=1}^{n-1}\frac{1}{\left( N-s-i\alpha \right) }+\sum_{s=n}^{N-1}%
\frac{1}{\left( s+i\alpha \right) }\right] \left( \prod_{l=1}^{N-1}\frac{1}{%
C_{l}+i\alpha D_{nl}}\right)   \nonumber \\
&=&\frac{1}{N\left( 2\pi G\beta m^{2}\mbox{ }\right) ^{N}}%
\sum_{n=1}^{N}\left\{ \left[ \Psi \left( 1-N+i\alpha \right) -\Psi \left(
n-N+i\alpha \right) +\Psi \left( N+i\alpha \right) -\Psi \left( n+i\alpha
\right) \right] \right.   \nonumber \\
&&\mbox{ \ \ \ \ \ \ \ \ \ \ \ \ \ \ \ \ \ \ \ \ \ \ \ \ \ \ \ \ \ \ \ }%
\left. \times \left( \frac{1}{\left( n-1\right) !(N-n)!}\frac{\Gamma \left(
N-n+1-i\alpha \right) }{\Gamma \left( N-i\alpha \right) }\frac{\Gamma \left(
n+i\alpha \right) }{\Gamma \left( N+i\alpha \right) }\right) \right\}  
\nonumber \\
&=&\frac{1}{N\left( 2\pi G\beta m^{2}\mbox{ }\right) ^{N}}%
\sum_{n=1}^{N}\left\{ \left[ \Psi \left( N-i\alpha \right) +\Psi \left(
N+i\alpha \right) -\Psi \left( N-n+1-i\alpha \right) -\Psi \left( n+i\alpha
\right) \right] \right.   \nonumber \\
&&\mbox{ \ \ \ \ \ \ \ \ \ \ \ \ \ \ \ \ \ \ \ \ \ \ \ \ \ \ \ \ \ \ \ }%
\left. \times \left( \frac{1}{\left( n-1\right) !(N-n)!}\frac{\Gamma \left(
N-n+1-i\alpha \right) }{\Gamma \left( N-i\alpha \right) }\frac{\Gamma \left(
n+i\alpha \right) }{\Gamma \left( N+i\alpha \right) }\right) \right\}  
\nonumber \\
&=&\frac{1}{N\left( 2\pi G\beta m^{2}\mbox{ }\right) ^{N}}\sum_{n=1}^{N}%
\left[ \left[ \Psi \left( N-i\alpha \right) +\Psi \left( N+i\alpha \right) %
\right] \left( \frac{1}{\left( n-1\right) !(N-n)!}\frac{\Gamma \left(
N-n+1-i\alpha \right) }{\Gamma \left( N-i\alpha \right) }\frac{\Gamma \left(
n+i\alpha \right) }{\Gamma \left( N+i\alpha \right) }\right) \right.  
\nonumber \\
&&\mbox{ \ \ \ \ \ \ \ \ \ \ \ \ \ \ \ \ \ \ \ \ \ \ \ \ \ \ \ \ \ \ \ \ \ \
\ }\left. -\frac{d}{d\sigma }\left( \frac{1}{\left( n-1\right) !(N-n)!}\frac{%
\Gamma \left( N-n+1-i\alpha +\sigma \right) }{\Gamma \left( N-i\alpha
\right) }\frac{\Gamma \left( n+i\alpha +\sigma \right) }{\Gamma \left(
N+i\alpha \right) }\right) \right] _{\sigma =0}  \nonumber \\
&=&\frac{1}{\left( 2\pi G\beta m^{2}\mbox{ }\right) ^{N}}\left[ \Psi \left(
N-i\alpha \right) +\Psi \left( N+i\alpha \right) \right] \frac{\Gamma \left(
1-i\alpha \right) }{\Gamma \left( N-i\alpha \right) }\frac{\Gamma \left(
1+i\alpha \right) }{\Gamma \left( N+i\alpha \right) }  \nonumber \\
&&\mbox{ \ \ \ \ \ \ \ \ \ \ \ \ \ \ \ \ \ \ \ \ \ \ \ \ \ \ \ \ \ \ \ \ \ }-%
\left[ \frac{1}{\Gamma \left( N+1\right) }\frac{d}{d\sigma }\left( \frac{%
\Gamma \left( N+1+2\sigma \right) \Gamma \left( 1-i\alpha +\sigma \right)
\Gamma \left( 1+i\alpha +\sigma \right) }{\Gamma \left( 2+2\sigma \right)
\Gamma \left( N-i\alpha \right) \Gamma \left( N+i\alpha \right) }\right) %
\right] _{\sigma =0}  \nonumber \\
&=&\frac{1}{\left( 2\pi G\beta m^{2}\mbox{ }\right) ^{N}}\frac{\Gamma \left(
1-i\alpha \right) }{\Gamma \left( N-i\alpha \right) }\frac{\Gamma \left(
1+i\alpha \right) }{\Gamma \left( N+i\alpha \right) }  \nonumber \\
&&\times \left[ \Psi \left( N-i\alpha \right) +\Psi \left( N+i\alpha \right)
-\Psi \left( 1-i\alpha \right) -\Psi \left( 1+i\alpha \right) -2\Psi \left(
N+1\right) +2\Psi \left( 2\right) \right]   \nonumber \\
&=&\frac{1}{\left( 2\pi G\beta m^{2}\mbox{ }\right) ^{N}}\frac{\Gamma \left(
1-i\alpha \right) }{\Gamma \left( N-i\alpha \right) }\frac{\Gamma \left(
1+i\alpha \right) }{\Gamma \left( N+i\alpha \right) }\left[ \sum_{s=1}^{N-1}%
\frac{1}{s-i\alpha }+\sum_{s=1}^{N-1}\frac{1}{s+i\alpha }-2\sum_{s=1}^{N-1}%
\frac{1}{s+1}\right]   \nonumber \\
&=&\frac{1}{\left( 2\pi G\beta m^{2}\mbox{ }\right) ^{N}}\frac{\Gamma \left(
1-i\alpha \right) }{\Gamma \left( N-i\alpha \right) }\frac{\Gamma \left(
1+i\alpha \right) }{\Gamma \left( N+i\alpha \right) }\left[ \sum_{s\neq
n}^{N-1}\frac{1}{s-i\alpha }+\sum_{s\neq n}^{N-1}\frac{1}{s+i\alpha }%
-2\sum_{s=1}^{N-1}\frac{1}{s+1}+\left\{ \frac{2n}{\alpha ^{2}+n^{2}}\right\} %
\right]   \label{appfz17}
\end{eqnarray}%
where the curly bracket contains the double-pole at $\alpha =in$, and all
other poles are in the same locations as before.

So we obtain 
\begin{eqnarray}
&&\frac{1}{N}\sum_{n=1}^{N}\int \frac{dk}{2\pi }\int d{\bf u}\exp \left(
-ikz-\lambda \beta \sum_{l=1}^{N-1}\left( C_{l}+i\alpha D_{nl}\right)
u_{l}\right) \left( \sum_{l=1}^{n-1}lu_{l}+\sum_{l=n}^{N-1}(N-l)u_{l}\right) 
\nonumber \\
&=&\frac{N\beta \lambda }{\left( 2\pi G\beta m^{2}\mbox{ }\right) ^{N}}\int 
\frac{d\alpha }{2\pi }e^{-i\alpha N\beta \lambda z}\frac{\Gamma \left(
1-i\alpha \right) }{\Gamma \left( N-i\alpha \right) }\frac{\Gamma \left(
1+i\alpha \right) }{\Gamma \left( N+i\alpha \right) }  \label{appfz18} \\
&&\times \left[ \sum_{s\neq n}^{N-1}\frac{1}{s-i\alpha }+\sum_{s\neq n}^{N-1}%
\frac{1}{s+i\alpha }-2\sum_{s=1}^{N-1}\frac{1}{s+1}+\left\{ \frac{2n}{\alpha
^{2}+n^{2}}\right\} \right]   \nonumber
\end{eqnarray}%
The total residue at $\alpha =in$ from (\ref{appfz18}) is 
\begin{eqnarray}
&&\frac{in\left( -1\right) ^{n}e^{N\beta \lambda nz}}{\Gamma \left(
N-n\right) \Gamma \left( N+n\right) }\left[ 2\sum_{s\neq n}^{N-1}\left( 
\frac{s}{s^{2}-n^{2}}\right) -2\sum_{s=1}^{N-1}\frac{1}{s+1}+\frac{1}{2n}%
\right.   \nonumber \\
&&\mbox{ \ \ \ \ \ \ \ \ \ }\left. +\left( \Psi \left( N+n\right) -\Psi
\left( N-n\right) \right) -\frac{1+N\beta \lambda nz}{n}\right]   \nonumber
\\
&=&\frac{-i\exp (N\beta \lambda nz)}{\left[ \Gamma (N)\right] ^{2}}\left(
C_{n}^{N}-\frac{1}{n}A_{n}^{N}\left( 1+N\beta \lambda nz\right) \right) 
\label{appfz19}
\end{eqnarray}%
provided $z<0$\ (otherwise it vanishes) and so 
\begin{eqnarray}
&&\frac{1}{N}\sum_{n=1}^{N}\int \frac{dk}{2\pi }\int d{\bf u}\exp \left(
-ikz-\lambda \beta \sum_{l=1}^{N-1}\left( C_{l}+i\alpha D_{nl}\right)
u_{l}\right) \left( \sum_{l=1}^{n-1}lu_{l}+\sum_{l=n}^{N-1}(N-l)u_{l}\right) 
\nonumber \\
&=&\frac{2N\beta \pi Gm^{2}}{\left( 2\pi G\beta m^{2}\mbox{ }\right) ^{N}%
\left[ \Gamma (N)\right] ^{2}}\sum_{l=1}^{N-1}\left[ C_{l}^{N}-\frac{1}{l}%
A_{l}^{N}\left( 1-2N\left( \beta \pi Gm^{2}\right) l\left| z\right| \right) %
\right] \exp \left( -2N\left( \beta \pi Gm^{2}\right) l\left| z\right|
\right) \mbox{ \ \ \ \ \ \ \ \ \ \ \ }  \label{appfz20}
\end{eqnarray}%
where 
\begin{equation}
C_{n}^{N}=\frac{n\left( -1\right) ^{n+1}\left[ \Gamma (N)\right] ^{2}}{%
\Gamma \left( N-n\right) \Gamma \left( N+n\right) }\left[ 2\sum_{s\neq
n}^{N-1}\left( \frac{s}{s^{2}-n^{2}}\right) -2\sum_{s=1}^{N-1}\frac{1}{s+1}+%
\frac{1}{2n}+\left( \Psi \left( N+n\right) -\Psi \left( N-n\right) \right) %
\right]   \label{appfz21}
\end{equation}

\bigskip

\bigskip Finally, we consider the expression 
\begin{eqnarray}
&&\frac{1}{N}\sum_{n=1}^{N}\int d{\bf u}\exp \left( -\lambda \beta
\sum_{l=1}^{N-1}\left( C_{l}+i\alpha D_{nl}\right) u_{l}\right)
\sum_{s=1}^{N-1}\sum_{t=s+1}^{N-1}\left( N-t\right) \left( t-s\right)
su_{s}u_{t}  \nonumber \\
&=&\frac{1}{N}\sum_{n=1}^{N}\sum_{s=1}^{N-1}\sum_{t=s+1}^{N-1}\left(
N-t\right) \left( t-s\right) s\;I_{n,s,t}^{2}  \nonumber \\
&=&\frac{1}{\left( 2\pi G\beta m^{2}\mbox{ }\right) ^{N+1}}\frac{1}{N}%
\sum_{n=1}^{N}\sum_{s=1}^{N-1}\sum_{t=s+1}^{N-1}\frac{\left( t-s\right) s}{%
C_{s}+i\alpha D_{ns}}\frac{\left( N-t\right) }{C_{t}+i\alpha D_{nt}}%
\prod_{l=1}^{N-1}\frac{1}{C_{l}+i\alpha D_{nl}}  \label{appfz22}
\end{eqnarray}%
Interchanging the order of the sums gives 
\begin{eqnarray}
&&\sum_{n=1}^{N}\sum_{s=1}^{N-1}\sum_{t=s+1}^{N-1}\frac{\left( t-s\right) s}{%
C_{s}+i\alpha D_{ns}}\frac{\left( N-t\right) }{C_{t}+i\alpha D_{nt}}%
\prod_{l=1}^{N-1}\frac{1}{C_{l}+i\alpha D_{nl}}  \nonumber \\
&=&\sum_{s=1}^{N-1}\sum_{t=s+1}^{N-1}\sum_{n=1}^{N}\left( \frac{\left(
t-s\right) s}{C_{s}+i\alpha D_{ns}}\frac{\left( N-t\right) }{C_{t}+i\alpha
D_{nt}}\frac{\left( \Gamma \left( N\right) \right) ^{2}}{\left( n-1\right)
!(N-n)!}\frac{\Gamma \left( N-n+1-i\alpha \right) }{\Gamma \left( N-i\alpha
\right) }\frac{\Gamma \left( n+i\alpha \right) }{\Gamma \left( N+i\alpha
\right) }\right)   \nonumber \\
&=&\frac{\left( \Gamma \left( N\right) \right) ^{2}}{\Gamma \left( N+i\alpha
\right) \Gamma \left( N-i\alpha \right) }\sum_{s=1}^{N-1}\sum_{t=s+1}^{N-1}%
\left[ \sum_{n=1}^{s}\left( \frac{\left( t-s\right) s}{\left( N-s\right)
\left( t+i\alpha \right) \left( s+i\alpha \right) }\frac{\Gamma \left(
N-n+1-i\alpha \right) }{\Gamma \left( N-n+1\right) }\frac{\Gamma \left(
n+i\alpha \right) }{\Gamma \left( n\right) }\right) \right.   \nonumber \\
&&+\sum_{n=s+1}^{t}\left( \frac{\left( t-s\right) }{\left( N-s-i\alpha
\right) \left( t+i\alpha \right) }\frac{\Gamma \left( N-n+1-i\alpha \right) 
}{\Gamma \left( N-n+1\right) }\frac{\Gamma \left( n+i\alpha \right) }{\Gamma
\left( n\right) }\right)   \nonumber \\
&&\left. +\sum_{n=t+1}^{N}\left( \frac{\left( N-t\right) \left( t-s\right) }{%
t\left( N-s-i\alpha \right) \left( N-t-i\alpha \right) }\frac{\Gamma \left(
N-n+1-i\alpha \right) }{\Gamma \left( N-n+1\right) }\frac{\Gamma \left(
n+i\alpha \right) }{\Gamma \left( n\right) }\right) \right]   \label{appfz23}
\end{eqnarray}%
When summed over $n$, we obtain 
\begin{equation}
\frac{1}{N}\sum_{n=1}^{N}\sum_{s=1}^{N-1}\sum_{t=s+1}^{N-1}\frac{\left(
t-s\right) s}{C_{s}+i\alpha D_{ns}}\frac{\left( N-t\right) }{C_{t}+i\alpha
D_{nt}}\prod_{l=1}^{N-1}\frac{1}{C_{l}+i\alpha D_{nl}}=\frac{%
\sum_{k=1}^{N-1}a_{k}\alpha ^{2k}}{\prod_{l=1}^{N-1}\left( \alpha
^{2}+l^{2}\right) ^{2}}  \label{appfz24}
\end{equation}%
which has double poles and single-pole residues at $\alpha =\pm in$ for
every nonzero value of $n<N$. \ The coefficients of the polynomial in the
numerator are calculable, but we have not found any closed-form expression
for them. The table below contains results for values up to $N=10$.

\begin{center}
. \ 
\begin{tabular}{ll}
$N$ & $\sum_{k=1}^{N-1}a_{k}\alpha ^{2k}$ \\ 
&  \\ 
$3$ & $2(a^{2}-1)(a^{2}-2)$ \\ 
$4$ & $4(180-109a^{2}+10a^{4}+11a^{6})$ \\ 
$5$ & $8(36576-3820a^{2}-75a^{4}+1590a^{6}+149a^{8})$ \\ 
$6$ & $48(5263200+1132124a^{2}+162455a^{4}+170877a^{6}+25445a^{8}+899a^{10})$
\\ 
$7$ & $%
288(1455926400+635262768a^{2}+123441248a^{4}+43494899a^{6}+6982689a^{8}+399833a^{10}+7163a^{12})
$ \\ 
$8$ & $\left\{ 
\begin{array}{c}
1152(1067349830400+638760596688a^{2}+149678407480a^{4}+34350170141a^{6}+5098185940a^{8}
\\ 
+351491854a^{10}+10505180a^{12}+110317a^{14})%
\end{array}%
\right. $ \\ 
$9$ & $\left\{ 
\begin{array}{c}
41472(143590977331200+103719257351424a^{2}+27508922447056a^{4}+5307728339928a^{6}
\\ 
+709200726957a^{8}+52255261172a^{10}+1966842318a^{12}+35272476a^{14}+237469a^{16})%
\end{array}%
\right. $ \\ 
$10$ & $\left\{ 
\begin{array}{c}
414720(108807366682828800+89224919703007488a^{2}+25843463092699920a^{4} \\ 
+4662765631167688a^{6}+573351004521465a^{8}+42828568506933a^{10}+1816855681610a^{12}
\\ 
+42140421618a^{14}+493478205a^{16}+2266273a^{18})%
\end{array}%
\right. $%
\end{tabular}
\end{center}

\bigskip

Hence we obtain 
\begin{eqnarray}
&&\frac{1}{N}\sum_{n=1}^{N}\int d{\bf u}\exp \left( -\lambda \beta
\sum_{l=1}^{N-1}\left( C_{l}+i\alpha D_{nl}\right) u_{l}\right)
\sum_{s=1}^{N-1}\sum_{t=s+1}^{N-1}\left( N-t\right) \left( t-s\right)
su_{s}u_{t}  \nonumber \\
&=&\frac{\mbox{2}\times 2N\beta \pi Gm^{2}}{\left( 2\pi G\beta m^{2}\mbox{ }%
\right) ^{N+1}\left[ \Gamma (N)\right] ^{2}}\sum_{l=1}^{N-1}\left[
D_{l}^{N}+K_{l}^{N}\left( 1-2N\left( \beta \pi Gm^{2}\right) l\left|
z\right| \right) \right] \exp \left( -2N\left( \beta \pi Gm^{2}\right)
l\left| z\right| \right)  \label{appfz25}
\end{eqnarray}
where the coefficients $D_{l}^{N}$ and $K_{l}^{N}$ are determined by the
residues given above. \ These are given in the next two tables.

\begin{center}
\begin{tabular}{ccccccccc}
$D_{l}^{N}$ & $N=3$ & $N=4$ & $N=5$ & $N=6$ & $N=7$ & $N=8$ & $N=9$ & $N=10$
\\ 
$l=1$ & $-\frac{1}{3}$ & $-\frac{1}{4}$ & $-\frac{2}{45}$ & $\frac{31}{144}$
& $\frac{2117}{4200}$ & $-\frac{2917}{3600}$ & $\frac{24869}{22050}$ & $%
\frac{113931}{78400}$ \\ 
$l=2$ & $\frac{11}{12}$ & $\frac{209}{125}$ & $\frac{169}{90}$ & $\frac{5599%
}{3087}$ & $-\frac{2131}{1344}$ & $\frac{30347}{24300}$ & $\frac{941}{1125}$
& $\frac{601906}{1630475}$ \\ 
$l=3$ &  & $-\frac{1387}{1500}$ & $-\frac{9278}{5145}$ & $-\frac{12781}{5488}
$ & $-\frac{1468}{567}$ & $-\frac{85427}{32400}$ & $-\frac{836537}{332750}$
& $-\frac{3601223}{1597200}$ \\ 
$l=4$ &  &  & $\frac{72917}{123480}$ & $-\frac{431581}{333396}$ & $\frac{%
214607}{113400}$ & $\frac{18987467}{8085285}$ & $\frac{3170471}{1197900}$ & $%
\frac{407903579}{146210350}$ \\ 
$l=5$ &  &  &  & $-\frac{1005251}{3333960}$ & $-\frac{22515781}{30187080}$ & 
$-\frac{31621187}{25874640}$ & $-\frac{440479867}{263178630}$ & $-\frac{%
4721124703}{2292578288}$ \\ 
$l=6$ &  &  &  &  & $\frac{162182479}{1207483200}$ & $\frac{8830831883}{%
23686076700}$ & $\frac{2430681577}{3582153575}$ & $\frac{10937730746}{%
10746460725}$ \\ 
$l=7$ &  &  &  &  &  & $-\frac{108510757181}{1989630442800}$ & $-\frac{%
7762820713}{46056260250}$ & $-\frac{22140852323}{65502236800}$ \\ 
$l=8$ &  &  &  &  &  &  & $\frac{53405900137}{2579150574000}$ & $\frac{%
9952938913257}{140792964111800}$ \\ 
$l=9$ &  &  &  &  &  &  &  & $-\frac{151474036840183}{20274186832099200}$%
\end{tabular}

\bigskip

\begin{tabular}{ccccccccc}
$K_{l}^{N}$ & $N=3$ & $N=4$ & $N=5$ & $N=6$ & $N=7$ & $N=8$ & $N=9$ & $N=10$
\\ 
$l=1$ & $-\frac{1}{3}$ & $-\frac{1}{2}$ & $-\frac{3}{5}$ & $-\frac{2}{3}$ & $%
-\frac{5}{7}$ & $-\frac{3}{4}$ & $-\frac{7}{9}$ & $-\frac{4}{5}$ \\ 
$l=2$ & $-\frac{5}{24}$ & $-\frac{1}{25}$ & $\frac{1}{10}$ & $\frac{31}{147}$
& $\frac{67}{224}$ & $\frac{10}{27}$ & $\frac{193}{450}$ & $\frac{289}{605}$
\\ 
$l=3$ &  & $\frac{7}{50}$ & $-\frac{27}{245}$ & $\frac{43}{784}$ & $-\frac{11%
}{1512}$ & $-\frac{37}{540}$ & $-\frac{381}{3025}$ & $-\frac{1297}{7260}$ \\ 
$l=4$ &  &  & $-\frac{257}{3920}$ & $-\frac{829}{10584}$ & $-\frac{361}{5040}
$ & $-\frac{853}{16335}$ & $-\frac{551}{21780}$ & $\frac{1153}{204490}$ \\ 
$l=5$ &  &  &  & $-\frac{2761}{10584}$ & $\frac{12431}{304920}$ & $\frac{3163%
}{65340}$ & $\frac{44893}{920205}$ & $\frac{24587}{572572}$ \\ 
$l=6$ &  &  &  &  & $-\frac{34541}{3659040}$ & $-\frac{50012}{2760615}$ & $-%
\frac{368191}{1431430}$ & $-\frac{66629}{2147145}$ \\ 
$l=7$ &  &  &  &  &  & $\frac{248029}{77297220}$ & $\frac{234797}{32207175}$
& $\frac{274399}{22902880}$ \\ 
$l=8$ &  &  &  &  &  &  & $-\frac{1075199}{1030629600}$ & $-\frac{1812235}{%
661893232}$ \\ 
$l=9$ &  &  &  &  &  &  &  & $\frac{6514549}{19856796960}$%
\end{tabular}%
\bigskip 
\end{center}

\bigskip Summarizing:

\begin{eqnarray}
&&\frac{1}{N}\sum_{n=1}^{N}\int \frac{dk}{2\pi }\int d{\bf u\;}e^{\left(
-ikz-\lambda \beta \sum_{l=1}^{N-1}\left( C_{l}+i\alpha D_{nl}\right)
u_{l}\right) }  \nonumber \\
&=&\frac{2N\beta \pi Gm^{2}}{\left( 2\pi G\beta m^{2}\mbox{ }\right) ^{N-1}%
\left[ \Gamma (N)\right] ^{2}}\sum_{l=1}^{N-1}A_{l}^{N}\exp \left( -2N\left(
\beta \pi Gm^{2}\right) l\left| z\right| \right)   \label{appfz26} \\
&&\frac{1}{N}\sum_{n=1}^{N}\int \frac{dk}{2\pi }\int d{\bf u}\left(
\sum_{s=1}^{N-1}s\left( N-s\right) u_{s}\right) {\bf \;}e^{\left(
-ikz-\lambda \beta \sum_{l=1}^{N-1}\left( C_{l}+i\alpha D_{nl}\right)
u_{l}\right) }  \nonumber \\
\mbox{ \ \ \ \ } &=&\frac{2N\beta \pi Gm^{2}}{\left( 2\pi G\beta m^{2}\mbox{ 
}\right) ^{N}\left[ \Gamma (N)\right] ^{2}}\sum_{l=1}^{N-1}\left[
B_{l}^{N}-A_{l}^{N}\left( 1-2N\left( \beta \pi Gm^{2}\right) l\left|
z\right| \right) \right] \exp \left( -2N\left( \beta \pi Gm^{2}\right)
l\left| z\right| \right)   \label{appfz27} \\
&&\frac{1}{N}\sum_{n=1}^{N}\int \frac{dk}{2\pi }\int d{\bf u}\left(
\sum_{s=1}^{n-1}sI_{n,s}^{1}+\sum_{s=n}^{N-1}(N-s)I_{n,s}^{1}\right) {\bf \;}%
e^{\left( -ikz-\lambda \beta \sum_{l=1}^{N-1}\left( C_{l}+i\alpha
D_{nl}\right) u_{l}\right) }  \nonumber \\
\mbox{ \ \ \ \ } &=&\frac{2N\beta \pi Gm^{2}}{\left( 2\pi G\beta m^{2}\mbox{ 
}\right) ^{N}\left[ \Gamma (N)\right] ^{2}}\sum_{l=1}^{N-1}\left[ C_{l}^{N}-%
\frac{1}{l}A_{l}^{N}\left( 1-2N\left( \beta \pi Gm^{2}\right) l\left|
z\right| \right) \right] \exp \left( -2N\left( \beta \pi Gm^{2}\right)
l\left| z\right| \right) \mbox{ \ \ \ \ }  \label{appfz28} \\
&&\frac{1}{N}\sum_{n=1}^{N}\int \frac{dk}{2\pi }\int d{\bf u}\left(
\sum_{s=1}^{N-1}\sum_{t=s+1}^{N-1}\left( N-t\right) \left( t-s\right)
su_{s}u_{t}\right) {\bf \;}e^{\left( -ikz-\lambda \beta
\sum_{l=1}^{N-1}\left( C_{l}+i\alpha D_{nl}\right) u_{l}\right) }  \nonumber
\\
\mbox{ \ \ \ \ } &=&\frac{2N\beta \pi Gm^{2}}{\left( 2\pi G\beta m^{2}\mbox{ 
}\right) ^{N+1}\left[ \Gamma (N)\right] ^{2}}\sum_{l=1}^{N-1}\left[
D_{l}^{N}+K_{l}^{N}\left( 1-2N\left( \beta \pi Gm^{2}\right) l\left|
z\right| \right) \right] \exp \left( -2N\left( \beta \pi Gm^{2}\right)
l\left| z\right| \right) \mbox{ \ \ \ \ \ }  \label{appfz29}
\end{eqnarray}
.

The final expression for the one-particle distribution function is 
\begin{eqnarray}
&&f_{cn}(p,z)  \nonumber \\
&=&\sqrt{N}\left( \sqrt{2\pi }G/c^{3}\right) ^{\left( N-1\right) }\left[
\left( N-1\right) !\right] ^{2}\left( \frac{2\pi m}{\beta }\right) ^{\left(
N-2\right) /2}\frac{2N\beta \pi Gm^{2}\left( \beta mc^{2}\right) ^{\frac{%
3\left( N-1\right) }{2}}}{\sqrt{N-1}\left( 2\pi G\beta m^{2}\mbox{ }\right)
^{N-1}\left[ \Gamma (N)\right] ^{2}}  \nonumber \\
&&\times \exp \left[ -\frac{1}{\beta mc^{2}}\left\{ \frac{3\left( N-1\right)
^{2}}{8N}\lambda _{1}-\frac{\left( N-1\right) \left( \lambda _{2}\left(
N-1\right) +\lambda _{3}\right) }{N}-\lambda
_{4}\sum_{k=1}^{N-1}\sum_{l=k+1}^{N-1}\frac{\left( l-k\right) }{l\left(
N-k\right) }\right\} \right]   \nonumber \\
&&\times \sum_{l=1}^{N-1}\left[ \left\{ A_{l}^{N}\left( 1+\frac{\lambda _{1}%
}{2\beta mc^{2}}\left( \frac{\beta ^{2}p^{4}\left( 1+\left( N-1\right)
^{3}\right) }{\left( 2m\right) ^{2}\left( N-1\right) ^{3}}+\frac{3\left(
N-2\right) \beta p^{2}}{2m\left( N-1\right) ^{2}}+\frac{3\left( N-2\right)
^{2}}{4\left( N-1\right) }\right) \right) \right. \right.   \nonumber \\
&&\mbox{ \ \ \ \ }+\frac{1}{\beta mc^{2}}\left( \frac{\left( \lambda
_{3}-\lambda _{2}\right) \beta p^{2}}{m\left( N-1\right) ^{2}}-\frac{\left(
\lambda _{3}+(N-2)\lambda _{2}\right) }{\left( N-1\right) }\right) \left[
B_{l}^{N}-A_{l}^{N}\left( 1-2N\left( \beta \pi Gm^{2}\right) l\left|
z\right| \right) \right]   \nonumber \\
&&+\frac{2\left( 2\lambda _{3}+(N-2)\lambda _{2}\right) }{\beta mc^{2}}%
\left( \frac{1}{2\left( N-1\right) }-\frac{N\beta p^{2}}{2m}\left[ \frac{1}{%
\left( N-1\right) ^{2}}\right] \right) \mbox{\ }\left[ C_{l}^{N}-\frac{1}{l}%
A_{l}^{N}\left( 1-2N\left( \beta \pi Gm^{2}\right) l\left| z\right| \right) %
\right]   \nonumber \\
&&\mbox{ \ \ \ \ \ \ \ \ \ \ \ \ }\left. \left. -\frac{\lambda _{4}}{\beta
mc^{2}}\left[ D_{l}^{N}+K_{l}^{N}\left( 1-2N\left( \beta \pi Gm^{2}\right)
l\left| z\right| \right) \right] \right\} \exp \left( -\frac{N\beta p^{2}}{%
2m\left( N-1\right) }-2\pi GN\beta m^{2}l\left| z\right| \right) \right]  
\nonumber \\
&=&\frac{\left( 2\pi Gm^{2}\right) \left( N\beta \right) ^{3/2}}{\sqrt{2\pi
m\left( N-1\right) }}  \nonumber \\
&&\times \exp \left[ -\frac{1}{\beta mc^{2}}\left\{ \frac{3\left( N-1\right)
^{2}}{8N}\lambda _{1}-\frac{\left( N-1\right) \left( \lambda _{2}\left(
N-1\right) +\lambda _{3}\right) }{N}-\lambda
_{4}\sum_{k=1}^{N-1}\sum_{l=k+1}^{N-1}\frac{\left( l-k\right) }{l\left(
N-k\right) }\right\} \right]   \nonumber \\
&&\mbox{ \ }\times \sum_{l=1}^{N-1}\left[ \left\{ A_{l}^{N}\left( 1+\frac{%
\lambda _{1}}{2\beta mc^{2}}\left( \frac{\beta ^{2}p^{4}\left( 1+\left(
N-1\right) ^{3}\right) }{\left( 2m\right) ^{2}\left( N-1\right) ^{3}}+\frac{%
3\left( N-2\right) \beta p^{2}}{2m\left( N-1\right) ^{2}}+\frac{3\left(
N-2\right) ^{2}}{4\left( N-1\right) }\right) \right) \right. \right.  
\nonumber \\
&&\mbox{\ }+\frac{1}{\beta mc^{2}}\left( \frac{\left( \lambda _{3}-\lambda
_{2}\right) \beta p^{2}}{m\left( N-1\right) ^{2}}-\frac{\left( \lambda
_{3}+(N-2)\lambda _{2}\right) }{\left( N-1\right) }\right) \left[
B_{l}^{N}-A_{l}^{N}\left( 1-2N\left( \beta \pi Gm^{2}\right) l\left|
z\right| \right) \right]   \nonumber \\
&&+\frac{2\left( 2\lambda _{3}+(N-2)\lambda _{2}\right) }{\beta mc^{2}}%
\left( \frac{1}{2\left( N-1\right) }-\frac{N\beta p^{2}}{2m}\left[ \frac{1}{%
\left( N-1\right) ^{2}}\right] \right) \mbox{\ }\left[ C_{l}^{N}-\frac{1}{l}%
A_{l}^{N}\left( 1-2N\left( \beta \pi Gm^{2}\right) l\left| z\right| \right) %
\right]   \nonumber \\
&&\mbox{ \ \ }\left. \left. -\frac{\lambda _{4}}{\beta mc^{2}}\left[
D_{l}^{N}+K_{l}^{N}\left( 1-2N\left( \beta \pi Gm^{2}\right) l\left|
z\right| \right) \right] \right\} \exp \left( -\frac{N\beta p^{2}}{2m\left(
N-1\right) }-2\pi GN\beta m^{2}l\left| z\right| \right) \right]   \nonumber
\\
&&  \label{appfz30}
\end{eqnarray}

The canonical density distribution function is given by integration of $\
f_{cn}(p,z)$ over $p$ 
\begin{eqnarray}
\rho _{c}\left( z\right)  &=&\int_{-\infty }^{\infty }dpf_{cn}(p,z) 
\nonumber \\
&=&\frac{\left( 2\pi Gm^{2}\right) \left( N\beta \right) ^{3/2}}{\sqrt{2\pi
m\left( N-1\right) }}\sqrt{\frac{2\pi m(N-1)}{N\beta }}  \nonumber \\
&&\times \exp \left[ -\frac{1}{\beta mc^{2}}\left\{ \frac{3\left( N-1\right)
^{2}}{8N}\lambda _{1}-\frac{\left( N-1\right) \left( \lambda _{2}\left(
N-1\right) +\lambda _{3}\right) }{N}-\lambda
_{4}\sum_{k=1}^{N-1}\sum_{l=k+1}^{N-1}\frac{\left( l-k\right) }{l\left(
N-k\right) }\right\} \right]   \nonumber \\
&&\mbox{ \ }\times \sum_{l=1}^{N-1}\left[ \left\{ A_{l}^{N}\left( 1+\frac{%
\lambda _{1}}{2\beta mc^{2}}\left( \frac{3\left( 1+\left( N-1\right)
^{3}\right) }{4N^{2}\left( N-1\right) }+\frac{3\left( N-2\right) }{2N\left(
N-1\right) }+\frac{3\left( N-2\right) ^{2}}{4\left( N-1\right) }\right)
\right) \right. \right.   \nonumber \\
&&+\frac{1}{\beta mc^{2}}\left( \frac{\left( \lambda _{3}-\lambda
_{2}\right) }{N\left( N-1\right) }-\frac{\left( \lambda _{3}+(N-2)\lambda
_{2}\right) }{\left( N-1\right) }\right) \left[ B_{l}^{N}-A_{l}^{N}\left(
1-2N\left( \beta \pi Gm^{2}\right) l\left| z\right| \right) \right]  
\nonumber \\
&&+\frac{2\left( 2\lambda _{3}+(N-2)\lambda _{2}\right) \pi Gm}{\beta mc^{2}}%
\left( \frac{1}{2\left( N-1\right) }-\left[ \frac{1}{2\left( N-1\right) }%
\right] \right) \mbox{\ }\left[ C_{l}^{N}-\frac{1}{l}A_{l}^{N}\left(
1-2N\left( \beta \pi Gm^{2}\right) l\left| z\right| \right) \right]  
\nonumber \\
&&\mbox{\ \ }\left. \left. -\frac{\lambda _{4}}{\beta mc^{2}}\left[
D_{l}^{N}+K_{l}^{N}\left( 1-2N\left( \beta \pi Gm^{2}\right) l\left|
z\right| \right) \right] \right\} \exp \left( -2\pi GN\beta m^{2}l\left|
z\right| \right) \right]   \nonumber \\
&=&\left( 2\pi Gm^{2}N\beta \right)   \nonumber \\
&&\times \exp \left[ -\frac{1}{\beta mc^{2}}\left\{ \frac{3\left( N-1\right)
^{2}}{8N}\lambda _{1}-\frac{\left( N-1\right) \left( \lambda _{2}\left(
N-1\right) +\lambda _{3}\right) }{N}-\lambda
_{4}\sum_{k=1}^{N-1}\sum_{l=k+1}^{N-1}\frac{\left( l-k\right) }{l\left(
N-k\right) }\right\} \right]   \nonumber \\
&&\times \sum_{l=1}^{N-1}\left\{ A_{l}^{N}+\frac{1}{\beta mc^{2}}\left( 
\frac{3\lambda _{1}}{8}\frac{\left( N-1\right) ^{2}}{N}A_{l}^{N}-\left( 
\frac{\left( \lambda _{2}\left( N-1\right) +\lambda _{3}\right) }{N}\right)
B_{l}^{N}-\lambda _{4}D_{l}^{N}\right) \right.   \nonumber \\
&&\left. +\frac{1}{\beta mc^{2}}\left[ \left( \frac{\left( \lambda
_{2}\left( N-1\right) +\lambda _{3}\right) }{N}\right) A_{l}^{N}-\lambda
_{4}K_{l}^{N}\right] \left( 1-2N\left( \beta \pi Gm^{2}\right) l\left|
z\right| \right) \right\} \exp \left( -2\pi Gm^{2}N\beta l\left| z\right|
\right)   \nonumber \\
&&  \label{appfz31}
\end{eqnarray}%
It is straightforward to show that the coefficients $A_{l}^{N}$ and $%
B_{l}^{N}$ obey the sum rules 
\begin{eqnarray}
\sum_{l=1}^{N-1}A_{l}^{N} &=&\frac{1}{2}\frac{N-1}{2N-3}  \label{Asumrule} \\
\sum_{l=1}^{N-1}B_{l}^{N} &=&\frac{1}{2}\frac{\left( N-1\right) ^{2}}{2N-3}
\label{Bsumrule}
\end{eqnarray}%
where eq. (\ref{Asumrule}) was previously derived in the $c\rightarrow
\infty $ limit \cite{Rybicki} . Since $\int_{-\infty }^{\infty }dz$ $\rho
_{c}\left( z\right) =1$, we must have 
\begin{eqnarray}
&&2\sum_{l=1}^{N-1}\frac{1}{l}\left\{ A_{l}^{N}+\frac{1}{\beta mc^{2}}\left( 
\frac{3\lambda _{1}}{8}\frac{\left( N-1\right) ^{2}}{N}A_{l}^{N}-\frac{\mbox{%
2}}{2}\left( \frac{\left( \lambda _{2}\left( N-1\right) +\lambda _{3}\right) 
}{N}\right) B_{l}^{N}-\lambda _{4}D_{l}^{N}\right) \right\}   \nonumber \\
&=&\exp \left[ \frac{1}{\beta mc^{2}}\left\{ \frac{3\left( N-1\right) ^{2}}{%
8N}\lambda _{1}-\frac{\left( N-1\right) \left( \lambda _{2}\left( N-1\right)
+\lambda _{3}\right) }{N}-\lambda _{4}\sum_{k=1}^{N-1}\sum_{l=k+1}^{N-1}%
\frac{\left( l-k\right) }{l\left( N-k\right) }\right\} \right] 
\label{appfz32}
\end{eqnarray}%
or alternatively, to the relevant order in $c$, 
\begin{eqnarray}
\sum_{l=1}^{N-1}\frac{1}{l}A_{l}^{N} &=&\frac{1}{2}  \label{Alsum} \\
\sum_{l=1}^{N-1}\frac{1}{l}B_{l}^{N} &=&\frac{\mbox{1}}{\mbox{2}}\left(
N-1\right)   \label{Blsum} \\
\sum_{l=1}^{N-1}\frac{1}{l}D_{l}^{N} &=&\frac{1}{2}\sum_{k=1}^{N-1}%
\sum_{l=k+1}^{N-1}\frac{\left( l-k\right) }{l\left( N-k\right) }
\label{Dlsum}
\end{eqnarray}%
each of which can be straightforwardly verified. We also can show 
\begin{eqnarray}
\sum_{l=1}^{N-1}\frac{1}{l}C_{l}^{N} &=&C_{1}^{N}=\frac{\left( N-1\right) }{N%
}  \label{Clsum} \\
\sum_{l=1}^{\infty }C_{l}^{N} &=&\frac{\pi ^{2}}{12}  \label{Csuminf}
\end{eqnarray}

\bigskip

\hspace*{5cm}--------------------------

\end{document}